\documentclass[11pt]{article}
\usepackage{pix}
\usepackage{epsfig}
\usepackage{amsfonts}
\usepackage{amsmath,amssymb}
\usepackage{bbm,bm}
\usepackage{caption}
\usepackage{subcaption}
\usepackage{multirow}
\usepackage{cite}
\allowdisplaybreaks[1]

\usepackage{tikz}
\usetikzlibrary{shapes,arrows,shadows,automata,positioning,arrows.meta,decorations.markings}
\newdimen\nodeDist
\usepackage{comment}
\usepackage[normalem]{ulem} 

\usepackage[
  colorlinks=true,
  linkcolor={red!50!black},
  citecolor={blue!50!black},
  filecolor=black,
  urlcolor=black,
  breaklinks=true
  ]{hyperref}
\usepackage{orcidlink}

\newcommand{\newnewtext}[1]{{\color{blue} #1}}
\newcommand{\vtwo}[1]{#1}
\newcommand{\vthree}[1]{#1}
\newcommand{\vfour}[1]{#1}

\renewcommand{\vec}[1]{{\bf #1}}

\renewcommand{\nn}{\nonumber \\}
\renewcommand{\rmi}[1]{{\mbox{\scriptsize #1}}}
\newcommand{\rmii}[1]{{\mbox{\tiny\rm{#1}}}}

\newcommand{\mD}{m_\rmii{D}}
\newcommand{\mh}{m_\rmii{$h$}}
\newcommand{\mA}{m_\rmii{$A$}}

\newcommand{\mH}{m_\rmii{$H$}}
\newcommand{\mHpm}{m_\rmii{$H^\pm$}}
\newcommand{\gY}{y_t}

\newcommand{\Gf}{G_f}
\newcommand{\alphas}{\alpha_\rmi{s}}
\newcommand{\LamdUS}{\bar{\mu}_{\rmii{3US}}}
\newcommand{\Lamd}{\bar{\mu}_{\rmii{3}}}
\newcommand{\LamD}{\bar{\mu}_{\rmii{4}}}

\newcommand{\mW}{m_\rmii{$W$}}
\newcommand{\mZ}{m_\rmii{$Z$}}
\newcommand{\mt}{m_t}

\newcommand{\gammaE}{\gamma_{\rmii{E}}}

\newcommand{\Tc}{T_{\rm c}}
\newcommand{\Tn}{T_{\rm n}}
\newcommand{\Tp}{T_p}

\newcommand{\vw}{v_{w}}
\newcommand{\vCJ}{v_\rmii{CJ}}
\newcommand{\cs}{c_{s}}

\newcommand{\vc}{v_{\rm c}}

\newcommand{\Veff}{V_{\rmi{eff}}}
\def\backtick{\char18}

\newcommand{\Tint}[1]{{\hbox{$\sum$}\!\!\!\!\!\!\!\int\,}_{\!\!\!\!\raise-0.9ex\hbox{$\scriptstyle{#1}$}}}
\newcommand{\Tinti}[1]{{{\Sigma}\!\!\!\!\raise0.3ex\hbox{$\int$}_\rmii{${#1}$}}}
\newcommand{\Tintip}[1]{{{\Sigma'}\!\!\!\!\!\raise0.3ex\hbox{$\int$}_\rmii{${#1}$}}}

\newcommand\MSbar{$\overline{\rm MS}$}
\newcommand{\downmapsto}{\rotatebox[origin=c]{-90}{$\mapsto$}\mkern2mu}

\def\scfc{0.7}  
\def\phgt{21}   
\def\pwc{21}    
\def\pwcb{31.5} 
\def\pwcc{42} 
\newcommand{\PIC}[4]{\;\parbox[c]{#2 pt}{\begin{picture}(#2,#3)(0,0)
\SetWidth{1.0}\SetScale{#4} #1 \end{picture}}\;}
\renewcommand{\pic}[1]{\PIC{#1}{\pwc}{\phgt}{\scfc}}
\renewcommand{\picb}[1]{\PIC{#1}{\pwcb}{\phgt}{\scfc}}
\renewcommand{\picc}[1]{\PIC{#1}{\pwcc}{\phgt}{\scfc}}

\makeatletter \@addtoreset{equation}{section} \makeatother
\renewcommand{\theequation}{\arabic{section}.\arabic{equation}}
\makeatletter
\renewcommand\section{\@startsection{section}{1}{\z@}%
  {-5.5ex \@plus -1ex \@minus -.2ex}
  {2.3ex \@plus.2ex}%
  {\normalfont\large\bfseries}}
\renewcommand\subsection{\@startsection{subsection}{2}{\z@}%
  {-3.25ex\@plus -1ex \@minus -.2ex}%
  {1.5ex \@plus .2ex}%
  {\normalfont\normalsize\bfseries}}
\renewcommand\thesection{\@arabic\c@section}
\renewcommand\thesubsection{\thesection.\@arabic\c@subsection}
\renewcommand{\@seccntformat}[1]{%
  \csname the#1\endcsname.\hspace{1.0em}}
\makeatother

\begin{document}

\flushbottom

\begin{titlepage}

\begin{flushright}
IPPP/25/39
\end{flushright}
\begin{centering}

\vfill

{\Large{\bf
Interpreting
the 95~GeV resonance\\
in
the Two Higgs Doublet Model\\[2mm]
\large Implications for the Electroweak Phase Transition
}}

\vspace{0.8cm}

\renewcommand{\thefootnote}{\fnsymbol{footnote}}
Ansh Bhatnagar%
\orcidlink{0000-0003-2825-6350}%
,$^{\rm a,}$%
\footnote{ansh.bhatnagar@durham.ac.uk}
Djuna Croon%
\orcidlink{0000-0003-3359-3706},%
$^{\rm a,}$%
\footnote{djuna.l.croon@durham.ac.uk}
and
Philipp Schicho%
\orcidlink{0000-0001-5869-7611} %
$^{\rm b,}$%
\footnote{philipp.schicho@unige.ch}

\vspace{0.8cm}

$^\rmi{a}$%
{\em
Institute for Particle Physics Phenomenology, Department of Physics,
\\
Durham University, Durham DH1 3LE, U.K.}

\vspace{0.3cm}

$^\rmi{b}$%
{\em
  D\'epartement de Physique Th\'eorique, Universit\'e de Gen\`eve,\\
  24 quai Ernest Ansermet, CH-1211 Gen\`eve 4, Switzerland
}

\vspace*{0.8cm}

\mbox{\bf Abstract}

\end{centering}

\vspace*{0.3cm}

\noindent

We investigate if the recent mass resonance excesses seen around 95~GeV at
the Large Hadron Collider (LHC) can
be reconciled
with a first-order electroweak phase transition.
Performing the first large-scale parameter scan of
the \vtwo{Type~I} Two Higgs Doublet Model (2HDM) using high-temperature
dimensionally reduced effective field theory,
we focus on regions of parameter space consistent with interpreting
the excess as an additional pseudoscalar state.
We find that, in contrast to the Standard Model,
the electroweak transition pattern in the 2HDM is generically first-order,
proceeding either in a single or in two steps.
While transition strengths can reach up to
$\vc/\Tc \sim 1.3$,
\vfour{the viable, collider-constrained parameter space
yields $\vc/\Tc \lesssim 1$.
Thus,}
the gravitational wave signals lie below the projected reach of
future interferometer experiments and
are likely insufficient to support successful electroweak baryogenesis.

\vfill
\end{titlepage}

{\hypersetup{hidelinks}
\tableofcontents
}
\clearpage

\renewcommand{\thefootnote}{\arabic{footnote}}
\setcounter{footnote}{0}

%

\section{Introduction}
\label{sec:intro}

Following the discovery of the Higgs boson~\cite{ATLAS:2012yve,CMS:2012qbp},
searches at the Large Hadron Collider (LHC) have increasingly focused on exploring
the structure of the Higgs sector.
Motivated by numerous Beyond the Standard Model (BSM) scenarios featuring extended scalar sectors,
the CMS collaboration extended its Higgs-like particle searches to include
invariant masses below 110~GeV.
While CMS
reported an excess near 95~GeV in the diphoton channel
by combining data from 8 and 13~TeV runs
with a local significance of $2.8\sigma$~\cite{CMS:2018cyk},
recently, this result was updated using the full 13~TeV dataset.
The latter shifted
the excess to 95.4~GeV with a local significance of $2.9\sigma$~\cite{CMS:2024yhz}.
The presence of a neutral scalar decaying into two photons around 95~GeV remains
compatible with the latest ATLAS results~\cite{ATLAS-CONF-2023-035}.

A resonance of a similar mass has also been reported by CMS in
$\tau\tau$ final state searches at
around 100~GeV~\cite{CMS:2022goy}, and
around 98~GeV in $b\bar{b}$ final state searches at
the Large Electron-Positron (LEP) collider in 2006~\cite{LEP_2006}.

Due to the limited resolution of the CMS and LEP measurements, these measurements seem to be compatible and could point to a new scalar particle at a mass of around 95~GeV.
The possibility of a lighter Higgs-like particle explaining these excesses has been
explored in numerous models (e.g. ~\cite{Belyaev:2023xnv,Cao:2016uwt,Heinemeyer:2021msz,Biekotter:2019kde,Biekotter:2021ovi,Biekotter:2021qbc,Biekotter:2022jyr,Biekotter:2022abc,Biekotter:2023jld,Biekotter:2023oen,Cao:2019ofo,Iguro:2022dok,Li:2022etb,Cline:2019okt,Crivellin:2017upt,Cacciapaglia:2016tlr,Abdelalim:2020xfk,Azevedo:2023zkg,Benbrik:2024ptw,Arhrib:2025pxy,Aguilar-Saavedra:2023tql,Belyaev:2024lah,Banik:2024ugs,Lian:2024smg,Cao:2023gkc,Cao:2024axg,Coutinho:2024zyp}).
A model that has been studied extensively in the context of new Higgs-like particles is
the Two Higgs Doublet Model (2HDM),
a minimal extension to the Standard Model that requires the addition of
an extra Higgs doublet~\cite{%
  Heinemeyer:2021msz,Biekotter:2019kde,Biekotter:2021qbc,Aguilar-Saavedra:2023tql,
  Belyaev:2024lah,Azevedo:2023zkg,Cacciapaglia:2016tlr,Arhrib:2025pxy,Benbrik:2024ptw,
  Goncalves:2021egx,Su:2020pjw
  }.
This model predicts additional scalar particles,
which may account for a 95~GeV resonance.
\vtwo{While the 2HDM can reproduce the CMS diphoton and ditau excesses through a pseudoscalar near 95~GeV, the preferred parameter region (with light charged scalars and small $\tan\beta$) is in tension with flavour observables, particularly the measured branching ratio of $b \to s\gamma$ at about the $2.5\sigma$ level \cite{Deschamps:2009rh,Mahmoudi:2009zx,Hermann:2012fc,Misiak:2015xwa,Misiak:2017bgg,Misiak:2020vlo}.
However, the 2HDM remains a particularly economical framework for interpreting a possible 95~GeV scalar resonance.}
Phenomenological investigations of the 2HDM in this context have largely
focused on collider observables~\cite{Heinemeyer:2021msz,Belyaev:2024lah},
the existence of new scalars coupled to
the Higgs can also alter the dynamics of cosmological evolution.

In this work, we explore the possibility of
a first-order electroweak phase transition (EWPT) in
the Type~I 2HDM model,
identifying the 95~GeV excess with an additional pseudoscalar state.
Employing state-of-the-art dimensional reduction to
a three-dimensional effective field theory (3D~EFT)~\cite{Kajantie:1995dw,Braaten:1995cm},
we perform a broad
finite-temperature scan over
the parameter region compatible with current collider limits.

Previous studies have explored the EWPT in
the 2HDM using both perturbative and non-perturbative approaches.
Early work using
one-loop finite-temperature effective potentials
(e.g.~\cite{%
  Cline:1995dg,Fromme:2006cm,Ivanov:2008er,Ginzburg:2009dp,Dorsch:2013wja,
  Basler:2016obg,Dorsch:2016nrg,
  Goncalves:2023svb,Su:2020pjw
  })
showed that
a strong first-order EWPT is possible in Type-I and Type-II 2HDMs,
typically requiring sizable mass splittings among the scalar states to enhance thermal barriers generated by gauge bosons or scalar loops
(e.g.~\cite{Cline:1995dg,Dorsch:2013wja}).
These studies often focused on parameter regions with heavy new scalars
with $M \gtrsim 300$~GeV,
motivated by electroweak precision tests~\cite{Grimus:2007if,Haber:2010bw,Funk:2011ad,Haber:1999zh,Jung:2010ik} and Higgs signal strength measurements.

More recently, non-perturbative studies of the dimensionally reduced 3D~EFT have been
employed to assess the nature of the electroweak phase transition more reliably;
see e.g.~\cite{Andersen:2017ika,Helset:2017esj,Gorda:2018hvi,Kainulainen:2019kyp} for
recent analyses, or~\cite{Losada:1996ju,Andersen:1998br,Laine:2000rm} for earlier foundational work.
The model has also been studied perturbatively, including in its inert doublet
realisation~\cite{Laine:2017hdk,Banerjee:2024fam,Jiang:2022btc}.

These studies have demonstrated that certain regions of 2HDM parameter space
can indeed support a strong first-order electroweak phase
transition, particularly when thermally induced cubic terms in the potential are
sufficiently enhanced. However, such analyses typically do not account for the
presence of a light scalar or pseudoscalar near 95~GeV, nor the associated
phenomenological constraints.

The potential existence of such a light state can substantially alter the structure of
the finite-temperature potential.
In particular, it can introduce new phase transition pathways or
weaken the strength of the transition by reducing the need for large mass splittings. 

We find that unlike in the Standard Model, where the EWPT is
a crossover~\cite{Kajantie:1996mn,Kajantie:1996qd,Gurtler:1997hr,Csikor:1998eu,DOnofrio:2015gop},
the transition is first order in the majority of the parameter space.
Moreover, depending on the parameters in the 2HDM Lagrangian,
the transition can occur in a single or in two steps;
see~\cite{Niemi:2020hto} for a non-perturbative analysis of two-step transitions
in the triplet extension of the SM.
We find that the transition strength remains modest across the viable parameter space,
with the order parameter not exceeding $\vc/\Tc \lesssim 1.3$.

A strong first-order phase transition can leave behind a
stochastic gravitational wave (GW) background.
However, due to the relatively modest values of
the order parameter in the 2HDM,
the resulting GW signal lies well below the projected sensitivity of
LISA~\cite{Caprini:2015zlo, Caprini:2019egz, LISACosmologyWorkingGroup:2022jok} and
other planned (space-based) observatories~\cite{Harry:2006fi,Kawamura:2011zz,Ruan:2018tsw};
see~\cite{Ramsey-Musolf:2024zex, Lee:2025hgb,Goncalves:2023svb,Zhou:2020irf,Biekotter:2022kgf}
for recent discussions of GW prospects in the 2HDM context.

Furthermore, the moderate strength of the transition also appears insufficient to support
another potential remnant of a strong first-order transition,
electroweak baryogenesis.
Achieving successful baryogenesis would likely require additional model ingredients,
such as tree-level barriers in the effective potential,
to sufficiently enhance the strength of the transition 
and realise a sharp turn-off of the electroweak sphaleron rates inside the bubbles of true vacuum.
We discuss possible extensions,
including singlet scalars or higher-dimensional operators,
that could enhance the transition.

%
\section{Two Higgs Doublet Model}
\label{sec:model}

The real Type~I Two Higgs Doublet Model (2HDM)
is \vtwo{economical and relatively} well suited to accommodate a new $\sim95$~GeV state while
respecting existing flavour and collider bounds.
\vtwo{As mentioned in the introduction, the main tension of 2HDMs with small $\tan\beta$ is in the flavour sector.}
In Type~I, all fermions couple to the same Higgs doublet,
so the stringent $b\to s\gamma$ constraints that force charged Higgs masses above a few hundred GeV
in Type~II and Flipped models are much weaker --
allowing $\mHpm\sim150-350$~GeV alongside
a light neutral scalar or pseudoscalar.
Moreover, imposing the alignment limit
$|c_{\beta-\alpha}|\ll1$ keeps
the 125~GeV Higgs SM-like
while still permitting a new state at 95~GeV with
suppressed couplings to $VV$ and enhanced loop-induced couplings to
$\gamma\gamma$ and $\tau\tau$.
\vtwo{Below we focus exclusively on the Type~I 2HDM and often refer
to it simply as the 2HDM.}

In this work,
we explore the scenario in which the pseudoscalar $A$ is identified with
the new 95~GeV resonance.
This assignment provides a better fit to existing collider data than
associating the scalar $H$ with the resonance.
The reason is as follows:
In the $\mH = 95$~GeV scenario,
the CP-even scalar $H$ couples to electroweak vector bosons at tree level.
However, these couplings are constrained by
the Higgs signal-strength sum rule and by global fits to Higgs data (see, e.g.,~\cite{Azevedo:2023zkg}),
which restrict the scalar mixing via
$|c_{\beta-\alpha}| \lesssim 0.2$.
This suppression limits both the production cross section and
the branching ratio to $\gamma\gamma$,
making it difficult to account for either the LEP excess in the $b\bar{b}$ channel~\cite{LEP_2006}
or the diphoton excess observed by CMS~\cite{Azevedo:2023zkg}.

The Higgs sector of the \vtwo{Type~I} 2HDM consists of two
${\rm SU}(2)_\rmii{L}$ doublets,
$\Phi_1$ and $\Phi_2$, with opposite charge under a $\mathbb{Z}_2$ symmetry and
hypercharge $Y = 1/2$.
All right-handed SM fermions are taken to be even under $\mathbb{Z}_2$, while
$\Phi_1$ is conventionally chosen to be odd, making it fermiophobic.
The resulting Yukawa sector of the \vtwo{Type~I} 2HDM is then identical to that of the SM,
except with $\Phi_2$ replacing the SM Higgs doublet and coupling to all fermions.
After electroweak symmetry breaking, the Higgs doublets can be parametrised in terms 
of the physical scalar degrees of freedom as follows~\cite{Fox:2017uwr}:
\begin{align}
\label{eq:doublet-parametrisation}
\Phi_1 &=
  \begin{pmatrix}
    -H^+ \sin{\beta} + G^+\cos{\beta}\\
    \frac{1}{\sqrt{2}} (v\cos{\beta} - h \sin{\alpha} + H\cos{\alpha}-iA\sin{\beta}+iG^0\cos{\beta})
  \end{pmatrix}
  \,,\\[2mm]
\Phi_2 &=
  \begin{pmatrix}
    H^+ \cos{\beta} + G^+\sin{\beta}\\
    \frac{1}{\sqrt{2}} (v\sin{\beta} + h \cos{\alpha} + H\sin{\alpha}+iA\cos{\beta}+iG^0\sin{\beta})
  \end{pmatrix}
  \,.
\end{align}
Here,
$h$ denotes the SM-like Higgs boson and is a CP-even scalar,
accompanied by the 
conventionally heavier CP-even state $H$.
The field $A$ is the neutral CP-odd pseudoscalar,
which we later associate with the new 95~GeV resonance.

The $H^\pm$ are a pair of charged
Higgs bosons, while $G^\pm$ and $G^0$ are the Goldstone bosons.
The SM Higgs vacuum expectation value (VEV) is $v = 246~\text{GeV}$.
The VEVs of the two 
doublets, $v_1$ and $v_2$,
are constrained by the requirement that  $v_1^2 + v_2^2 = v^2$, and
the angle $\beta$ is defined via $\tan\beta = v_2 / v_1$.
The angle $\alpha$ diagonalises
the CP-even scalar mass matrix and determines the physical mass eigenstates~\cite{Branco:2011iw}.

We can define a new angle
$\delta = \beta - \alpha -\pi/2$ to relate the tree-level couplings between
the physical resonances and the fermions,
\begin{align}
  c^h_f &= \frac{\cos{\alpha}}{\sin{\beta}} = \cos{\delta}-\frac{\sin{\delta}}{\tan{\beta}}
  \,,&
  c^\rmii{$H$}_f &= \frac{\sin{\alpha}}{\sin{\beta}} = -\sin{\delta}-\frac{\cos{\delta}}{\tan{\beta}}
  \,,&
  c^{A^0}_u &=-c^{A^0}_{d,l} = \cot{\beta}
  \,,
\end{align}
  as well as gauge bosons,
\begin{align}
  c^h_\rmii{$V$} &= \sin{(\beta-\alpha)} = \vtwo{\cos{\delta}}
  \,,&
  c^\rmii{$H$}_\rmii{$V$} &= \cos{(\beta-\alpha)} = -\sin{\delta}
\end{align}
where $V = W^\pm,Z$
and we henceforth denote
\vtwo{
$\cos(\beta - \alpha) \equiv c_{\beta-\alpha}$ and
$\tan\beta \equiv t_\beta$.
}

The couplings of the light Higgs boson $h$ approach their SM values
($c^h_f \to 1$) in the limit $\alpha \to 0$ and $\beta \to \pi/2$.
In contrast, the heavier scalar $H$ does not couple to fermions at $\alpha = 0$
and decouples from gauge bosons when $\alpha = \beta \pm \pi/2$.
The latter condition corresponds to $c_{\beta-\alpha} = 0$,
known as the {\em alignment limit} of the model.
In this limit, the light Higgs $h$ has SM-like tree-level couplings to gauge bosons.

The tree-level potential for the Type~I 2HDM is given by
\begin{align}
\label{eq:Veff:tree}
V_H &=
    m_{11}^2\Phi_1^\dagger \Phi_1
  + m_{22}^2\Phi_2^\dagger \Phi_2
  - m_{12}^2\bigl(\Phi_1^\dagger \Phi_2 + \text{h.c.} \bigr)
  + \lambda_1(\Phi_1^\dagger \Phi_1)^2
  \nn &
  +  \lambda_2(\Phi_2^\dagger \Phi_2)^2
  + \lambda_3(\Phi_1^\dagger \Phi_1)(\Phi_2^\dagger \Phi_2)
  + \lambda_4(\Phi_1^\dagger \Phi_2)(\Phi_2^\dagger \Phi_1)
  + \frac{\lambda_5}{2} \Bigl[(\Phi_1^\dagger \Phi_2)^2+ \text{h.c.} \Bigr]
  \, ,
\end{align}
where
$m_{11}$ and $m_{22}$ are the doublet mass parameters,
$m_{12}$ is the mixing parameter, and
$\lambda_{1},\dots,\lambda_{5}$ are the quartic couplings.
A more general treatment of the potential would include the additional terms
\begin{equation}
    V_H\supset\left(
        \lambda_6\Phi_1^\dagger\Phi_1
      + \lambda_7\Phi_2^\dagger\Phi_2\right)\Phi_1^\dagger \Phi_2
      + \text{h.c.}
      \, ,
\end{equation}
However, these terms are forbidden by the $\mathbb{Z}_2$ symmetry,
setting $\lambda_6,\lambda_7=0$.
This softly broken discrete symmetry is essential because
the non-observation of flavour-changing neutral currents (FCNCs)
or interactions that change fermion flavour
requires that fermions couple to a single Higgs doublet only,
as mentioned previously.
This is achieved by charging the doublets and fermions
under $\mathbb{Z}_2$,
thereby forbidding
the $\lambda_6,\lambda_7$ terms~\cite{Bernal:2022wct,Glashow:1976nt,Paschos:1976ay,Bernon:2015qea}.
The $m_{12}^2$ parameter is permitted to be non-zero, softly breaking
$\mathbb{Z}_2$ without reintroducing FCNCs~\cite{Bernal:2022wct,Bernon:2015qea}.

In the \vtwo{Type~I} 2HDM,
the background fields $\phi_i$
are identified as the second,
real components of the two Higgs doublets respectively,
\begin{equation}
\label{eq:Phi:background}
  \Phi_i \to \frac{1}{\sqrt{2}} \phi_{i} \delta_{n,2} + \Phi_i
  \,,
\end{equation}
where $n$ is the component index of the $n$-tuplet.
After the two Higgs doublets take these background field values,
we arrive at
the scalar mass matrix for the fields 
\newnewtext{$\{H,A,H^\pm,h,G^0,G^\pm\}$,
which in Landau gauge reads:}
\begin{equation}
\label{eq:mass-matrix}
    M^2 = \begin{pmatrix}
        M_{11}^2 & 0 & 0 & 0 & M_{15}^2 & 0 & 0 & 0 \\
        0 & M_{22}^2 & 0 & 0 & 0 & M_{26}^2 & 0 & 0 \\
        0 & 0 & M_{33}^2 & 0 & 0 & 0 & M_{37}^2 & 0 \\
        0 & 0 & 0 & M_{44}^2 & 0 & 0 & 0 & M_{48}^2 \\
        M_{15}^2 & 0 & 0 & 0 & M_{55}^2 & 0 & 0 & 0 \\
        0 & M_{26}^2 & 0 & 0 & 0 & M_{66}^2 & 0 & 0 \\
        0 & 0 & M_{37}^2 & 0 & 0 & 0 & M_{77}^2 & 0 \\
        0 & 0 & 0 & M_{48}^2 & 0 & 0 & 0 & M_{88}^2
    \end{pmatrix}
    \,,
\end{equation}
where,
\begin{align}
  M_{11}^2 &= m_{11}^2 + \frac{1}{2}(2\lambda_1 \phi_1^2 + \lambda_{-}\phi_2^2)
  \,, &
  M_{22}^2 &= m_{11}^2 + \frac{1}{2}(2\lambda_1 \phi_1^2 + \lambda_3 \phi_2^2)
  \,,\nn
  M_{33}^2 &= m_{11}^2 + \frac{1}{2}(6\lambda_1 \phi_1^2 + \lambda_{+}\phi_2^2)
  \,, &
  M_{44}^2 &= M_{22}^2
  \,,\nn
  M_{55}^2 &= m_{22}^2 + \frac{1}{2}(2\lambda_2 \phi_2^2 + \lambda_{-}\phi_1^2)
  \,, &
  M_{66}^2 &= m_{22}^2 + \frac{1}{2}(2\lambda_2 \phi_2^2 + \lambda_3 \phi_1^2)
  \,,\nn
  M_{77}^2 &= m_{22}^2 + \frac{1}{2}(6\lambda_2 \phi_2^2 + \lambda_{+}\phi_1^2)
  \,, &
  M_{88}^2 &= M_{66}^2
  \,,\nn
  M_{15}^2 &= -m_{12}^2 + \lambda_5 \phi_1 \phi_2
  \,, &
  M_{26}^2 &= -m_{12}^2 + \frac{1}{2}(\lambda_4+\lambda_5)\phi_1 \phi_2
  \,,\nn
  M_{37}^2 &= -m_{12}^2 + \lambda_{+}\phi_1 \phi_2
  \,, &
  M_{48}^2 &= M_{26}^2
  \,,
\end{align}
and where we used the abbreviation
$\lambda_\pm = \lambda_3 + \lambda_4 \pm \lambda_5$.
To obtain the final mass eigenvalues,
the diagonalisation of the mass matrix in eq.~\eqref{eq:mass-matrix}
is conducted as in~\cite{Helset:2017esj,Gorda:2018hvi}.
\newnewtext{Since we are working in Landau gauge,
the mass matrix for the vector fields
$\{W^\pm,Z,\gamma\}$
is the SM one.}

\vtwo{Our analysis relies on the one-loop effective potential and the
dimensionally reduced three-dimensional theory formulated in the
\MSbar{} scheme.
To ensure ${\cal O}(g^4)$ accuracy in the thermal EFT matching relations
(cf.\ sec.~\ref{sec:higher:orders:Veff}),
the \MSbar{} running parameters of appendix~\ref{app:betafunctions}
must be expressed at the same order
in terms of physical observables such as pole masses and mixing angles.
Accordingly, we relate the input masses
$\{\mh,\, \mH,\, \mA,\, \mHpm\}$,
angles $\alpha$, $\beta$, and mixing mass $m_{12}$
to the Lagrangian parameters
$m_{11}^2,\, m_{22}^2,\, m_{12}^2,$ and $\lambda_{1},\dots,\lambda_{5}$.
Achieving this accuracy requires a one-loop renormalisation of the vacuum theory.%
\footnote{
  The tree-level relations for computing these parameters
  can be directly taken from appendix~B.2 of~\cite{Gorda:2018hvi}.
}
While an on-shell scheme could, in principle,
keep the physical masses and mixing angles fixed at
their tree-level values~\cite{Sirlin:1980nh,Sirlin:1983ys,Bohm:1986rj,Hollik:1988ii},
we adopt the \MSbar{} \emph{vacuum renormalisation} approach of,
e.g.,~\cite{Kajantie:1995dw,Laine:2017hdk,Croon:2020cgk,Niemi:2021qvp},
which treats divergent and finite parts consistently within dimensional regularisation and
ensures that the renormalised parameters entering the finite-temperature
potential are defined within the same scheme.
Our one-loop implementation of this procedure is described in detail in
sec.~\ref{sec:recipe} and
appendix~\ref{sec:vacuum:renormalisation}.}

\section{The \vtwo{Type~I} 2HDM thermal potential}
\label{sec:Veff:2hdm}

To compute the temperature-dependent effective potential for the Type~I 2HDM,
we make use of high-temperature dimensional reduction,
initially devised in~\cite{%
  Ginsparg:1980ef,Appelquist:1981vg,Nadkarni:1982kb,Landsman:1989be,Kajantie:1995dw,
  Braaten:1995cm,Braaten:1995jr}.
In the context of Lorentz scalar-driven transitions,
this approach is well-suited for the high-temperature regime relevant to the transitions we study.
Dimensional reduction constructs an effective field theory (EFT) for the static modes at
finite temperature and systematically incorporates thermal resummation to all orders.
In particular, this method allows for a consistent inclusion of all large thermal corrections
including two-loop thermal masses essential for
the renormalization scale independence~\cite{Croon:2020cgk,Gould:2021oba,Lewicki:2024xan}
which in turn is essential for theoretical consistency and ensuring physically meaningful results,
but commonly not achieved,
cf.\ e.g.~\cite{Grojean:2006bp,Delaunay:2007wb,Dorsch:2016nrg,Ellis:2018mja}.

As an EFT framework, dimensional reduction
utilises the thermal hierarchy of scales
\begin{equation}
    \frac{g^2}{\pi}T \ll gT \ll \pi T
    \,.
\end{equation}
The scale
$|p| \sim \pi T$ corresponds to the {\em hard} non-zero bosonic (fermionic) Matsubara modes~\cite{Matsubara:1955ws},
$|p| \sim g T \sim \mD$ to the {\em soft} Debye screening modes, and
$|p| \sim g^2 T$ to the non-perturbative {\em ultrasoft} modes in
the infrared (IR)~\cite{Linde:1980ts}.
By successively integrating out
ultraviolet (UV) modes---see e.g.~\cite{Hirvonen:2022jba}---%
one obtains a sequence of dimensionally reduced EFTs.
The final step yields a 3D EFT at
the scale of bubble nucleation corresponding to
$g^2 T/\pi \ll |p| \ll gT$~\cite{Gould:2021ccf},
the so-called {\em softer} scale~\cite{Gould:2023ovu,Lewicki:2024xan}.
Since in practice,
the mass of the phase-transition undergoing scalar will be parametrically larger
than the softer scale and therefore the actual EFT lies between the soft and the
softer scale.
In this final EFT%
\footnote{%
  In appendix~\ref{app:betafunctions},
  we discuss the choice of the scale for $\LamdUS$ which defines the softer scale.
}, the long-distance dynamics of the transition
is encoded in three spatial dimensions.

To compute the thermodynamics of a phase transition,
one typically makes use of the effective potential at finite temperature.
This potential is then computed in the final 3D EFT at the softer scale.
To obtain the potential and the 3D EFT Lagrangian in practice,
we utilise
the thermal EFT matching software
{\tt DRalgo}~\cite{Ekstedt:2022bff}
which we crosscheck via further in-house software in
{\tt FORM}~\cite{Ruijl:2017dtg} and applied
{\tt qgraph}~\cite{Nogueira:1991ex} for diagram generation.

%
\subsection{The dimensionally reduced effective potential}
\label{sec:Veff:T}

The starting point for our analysis is
the effective potential up to two-loop level in the softer 3D EFT.
This corresponds to the functionalities of {\tt DRalgo} {\tt v1.2.0}.
Since we are interested in tracking the phases in multi-field space
(cf.\ sec.~\ref{sec:results}),
we refrain from integrating out further, potentially heavy vector fields~\cite{Gould:2023ovu}.
By keeping the matching relations (which relate Lagrangian parameters between the hard, soft, and softer theories) at two-loop level,
the options to compute the effective potential amount to
its different loop orders, {\em viz.}
\begin{align}
  \label{eq:3DatNLO,VeffatLO}
  &
  \text{{\tt [3D@NLO~$V_3$@LO]}:} &
  \text{two-loop EFT matching,} &&
  \text{one-loop effective potential}
  \,,\\
  \label{eq:3DatNLO,VeffatNLO}
  &
  \text{{\tt [3D@NLO~$V_3$@NLO]}:} &
  \text{two-loop EFT matching,} &&
  \text{two-loop effective potential}
  \,.
\end{align}
Here, we will focus on the former,
namely
{\tt [3D@NLO~$V_3$@LO]}.
The reason for this is that the direct computation of
the two-loop effective potential contains
scalar contributions that, for some parts of the parameter space,
should be counted as higher-order in comparison to
the heavy vector contributions;
see~\cite{Ekstedt:2022zro,Ekstedt:2024etx}.
An artefact of this setup is the presence of
logarithmically divergent terms in the two-loop potential, which are the result of negative mass eigenvalues in the logarithms coming mostly from sunset diagrams. The divergences lead to numerical issues when calculating the transitions; unphysical vacua are identified due to sharp discontinuous local minima, leading to unphysical phase transition strength parameters (such as bubble walls moving super-luminously, $\vw>1$).
To avoid such issues we follow~\cite{Schicho:2022wty}
and compute the one-loop potential with two loop matching relations;
see also~\cite{Lewicki:2024xan} for a similar application.
In the following,
we refer to the setup~\eqref{eq:3DatNLO,VeffatLO} as
the {\em two-loop improved one-loop potential}.

The final effective potential at the transition scale,
is a function of the 3D effective fields
$\bm{\phi}^\rmii{3D} = \{\phi_1^\rmii{3D}, \phi_2^\rmii{3D}\}$\footnote{These background fields are the 3D analogues of the background fields defined in eq. \eqref{eq:Phi:background}.}
and the temperature $T$.

The 4D effective thermal potential, $V_4$, can be calculated from the 3D potential via
\begin{equation}
\label{eq:4Deffpot}
  V_4(\bm{\phi},T) = TV_3(\bm{\phi}^\rmii{3D},T)
\end{equation}
using the relation between field values in three- and four-dimensions
$\bm{\phi}^\rmii{3D}\to \bm{\phi} T^{-\frac12}$.
Henceforth, we suppress the 3D superscript for
the fields in the 3D~EFT.

%
\subsection{Higher-orders in the effective potential}
\label{sec:higher:orders:Veff}

In the effective potential at the softer scale,
we utilise the NLO matching relations from~{\tt DRalgo}
directly, based on
the example file~\href{https://github.com/DR-algo/DRalgo/blob/main/examples/2hdm.m}{\tt 2hdm.m}.%
\footnote{See the {\tt DRalgo} GitHub,
\url{https://github.com/DR-algo/DRalgo/blob/main/examples/2hdm.m}.
}
Similar matching relations can also be found in~\cite{Gorda:2018hvi}.%
\footnote{
  In comparison with the matching relations of~\cite{Gorda:2018hvi},
  scalar masses in our softer-scale matching relations
  are counted as higher-order inside of
  logarithmic terms for e.g.~$\overline{m}_{11}^2$.
  Also, in eq.~(3.15) of~\cite{Gorda:2018hvi}, $N_f$
  should be the number of fermions and not the number of families.
}
For the corresponding vacuum renormalisation,
see secs.~\ref{sec:recipe} and \ref{sec:vacuum:renormalisation}.

Up to two-loop order, the 3D effective potential is
\begin{align}
\label{eq:Veff:3d}
V_{3} &=
  V_{0,3} + V_{1,3} + V_{2,3}
  \,,
  \\
V_{1,3}^{ } &=
  \sum_i n_{i}^{ }\, J_3^{ }\bigl(\overline{m}_i^{2}(\bm{\phi},T)\bigr)
  \,,
\end{align}
where
$d = 3 - 2\epsilon$,
$V_{0,3}(\bm{\phi},T)$ is the three-dimensional version of
the tree-level potential~\eqref{eq:Veff:tree}~\cite{Gorda:2018hvi,Helset:2017esj}, and
the degrees of freedom, $n_i$, are $d$-dependent.
The corresponding mass eigenvalues $\overline{m}_i$ of
the dynamical fields $i\in \{W,Z,H,A,H^\pm,h,G^0,G^\pm\}$ in the 3D~EFT
depend on the background fields for the two Higgs doublets,
whose 4D analogues are seen in eq.~\eqref{eq:doublet-parametrisation}. We discussed the calculation of the mass eigenvalues from the mass matrix in sec. \ref{sec:model}.

Similarly,
to the 4D vacuum case and the corresponding Coleman-Weinberg potential~\cite{Coleman:1973jx},
the one-loop 3D EFT potential,
$V_{1,3}$,
takes a closed form.
The corresponding integrals are
UV-finite and three-dimensional
\begin{align}
  J_3(m^2) = \frac{1}{2} \int_{\vec{p}}  \ln (p^2 + m^2)
  \stackrel{d=3-2\epsilon}{=}
  -\frac{1}{12\pi}[m^2]^\frac{3}{2}
\,.
\end{align}
The two-loop contributions, $V_{2,3}$, to the potential~\eqref{eq:Veff:3d},
as well as the two-loop 3D~EFT matching relations
are directly adopted
from {\tt DRalgo}~\cite{Ekstedt:2022bff}
can also be taken
from~\cite{Niemi:2021qvp}.
The parameters of
this final 3D~EFT
are evolved to the 3D renormalisation group (RG) scale,
$\Lamd$,
which we set to
$\Lamd = T$ in our analysis.
The RG evolution and vacuum renormalisation procedures
are detailed in
secs.~\ref{sec:recipe} and \ref{sec:vacuum:renormalisation}.

In sec.~\ref{sec:Veff:T}, we argued that directly employing the two-loop effective potential,
without integrating out heavy vector and temporal modes to induce the transition,
can lead to pathological behaviour for some benchmark points along the transition path in the
multi-field field space.
Since the preferred approach is therefore to use the
{\tt [3D@NLO~$\Veff$@LO]} prescription~\eqref{eq:3DatNLO,VeffatLO},
one may wonder about the importance of omitting two-loop corrections
in the effective potential while keeping them in the matching.
These effects have been studied in~\cite{Schicho:2022wty,Lewicki:2024xan},
which concluded that the dominant uncertainties are associated
with higher-order corrections in the matching.
To investigate this and support our choice of using the setup~\eqref{eq:3DatNLO,VeffatNLO}
  for our scan,
in sec.~\ref{sec:results} we examine a few benchmark points to assess the magnitude of
uncertainty introduced by neglecting higher-order terms in the effective potential.
These benchmark points are summarised in table~\ref{tab:2loop-benchmarks}.

\section{Bubble Nucleation and Gravitational Waves}
To find the phases, calculate phase transition properties and gravitational wave spectra from our 4D potential, we use
the {\tt PhaseTracer2} package for {\tt C++}~\cite{Athron:2020sbe,Athron:2024xrh}. The package automates the pipeline from a 4D thermal potential to the gravitational wave spectrum parameters and signal-to-noise ratios (SNRs) for proposed gravitational wave interferometers,
such as
LISA~\cite{LISACosmologyWorkingGroup:2022jok,Danzmann:2000yvf} and
Taiji~\cite{Ruan:2018tsw}.
In this section, we provide a brief review of the computation performed by the package.

\subsection{Bubble nucleation}
\label{sec:bubble}
First order phase transitions occur through bubble nucleation, which describe tunnelling from a high temperature phase (typically symmetric under a symmetry) to a low temperature phase, which are separated by a potential barrier. 

Bubbles nucleate when there is a viable transition path between the minima with a low enough action. After switching to Euclidean space, with $t\to -i\tau$, the transition path can be parametrised as $\phi_1(\rho)$ and $\phi_2(\rho)$ with $\rho=\sqrt{\tau^2+\vec{x}^2}$ being a radial parameter in space, indicating a spherically symmetric bubble.
The associated Euclidean action can be written as 
\begin{equation}
    S_d[\phi_1,\phi_2](T) =
    \Omega_{d-1}\int_0^\infty {\rm d}\rho\,\rho^{d-1}\biggl\{
        \frac{1}{2}(\dot{\phi}_1^2
      + \dot{\phi}_2^2) + V(\phi_1, \phi_2, T)
    \biggr\} 
    \, ,
\end{equation}
with $d$ being the number of dimensions of the $O(d)$-symmetric field configuration, here $d=3$ for nucleation by thermal fluctuations.
The Euler-Lagrange equation provides the saddle point of this action, known as the bounce equation, given by \cite{Coleman:1977py}
\begin{equation}
\label{eq:bounce}
    \ddot{\phi_i}(\rho) + \frac{d-1}{\rho}\dot{\phi}_i(\rho) = \frac{\partial}{\partial \phi_i}V(\phi_1,\phi_2,T) \, .
\end{equation}
This equation is solved with the boundary conditions $\dot{\phi_i}(0)=\dot{\phi_i}(\rho\to\infty) = 0$, and $\phi_i(\rho\to\infty) = \phi_i^f$, where $\phi_i^t$ is the value of the fields at the true vacuum (see \cite{Andreassen:2016cvx,Croon:2023zay} for a discussion of the appropriate boundary conditions for tunneling). 
In our two-field model,
we employ {\tt PhaseTracer2}~\cite{Athron:2024xrh}
to calculate the bounce configuration using the path deformation method,
where an initial guess for the path is perturbed using normal forces to minimise the bounce action. 
The action $S_3(T)$ is then calculated for the bounce solution at temperature $T$.

We define the onset of the phase transition through
the following
nucleation criterion~\cite{Guth:1979bh,Guth:1981uk,Caprini:2019egz}
\begin{equation}
\label{eq:nucleation:condition}
    \frac{S_3}{\Tn} \simeq 141
    + \ln\frac{A}{\Tn^4}
    - 4 \ln\frac{\Tn}{100 \,\text{GeV}}
    -\ln\frac{\bar{\beta}}{100 H_*}
    \, ,
\end{equation}
which defines the nucleation temperature $\Tn$.
\vtwo{The pre-factor
  $A = A(T)$ of the nucleation rate
  $\Gamma \simeq A(T) e^{-S_3/T}$~\cite{langer1973hydrodynamic},
  accounts for the fluctuations around the critical bubble.
  It splits
  into a statistical and dynamical part,
  {\em viz.}
  $A = A_\rmi{stat} + A_\rmi{dyn}$ which on dimensional grounds
  scale as
  $A_\rmi{stat} \sim T^3$ and
  $A_\rmi{dyn} \sim T$.
}
\vthree{
While a practical
standard approximation~\cite{Linde:1980tt,Linde:1981zj}
is
$A(T) = T^4(\frac{S_3}{2\pi T})^{3/2}$,
we follow~\cite{Carrington:1993ng,Caprini:2019egz}
and employ
$\ln(A/T^4) \simeq -14$~\cite{Carrington:1993ng}
as an estimate
for the EWPT.}%
\footnote{%
  In practice, the nucleation rate receives higher-order corrections from fluctuations around
  the bounce solution, modifying the prefactor $A(T)$ and the interpretation of $S_3$, and
  have recently been studied in detail using
  functional determinant methods~\cite{Ekstedt:2021kyx,Gould:2021ccf}.
  Public tools such as
  {\tt bubbleDet}~\cite{Ekstedt:2023sqc} enable the automated inclusion of
  these corrections, providing a more accurate estimate of the nucleation rate.
} 

While the condition~\eqref{eq:nucleation:condition} is phrased in terms of the Euclidean action,
it is derived from the requirement that the nucleation rate per unit volume
becomes significant in an expanding universe. More precisely,
it approximates the point at which the probability of
bubble nucleation in a Hubble volume becomes order unity.
The full percolation criterion requires integrating the nucleation
rate over spacetime to track the volume fraction of
the false vacuum (see e.g.~\cite{Hindmarsh:2020hop}),
which is well approximated by the above condition in most cases.
Here 
\begin{equation}
\label{eq:beta:def}
    \bar{\beta} =
    \frac{{\rm d}(S_3/T)}{{\rm d}t} \biggr|_{t=t_*}= 
    H_*T_* \frac{{\rm d}(S_3/T)}{{\rm d}T}\biggr|_{T=T_*}
    \, ,
\end{equation}
is a measure of the inverse time it takes for the phase transition to complete,
where $H_*$ ($T_*$) refers to the Hubble constant (temperature) at the time $t_*$.%
\footnote{%
  Note, that we have added a bar to the conventional notation to
  avoid confusion with the Lagrangian parameter in the \vtwo{Type~I} 2HDM.
}
The percolation temperature, $\Tp$,
can be calculated by using the percolation criterion~\cite{Enqvist:1991xw},
which is
\begin{equation}
\label{eq:percol}
  \frac{S_3}{\Tp} \simeq 131
  + \ln\frac{A}{\Tp^4}
  - 4 \ln\frac{\Tp}{100 \,\text{GeV}}
  - 4 \ln\frac{\bar{\beta}}{100 H_*}
  + 3 \ln\vw
  \, ,
\end{equation}
where in our numerical studies,
we use
\vthree{$\ln( A/\Tp^4) \simeq -14$~\cite{Carrington:1993ng}
as above} and
an ansatz for $\bar{\beta}/H_* =10^4$,
justified a posteriori in
sec.~\ref{sec:results}.
A thorough calculation of $\vw$ can be achieved through hydrodynamical simulations of the phase transition, as in~\cite{Wang:2024slx,Tian:2024ysd},
however these are time consuming and $\vw$ is typically supplied
as an input parameter~\cite{Ai:2023see,Athron:2024xrh}.
A general perturbative determination of $\vw$
requires including
out-of-equilibrium effects~\cite{%
  Laurent:2022jrs,DeCurtis:2022hlx,DeCurtis:2023hil,DeCurtis:2024hvh}
as recently automated in~\cite{Ekstedt:2024fyq,vandeVis:2025plm}.
It has been argued by~\cite{Steinhardt:1981ct} that $\vw$ can be
approximated by the Chapman-Jouguet velocity,
typically used to describe explosive detonations:
\begin{equation}
    \vw \approx \vCJ =
      \frac{1}{1+\bar{\alpha}}
      \biggl(\cs + \sqrt{\bar{\alpha}^2 + \frac{2}{3} \bar{\alpha}} \biggr)
      \,,
      \label{eq:vCJ}
\end{equation}
where we take the speed of sound in the plasma to be $\cs = 1/\sqrt{3}$.
The parameter $\bar\alpha$ characterises the strength of the phase transition and will be defined below.
For the strongest transitions we find in sec.~\ref{sec:results} (which have $\bar{\alpha} \approx 4\times 10^{-3} $),
the Chapman-Jouguet velocity gives $\vw \approx 0.63$.
We use this value as an input to \eqref{eq:percol} in our numerical studies.
However, \eqref{eq:vCJ} is valid only in a restricted regime.
As noted in~\cite{Laine:1993ey,Ai:2023see}, the assumptions underlying
the Chapman-Jouguet condition do not strictly apply to cosmological phase transitions. More accurate treatments bracket the true value of $\vw$ between two physical limits: a ballistic limit, representing minimal interaction between the bubble wall and plasma~\cite{Ai:2023see}, and a local thermal equilibrium (LTE) limit, which assumes local entropy conservation and requires detailed thermodynamic input~\cite{Ai:2021kak,Ai:2024btx,Branchina:2025jou}.
The LTE expression has been shown to match well with numerical simulations~\cite{Krajewski:2024gma}, but is computationally prohibitive for parameter scans. For simplicity, we adopt the Chapman–Jouguet approximation in this work.

The trace anomaly difference of the energy-momentum tensor,
\begin{equation}
    \Delta \theta =
      \Bigl(
        V(\phi, T_*)
      - \frac{T_*}{4} \frac{\partial}{\partial T}V(\phi, T)\bigr|_{T_*}
    \Bigr) \Big|^{\phi_f}_{\phi_t}
    \, ,
\end{equation}
is taken in the relativistic plasma limit and
in practice receives further corrections if
the broken-phase speed of sound squared differs
from $\cs^2=1/3$~\cite{Giese:2020rtr,Tenkanen:2022tly}.
Here,
$T_*$ is the gravitational wave production temperature, typically identified with the percolation temperature of the phase transition $\Tp$. $\Delta \theta$ quantifies the amount of energy available for conversion to spacetime shear stress, which is represented by the off-diagonal components of the stress-energy tensor $T^{\mu\nu}$, and is responsible for the generation of gravitational waves \cite{Witten:1984rs,Hogan:1986qda}.

The ratio of the trace anomaly difference to the plasma energy density
(approximated by the radiation energy density at the GW production temperature,
$\rho_r = \pi^2 g_* \Tp^4 / 30$)
quantifies the energy available for gravitational wave production and
thus characterises the phase transition strength:%
\footnote{%
  As above, we use barred notation for this parameter to distinguish it
  from the Lagrangian parameters in the \vtwo{Type~I} 2HDM.
}
\begin{equation}
\label{eq:alpha:def}
    \bar{\alpha} \equiv \frac{\Delta \theta}{\rho_r}
    \, .
\end{equation}
Several alternative definitions of $\bar{\alpha}$ exist in the literature.
A commonly used definition takes the ratio of the vacuum energy difference to
the radiation energy density,
$\bar{\alpha} = \Delta \Veff / \rho_r$,
which neglects thermal effects~\cite{Caprini:2015zlo}.
Another popular definition comes from hydrodynamics, where
$\bar{\alpha} = \Delta \rho / w$, with $w = \rho + p$,
describing the energy injected into the plasma relative to
its enthalpy~\cite{Espinosa:2010hh}. 
A third definition is based on the \emph{pseudo} trace anomaly~\cite{Giese:2020znk},
which approximates the energy available for shear stress using
a simplified thermal treatment.
The definition used in this work includes both vacuum and
thermal contributions and is particularly appropriate for models
like the \vtwo{Type~I} 2HDM, where no tree-level barrier is present
and thermal corrections are essential for
realising a first-order phase transition (FOPT).

\subsection{Gravitational Wave spectrum}
\label{sec:GW}
Early universe FOPTs give three main sources of gravitational waves: sound waves induced by the expansion of bubbles into the surrounding plasma,
bubble collisions creating anisotropic stress directly \cite{Kosowsky:1992vn}, and
turbulence in the plasma \cite{Caprini:2009yp,RoperPol:2019wvy} caused by bubble collision energy.
Typically, the GW contributions from these three sources can be calculated from knowledge of the thermal parameters $\bar{\alpha}, \bar{\beta}/H_*$ and $\vw$.
In our case,
the sound wave contribution dominates throughout our parameter space, and
we will therefore limit our discussion to this effect.

Numerical simulations indicate that sound waves are typically
the dominant source of GWs from FOPTs~\cite{Hindmarsh:2013xza,Hindmarsh:2015qta}
(however, see~\cite{Ellis:2019oqb} for examples of exceptions).
Two length scales dictate the bounds of the power spectrum of the acoustic GW contribution, which are the mean distance between the bubbles~\cite{Caprini:2019egz} (cf.~\cite{Enqvist:1991xw}),
\begin{equation}
    R_*=(8\pi)^{1/3} \vw/\bar{\beta} \, ,
\end{equation}
and the thickness of the expanding sound shell in the plasma outside the bubble of the new phase,
\begin{equation}
    \Delta R_* = R_*\frac{|\vw-\cs|}{\cs} \, .
\end{equation}
The sound shell thickness $\Delta R_*$ dictates the position of the peak of the power spectrum \cite{Hindmarsh:2017gnf}.

An analytic fitting formula for the acoustic GW contribution was derived in\cite{Hindmarsh:2015qta,Hindmarsh:2017gnf,Caprini:2019egz},
which we write here in the form used by
{\tt PhaseTracer2}~\cite{Athron:2024xrh}, 
\begin{equation}
  \Omega_\text{sw}h^2(f) = 2.061 F_{\rmi{gw},0} \Gamma^2 \bar{U}_f^4
    S_\text{sw}(f/f_\text{sw}) \tilde{\Omega}_\text{gw} \times \min(H_*R_*/\bar{U}_f, 1) (H_*R_*)h^2
    \, ,
\end{equation}
where $\min(H_*R_*/\bar{U}_f, 1) (H_*R_*)$ accounts for the finite lifetime of the source.
Here,
\begin{align}
    F_{\rmi{gw},0} &= 3.57\times 10^{-5} \left(\frac{100}{g_*}\right)^{1/3} \, ,\\ 
    S_\text{sw}(x) &= x^3 \left( \frac{7}{4+3x^2}\right)^{7/2}
    \, ,
\end{align}
with
$S_\text{sw}(f)$ being the spectral shape
the peak frequency of the sound waves
\begin{align}
    \frac{f_\text{sw}}{1 \, \mu \text{Hz}} &= \frac{2.6}{H_*R_*} \left(\frac{z_p}{10}\right)\left(\frac{T_*}{100 \, \text{GeV}}\right)\left(\frac{g_*}{100}\right)^{1/6}
    \, .
\end{align}
$\Gamma$ is the ratio of enthalpy to the energy density of the plasma,
taken to be $4/3$ for the early universe, $\bar{U}_f$ is
the enthalpy weighted root mean square fluid velocity of the plasma, and
$z_p \sim 10$ and $\tilde{\Omega}_\text{gw} \sim 0.012$ are
parameters informed by the numerical simulations.
Thus, we see that $\Gamma \bar{U}_f^2 = K_\text{sw}$ is
the ratio of kinetic energy in the fluid to its energy density.
We can write this as 
\begin{equation}
    \Gamma \bar{U}_f^2 = K_\text{sw}= \frac{\kappa_\text{sw} \bar{\alpha}}{1+\bar{\alpha}} \, ,
\end{equation}
where $\kappa_\text{sw}$ is an efficiency factor for
the conversion of the latent energy of the phase transition into
the acoustic kinetic energy of the fluid. 
{\tt PhaseTracer2} employs the following fitting formula
for $\kappa_\text{sw}$ when the wall velocity is taken to be
the Chapman-Jouguet velocity:
\begin{equation}
    \kappa_\text{sw} = \frac{\sqrt{\bar{\alpha}}}{0.135+\sqrt{0.98+\bar{\alpha}}} \, .
\end{equation}
The signal to noise ratios (SNRs) for
GW experiments such as LISA can be calculated through~\cite{Smith:2019wny,Caprini:2019egz}
\begin{equation}
  \text{SNR}_\text{LISA} = \sqrt{\mathcal{T}
    \int^{f_\text{max}}_{f_\text{min}} {\rm d}f
    \left(\frac{h^2\Omega_\text{gw}(f)}{h^2\Omega_\text{sens}(f)}\right)^2}
    \, ,
    \label{eq:SNRLISA}
\end{equation}
where $\mathcal{T}$ is the length of time for data collection, and
$h^2\Omega_\text{sens}(f)$ is the frequency dependent sensitivity of the experiment.
It is given by~\cite{Smith:2019wny}
\begin{equation}
    h^2\Omega_\text{sens}(f) = \frac{4\pi^2}{3H_0^2} f^3 S_h(f)
\end{equation}
where the $S_h(f)$ is the inverse noise weighted sensitivity to the spectral density,
\begin{equation}
    S_h(f) \simeq  \frac{20\sqrt{2}}{3}
      \biggl( \frac{S_I(f)}{(2\pi f)^4} +S_{II}(f)\biggr)
      \biggl(1+\biggl(\frac{3f}{4f_*}\biggr)^2 \biggr) \, .
\end{equation}
For LISA, the term involving
\begin{equation}
    S_I(f) = 5.76 \times 10^{-48} \biggl(1+\biggl(\frac{f_1}{f}\biggr)^2\biggr) \,  \text{Hz}^{3}
\end{equation}
where $f_1 = 0.4 \, \text{mHz}$, gives the acceleration noise associated with spurious forces on the test masses, for example those that occur due to the build up of electrostatic charge \cite{Sumner:2019qmx}.
The term involving 
\begin{equation}
    S_{II}(f) = 3.6 \times 10^{-41} \,  \text{Hz}^{-1}
\end{equation}
corresponds to the noise from optical path length fluctuations.
The characteristic LISA frequency $f_* =c/(2\pi L)$, where $L = 2.5\times 10^6 \,\text{km}$ and $c$ is the speed of light, relates to the distance that light travels between LISA sensors.
Useful quantities that can be taken from $\Omega_\text{gw}h^2$ include the frequency $f_\text{gw}$ for which the amplitude is greatest (the `peak frequency'), and the peak amplitude $\Omega_\text{gw}h^2(f_\text{gw})$ at that frequency.

\subsection{From dimensionally reduced potential to gravitational waves}
\label{sec:recipe}

In the previous two sections, we have described our calculation of
the \vtwo{Type~I} 2HDM thermal potential and
the dynamics of the first order phase transition with
a light degree of freedom.
We summarise the calculation pipeline with the following steps:
\begin{enumerate}
\item First, we take the physical input parameters
    $\{\mh, \mH, \mA, \mHpm, t_\beta,c_{\beta-\alpha}, m_{12}\}$
    defined at the input RG scale $\Lambda_0 = \mZ$.
\item Next, we use these physical input parameters to calculate the one-loop renormalised 4D Lagrangian input parameters
  $\{m_{11}^2, m_{22}^2, m_{12}^2, \lambda_1,\dots,\lambda_5,\gY,g_{1-3}^2\}$, see also appendix~\ref{sec:vacuum:renormalisation}.
\item
  Next, we use the beta functions (provided in appendix~\ref{app:betafunctions})
  to RG-evolve these renormalised 4D Lagrangian parameters to
  the matching scale
  $\LamD = 4\pi e^{-\gammaE}T$,
  such that we have
  $\{\bar{m}_{11}^2, \bar{m}_{22}^2, \bar{m}_{12}^2, \bar{\lambda}_1,\dots,\bar{\lambda}_5,\bar{y}_t,\bar{g}_{1-3}^2\}$.
\item
  Using the soft 3D matching relations,
  we calculate the soft parameters
\begin{equation}
    \Bigl\{
      (m_{11}^\rmii{3D})^2,
      (m_{22}^\rmii{3D})^2,
      (m_{12}^\rmii{3D})^2,
      \lambda_{1}^\rmii{3D},
      \dots,
      \lambda_{7}^\rmii{3D},
      \gY^\rmii{3D},
      (g_{1}^\rmii{3D})^2,
      \dots,
      (g_{3}^\rmii{3D})^2
    \Bigr\}
\end{equation}
  at the scale $\Lamd = T$.
  In the soft theory,
  $\lambda_{6,7}^\rmii{3D}$ arise from integrating out the non-zero Matsubara modes, despite our model having as input
  $\lambda_{6,7}=0$.

\item
  Using the softer matching relations,
  we calculate the softer parameters
\begin{equation}
  \Bigl\{
    (\bar{m}_{11}^\rmii{3D})^2,
    (\bar{m}_{22}^\rmii{3D})^2,
    (\bar{m}_{12}^\rmii{3D})^2,
    \bar{\lambda}_{1}^\rmii{3D},
    \dots,
    \bar{\lambda}_{7}^\rmii{3D},
    \bar{y}_t^\rmii{3D},
    (\bar{g}_{1}^\rmii{3D})^2,
    \dots,
    (\bar{g}_{3}^\rmii{3D})^2
    \Bigr\}
\end{equation}
  at the softer matching scale $\LamdUS = \Lamd$.

\item
  The softer parameters are input to construct the 3D effective potential,
  $V_3(\phi_1^\text{3D}, \phi_2^\text{3D},T)$.
  As mentioned previously,
  we find the 4D effective potential $V_4(\phi_1,\phi_2,T)$ through eq.~\eqref{eq:4Deffpot}.
\item Through $V_4(\phi_1,\phi_2,T)$, we can then,
\begin{enumerate}
\item Find the minima of the potential, identifying when they co-exist and finding possible critical temperatures $\Tc$.
\item Find possible transition paths between the minima and calculate the action $S$ along the paths, comparing $S/T$ to the percolation criterion in \eqref{eq:percol}. This allows us to calculate the percolation temperature(s) $\Tp$.
\item Calculate the transition parameters $\bar{\alpha}, \bar{\beta}/H_*$, and
  then the peak gravitational wave frequencies and amplitudes associated with the transitions. From this, the signal-to-noise ratios (SNRs) for LISA or other experiments can be calculated.
\end{enumerate}

\end{enumerate}

%
\section{Parameter scans}
\label{sec:results}

As explained above, we work in the real \vtwo{Type~I} 2HDM,
associating the 95~GeV resonance with the pseudoscalar $A$.
\vtwo{To understand the structure of the EWPT in the 2HDM,
in sec.~\ref{sec:PT:scan},
we first perform an exploratory scan over the parameter space without
explicitly imposing constraints from colliders and electroweak precision tests.
In this scan, we limit ourselves to two free parameters,
fixing the others to values we motivate below.
The purpose of this preliminary scan is to identify the qualitative features of
the finite-temperature potential:
specifically, the regions that support FOPT and the characteristic transition patterns.
Such a ``phase-transition-driven'' exploration offers insight into how
the scalar mass spectrum and mixing angles influence the order and strength of
the transition before overlaying phenomenological constraints.}

\vtwo{%
Following this first scan,
we then consider in sec.~\ref{sec:collider:constraints}
how collider measurements reshape the viable region of parameter space.
Restricting the model to be consistent with current Higgs and scalar searches can
significantly alter the available regions that support FOPT.
In particular, the alignment limit,
$|c_{\beta - \alpha}| \simeq 0$, and
the mass splittings among the additional scalar states play a key role
in determining whether viable FOPT survive after collider bounds are applied.
We emphasise that we do not impose flavour constraints here, as
these are already known to be in tension with the parameter region relevant for
the 95~GeV excess~\cite{Azevedo:2023zkg}.
\vthree{Indeed,
in agreement with~\cite{Azevedo:2023zkg},
the region of parameter space relevant for the 95~GeV excess
is strongly disfavoured by $\mathrm{BR}(b \to s\gamma)$,
with the points in our scan lying at the level of $\mathcal{O}(3\sigma)$ tension.
}
Our goal is rather to examine which regions remain accessible once
direct collider exclusions and Higgs signal-strength measurements are accounted for.}

\subsection{Scanning the parameter space for first-order phase transitions}
\label{sec:PT:scan}

\begin{table}[t]
\centering
\small
\begin{tabular}{|l|l|l|}
  \hline
  Parameter & Value & Motivation \\
  \hline
  \hline
  $\mA$
  & 95~GeV
  & Resonance mass
  \\
  $\mh$
  & 125~GeV
  & SM-like Higgs
  \\
  $\mHpm$
  & \vtwo{$\mH$}
  & \vtwo{Satisfy constraints from oblique parameters~\eqref{eq:STUvalues}}
  \\
  $t_\beta$
  & 1.57
  & Fit $\gamma\gamma$ and $\tau\tau$ excesses
  \\
  $m_{12}^2$
  & 1~GeV$^2$
  & Minimal impact on EWPT strength
  \\
  $v = \sqrt{v_1^2+v_2^2}$
  & 246~GeV 
  & Electroweak VEV
  \\
  \hline
\end{tabular}
\caption{%
  Fixed inputs for our scan identifying
  $A$
  with
  the resonance respectively.
  }
\label{tab:scan1}
\end{table}
The Type~I 2HDM features eight free parameters,
six of which we fix to benchmark values informed by theoretical and experimental constraints;
see table~\ref{tab:scan1}.
\vtwo{We identify
the pseudoscalar $A$ with the $95$~GeV resonance and $h$ with the SM Higgs boson.
The geometric combination $v= \sqrt{v_1^2+v_2^2}$ is fixed at $v=246$~GeV.
We fix $t_\beta = 1.57$ in this scan to accommodate both the diphoton and ditau signals~\cite{Azevedo:2023zkg}.
A preliminary scan extending to
$m_{12} \sim 10^{2}$~GeV revealed that $\bar{\alpha}$
is largely insensitive to the value of $m_{12}$.}%
\footnote{
  A marginal peak was found at $m_{12} \approx 10^{1.35}$~GeV,
  but an analysis with $m_{12}$ fixed at this optimal value showed no qualitative differences and
  only minor quantitative variations from the results presented below,
  with maximum LISA SNR of $\mathcal{O}(10^{-6})$.
}
\vtwo{In the following, we fix $m_{12} \simeq 1$~GeV.}
\vtwo{We choose $\mHpm$ anticipating strong constraints from EW precision tests expressed through
the oblique parameters $S$, $T$, $U$~\cite{Altarelli:1990zd,Peskin:1990zt,Peskin:1991sw,Maksymyk:1993zm} at
the $2\sigma$ confidence level~\cite{Baak:2014ora},
\begin{align}
\label{eq:STUvalues}
    S &= 0.05 \pm 0.11
    \,, &
    T &= 0.09 \pm 0.13
    \,, &
    U &= 0.01 \pm 0.11
    \,,
\end{align}
in particular the oblique parameter $T$,
impose strong constraints on scalar mass splittings.
To this end, charged states must be close in mass with a neutral state
because large separations between
$\mH$ and $\mHpm$ violate custodial symmetry and
lead to unacceptably large contributions to $T$~\cite{Grimus:2007if,Dolle:2009fn}.
Therefore, we fix $\mHpm = \mH$.}
The remaining two parameters and the
temperature $T$ are scanned over
\begin{align}
\label{eq:range:scan1}
  c_{\beta-\alpha} &\in [-0.3,\,0.3]
  \,, &
  \mH &\in [130,\,300]\,\text{GeV}
  \,, &
  T &\in [20,\,300]\,\text{GeV}
  \,,
\end{align}
\vtwo{evading direct detection bounds on $c_{\beta-\alpha}$
which will be further discussed
in sec.~\ref{sec:collider:constraints} and
as inspired by~\cite{Azevedo:2023zkg}.}
As we will show in fig.~\ref{fig:multiplot}, allowing smaller values of $\mH$
does not lead to appreciably different vacuum structures
or stronger transitions.
Moreover, the implied quartic couplings in the scalar potential remain $\mathcal{O}(1)$,
well below the tree-level unitarity bounds and safely perturbative,
so all $2\to2$ scalar--scalar amplitudes satisfy $|a_0|<\tfrac12$.

\vtwo{
Before we proceed, we verify that the six fixed parameters in table~\ref{tab:scan1}
can reproduce the $\gamma\gamma$ and $\tau\tau$ excesses throughout
the whole scan range.
To this end, as in \cite{Azevedo:2023zkg}, we define
the signal strength normalised to that of a SM Higgs boson of the same mass
\begin{equation}
\mu_{\rmii{$XX$}} \equiv
  \frac{
    \sigma(gg\to A)\,\mathrm{BR}(A\to XX)}{
    \sigma_\rmii{SM}(gg\to H)\,\mathrm{BR}_\rmii{SM}(H\to XX)}
\,.
\end{equation}
In the CP-conserving Type~I 2HDM, neither
the signal strength $\mu_{\gamma\gamma}$ nor $\mu_{\tau\tau}$
depend strongly on the parameters we vary.
Both signal strengths represent the gluon-fusion production of
the 95~GeV CP-odd pseudoscalar $A$ followed by its decay to
$\gamma\gamma$ or
$\tau^+\tau^-$, and are to good approximation governed solely by
the axial Yukawa couplings (functions of $t_\beta$, which we fix) and
kinematic factors set by $\mA$ (also fixed),
with no CP-even mixing or heavy-scalar contributions entering the amplitudes.
We therefore expect this description to remain valid throughout the scan to good approximation.}

For supercooled transitions,
it is also essential to verify that the volume fraction in
the false vacuum is decreasing, in addition to satisfying
the percolation condition~\eqref{eq:percol},
since the latter alone does not guarantee irreversible progression of the transition. 
Since the degree of supercooling, quantified by the relative difference
$\delta_\rmii{SC} = (\Tc-\Tn)/\Tn $, never exceeds $\sim 20\% $ across the parameter space,
the transitions are not strongly supercooled.
This absence of significant supercooling ensures that
the false vacuum volume shrinks monotonically throughout the investigated
parameter space,
preventing subsequent thermal inflation~\cite{Lyth:1998xn,Sagunski:2023ynd}.
The observed mild values of supercooling
result also in relatively small field values (cf.\ sec.~\ref{sec:results})
and no large separation of temperature scales
in contrast to classically conformal models,
where supercooling is large and one needs to utilise a more refined
framework~\cite{Kierkla:2022odc,Kierkla:2023von,Kierkla:2025qyz,Kierkla:2025vwp}.
In turn, this gives credence to the validity of the high-temperature expansion
for computing both
the transition timescale and strength within the 3D EFT framework.

\begin{figure}[t]
  \centering
  \begin{subfigure}[t]{0.32\textwidth}
      \centering
    \begin{tikzpicture}[>=Stealth, thick, 
      every node/.style={font=\sffamily},
      dot/.style={circle, draw, minimum size=8pt, inner sep=0pt},
      ->/.style={decoration={markings, mark=at position 0.5 with {\arrow{>}}},
                  postaction={decorate}}
      ]
    \draw[thick] (-2,0) -- (2,0);
    \draw[thick] (0,-1) -- (0,2.5);
    \node[dot] at (0,0) {};
    \node[dot] at (0,1.2) {};
    \node[dot] at (1,2.3) {}; 
    \draw[->] (0,0.0) .. controls (-0.5,0.5) and (-0.5,1.1) .. (0,1.2);
    \draw[->] (0,1.2) .. controls (1,1.3) and (1.6,1.9) .. (1,2.3); 
    \node at (-0.7,1.0) {$1$};
    \node at (1.3,1.6) {$2$};
    \end{tikzpicture}
      \caption{Class 2S}
      \label{fig:2-step}
  \end{subfigure}%
  \begin{subfigure}[t]{0.32\textwidth}
      \centering
    \begin{tikzpicture}[>=Stealth, thick, 
      every node/.style={font=\sffamily},
      dot/.style={circle, draw, minimum size=8pt, inner sep=0pt},
      ->/.style={decoration={markings, mark=at position 0.5 with {\arrow{>}}},
                  postaction={decorate}}
      ]
    \draw[thick] (-2,0) -- (2,0);
    \draw[thick] (0,-1) -- (0,2.5);
    \node[dot] at (0,0) {};
    \node[dot] at (0,1.2) {};
    \node[dot] at (1,2.3) {}; 
    \draw[->] (0,0.0) .. controls (-0.5,0.5) and (-0.5,1.1) .. (0,1.2);
    \draw[->] (0,0.0) .. controls (1.8,1) and (1.6,1.9) .. (1,2.3); 
    \node at (-0.7,1.0) {$1'$};
    \node at (1.6,1.6) {$2'$};
    \end{tikzpicture}
      \caption{%
      Class 1S @ {\tt[$V_3$@LO]}}
      \label{fig:1-step-1-loop}
  \end{subfigure}%
  \begin{subfigure}[t]{0.32\textwidth}
      \centering
      \begin{tikzpicture}[>=Stealth, thick, 
        every node/.style={font=\sffamily},
        dot/.style={circle, draw, minimum size=8pt, inner sep=0pt},
        ->/.style={decoration={markings, mark=at position 0.5 with {\arrow{>}}},
                    postaction={decorate}}
        ]
      \draw[thick] (-2,0) -- (2,0);
      \draw[thick] (0,-1) -- (0,2.5);
      \node[dot] at (0,0) {};      
      \node[dot] at (0,1.2) {};    
      \node[dot] at (1,2.3) {};    
      \draw[->] (0,00) .. controls (-0.5,0.5) and (-0.5,1.1) .. (0,1.2);
      \draw[dashed] (-0.1,1.30) --  +(1,1.1);
      \draw[dashed] (+0.1,1.10) --  +(1,1.1);
      \node at (-0.7,1.0) {$1'$};
      \node at (1.1,1.6) {$2$};
      \end{tikzpicture}
      \caption{%
      Class 1S @ {\tt[$V_3$@NLO]}}
      \label{fig:1-step-2-loop}
  \end{subfigure}
    \caption{%
      Illustrative phase diagrams for
      classes 1S and 2S from
      bands in fig.~\ref{fig:multiplot} and
      benchmark points in table~\ref{tab:2loop-benchmarks}.
      Their corresponding regions are shown in fig.~\ref{fig:multiplot} all at
      {\tt [3D@LO]}.
      Class~(2S) exhibits a two-step transition.
      Class~(1S) at 1-loop features a direct transition $2'$ to the zero-temperature minimum,
      as well a weaker transition $1’$ at a lower $\Tp$.
      Class~(1S) at 2-loop features a direct transition $1’$,
      followed by a crossover to the zero-temperature minimum.
  }
  \label{fig:phase-diagram}
\end{figure}
\begin{figure}[t]
    \centering
    \includegraphics[width=\linewidth]{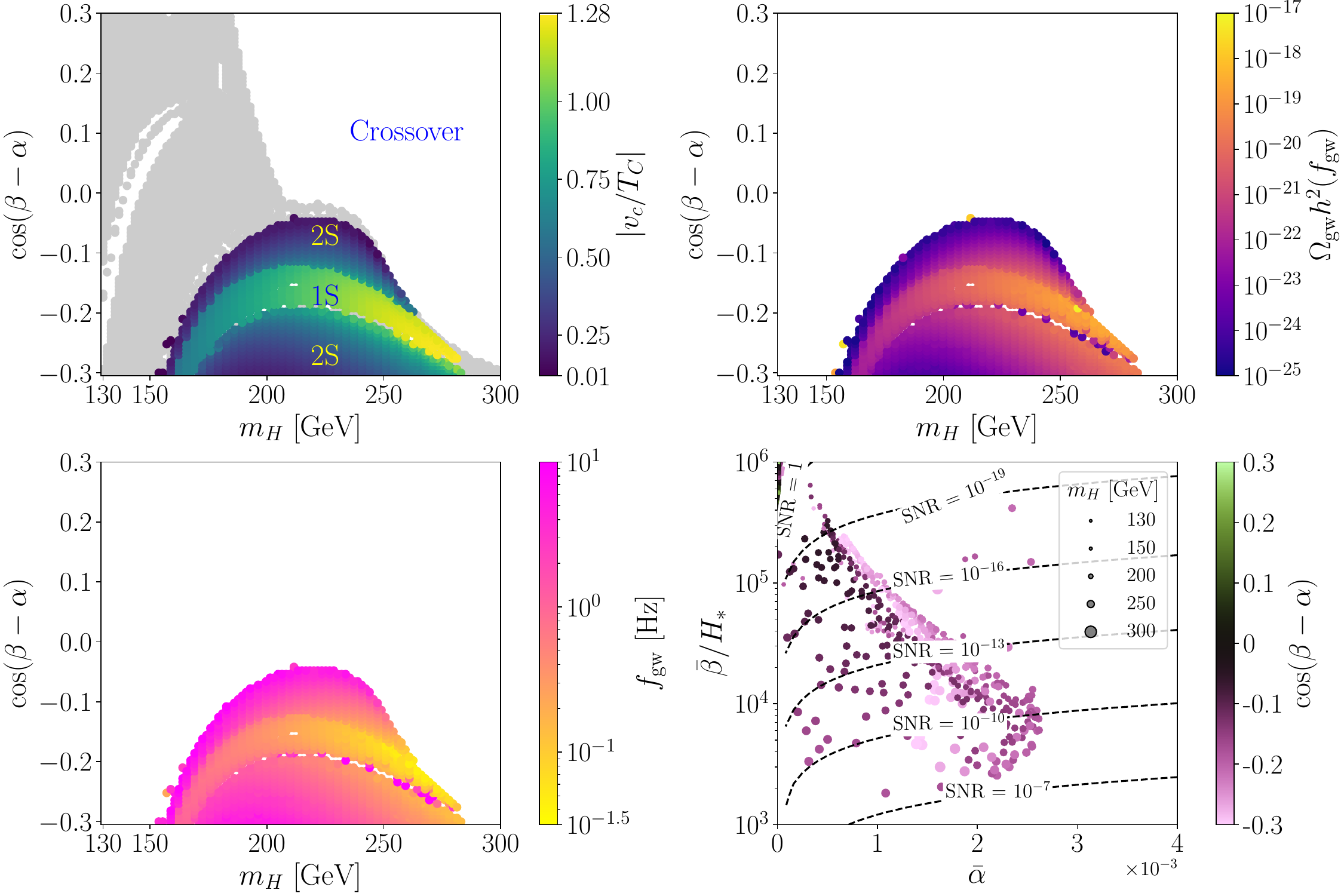}
    \caption{%
      \textbf{Top left:}
      First-order phase transitions in the $\mH$--$\cos(\beta-\alpha)$ plane.
      Color indicates the largest $\vc/\Tc$ at each point.
      Two-step (2S), one-step (1S), and crossover regions are labeled
      according to fig.~\ref{fig:phase-diagram}.
      \textbf{Top right:} Peak amplitude $\Omega_\text{gw}h^2(f_\text{gw})$.
      \textbf{Bottom left:} Peak frequency $f_\text{gw}$.
      \textbf{Bottom right:}
      Phase transitions plotted against $\bar{\alpha}$ and $\bar{\beta}/H_*$
      (randomly sampled subset, 1 in 4, shown for clarity).
      Color indicates $c_{\beta-\alpha}$; circle size indicates $\mH$.
      LISA SNR contours assume $T_\text{p} = 160$~GeV and $\vw = 0.63$.
      \vtwo{The gray region indicates weak 2S transitions.}
    }
    \label{fig:multiplot}
\end{figure}
The finite-temperature vacuum structure of
the scanned parameter space is classified in fig.~\ref{fig:phase-diagram} and
depends on $c_{\beta-\alpha}$ and $\mH$.
Figure~\ref{fig:multiplot} shows the results of our scan.
At intermediate $c_{\beta-\alpha} \sim -0.2$,
the LO potential {\tt[$V_3$@LO]} features two one-step (1S) transitions
labelled $1'$ and $2'$ in fig.~\ref{fig:1-step-1-loop} with 
EWSB proceeding via the stronger transition along path $2'$ to the SM-like vacuum.
In this regime, the electroweak gauge bosons
couple most strongly to the SM-like state, resulting in the strongest phase transitions.
\vtwo{Notably, when including two-loop corrections to the 3D potential,
some of these transitions will develop a second crossover step to the SM-like vacuum;
see fig.~\ref{fig:1-step-2-loop}.}
For smaller $\mH \lesssim 200$~GeV and larger $|c_{\beta-\alpha}| \gtrsim 0.1$,
the first vacuum away from the origin that develops does not give the SM-like
resonance $h$ a VEV. The SM-like vacuum appears only at lower temperature,
so EWSB proceeds through a two-step transition (2S; fig.~\ref{fig:2-step}).
Each step in this scenario is weaker than the direct 1S transition.
Finally, for large $\mH$ and large $|c_{\beta-\alpha}|$, no barrier develops
between vacua and EWSB proceeds via a smooth crossover.

In fig.~\ref{fig:multiplot},
we also show
the peak amplitude (top right) and
frequency (bottom left) of the GW spectrum from the strongest transition.
As expected, the largest amplitudes are found in the band where the transition proceeds directly to the SM-like vacuum.
However, the amplitudes throughout this parameter space remain below the sensitivity of anticipated GW experiments in the coming decades.
Moreover, the peak frequencies typically lie outside the range accessible to space-based interferometers,
reflecting the large $\bar{\beta}/H_*$ values found in our scan.

This is demonstrated explicitly in
fig.~\ref{fig:multiplot} (bottom right),
where we show the predicted
latent heat parameter $\bar{\alpha}$ and
inverse duration parameter $\bar{\beta}/H_*$ across the scanned model space.
We overlay contours of constant SNR for the LISA experiment
calculated via eq.~\eqref{eq:SNRLISA} as dashed contours,
assuming a characteristic transition temperature of $\Tp \simeq 160$~GeV
and a bubble wall speed of $\vw = 0.63$.
None of the models in our scan reach unit SNR.

\vtwo{
%
This preliminary study helps identify regions of strong transitions, and definite structure of the phase diagram.
However, as we will see in sec.~\ref{sec:collider:constraints}, only a small fraction of the parameter space survives the full set of collider constraints,
primarily through strong bounds on $c_{\beta-\alpha}$ from direct searches.} 

\subsubsection{%
  \vtwo{Uncertainties from higher-order corrections in the effective potential}
}

Finally, we estimate uncertainties from using
the two-loop improved one-loop potential rather than
the full two-loop potential, which we do not employ in our full parameter scan
due to issues outlined in sec.~\ref{sec:Veff:T}.

To this end,
\vthree{based on the scan results in fig.~\ref{fig:multiplot}},
we select the benchmark (BM) points
{\tt BM1}--{\tt BM7}
that lie within the two-step (2S) region 
at one-loop~{\tt [$\Veff$@LO]},
then reclassify them according to their transition patterns
at two-loop~{\tt [$\Veff$@NLO]}.
The two emergent cases,
\begin{align}
  \label{eq:case:1}
  \tag{case~1}
  \text{{\tt BM1}-{\tt BM3}}:&
  & \text{2S at 1-loop {\tt[$V_3$@LO]}}
  \,,
  &&
  & \text{2S at 2-loop {\tt[$V_3$@NLO]}}
  \,,\\
  \label{eq:case:2}
  \tag{case~2}
  \text{{\tt BM4}-{\tt BM7}}:&
  & \text{2S at 1-loop {\tt[$V_3$@LO]}}
  \,,
  &&
  & \text{1S at 2-loop {\tt[$V_3$@NLO]}}
  \,,
\end{align}
are classified according to fig.~\ref{fig:phase-diagram}.
In both cases,
two-loop 3D matching {\tt [3D@NLO]} is employed.
\begin{table}[t]
    \centering
    \resizebox{\textwidth}{!}{
      \begin{tabular}{|c|l|c|l|ccc|ccc|}
      \hline
            & BM
            & $\mH$[GeV]
            & \multicolumn{1}{c|}{$c_{\beta-\alpha}$}
            & $\bar{\alpha}_{1\ell}$
            & $\bar{\alpha}_{2\ell}$
            & $\delta \bar\alpha$
            & $(\bar{\beta}/H_*)_{1\ell}$
            & $(\bar{\beta}/H_*)_{2\ell}$
            & $\delta \bar\beta$
            \\
          \hline
          \hline
          \multirow{3}{*}{\rotatebox{90}{\ref{eq:case:1}}}
            & {\tt BM1} & 194 & -0.0715 & 0.00124~ & 0.00141 & 0.12 & $5.62\times 10^4$ & $2.07\times 10^5$ & 0.73 \\
            & {\tt BM2} & 195 & -0.0533 & 0.00103~ & 0.00163 & 0.37 & $7.99\times 10^4$ & $1.71 \times 10^5$ & 0.53 \\
            & {\tt BM3} & 176 & -0.0817 & 0.00103~ & 0.00214 & 0.52 & $9.26\times 10^4$ & $3.20\times10^4$ & 1.9\;\, \\
          \hline
          \multirow{4}{*}{\rotatebox{90}{\ref{eq:case:2}}}
            & {\tt BM4} & 195 & -0.00379 & 0.000696 & 0.00325 & 0.79 & $1.56\times 10^5$ & $1.53\times 10^4$ & 9.2\;\, \\
            & {\tt BM5} & 192 & -0.0286  & 0.000815 & 0.00268 & 0.70 & $1.24\times 10^5$ & $3.95\times 10^4$ & 2.1\;\, \\
            & {\tt BM6} & 174 & -0.103   & 0.00103~ & 0.00208 & 0.50 & $9.37\times 10^4$ & $4.78\times 10^4$ & 0.96 \\
            & {\tt BM7} & 186 & -0.0565  & 0.00101~ & 0.00225 & 0.55 & $9.42\times 10^4$ & $2.29\times 10^4$ & 3.11 \\
          \hline
      \end{tabular}
    }
    \caption{%
      The transition strength parameter $\bar{\alpha}$ and
      \vtwo{the inverse duration parameter $\bar{\beta}/H_*$} with
      relative uncertainties defined in eq.~\eqref{eq:deltaX}
      for benchmark (BM) points {\tt BM1}--{\tt BM7} at
      {\tt [3D@NLO~$V_3$@LO]} and
      {\tt [3D@NLO~$V_3$@NLO]} level.
      The benchmarks are classified into
      \ref{eq:case:1} and \ref{eq:case:2} given their {\tt [$V_3$@NLO]} transition pattern;
      see fig.~\ref{fig:phase-diagram}.
      {\tt BM1}--{\tt BM3} remain in class 2S while
      {\tt BM4}--{\tt BM7} shift to class 1S when going from
      1-loop {\tt [3D@NLO~$V_3$@LO]} to
      2-loop {\tt [3D@NLO~$V_3$@NLO]}.
      The remaining parameters are fixed according to table~\ref{tab:scan1}.
    }
    \label{tab:2loop-benchmarks}
\end{table}
The results for
{\tt BM1}--{\tt BM7} are summarised in
table~\ref{tab:2loop-benchmarks}.

At one-loop level in the effective potential, all benchmarks
start out as 2S transitions.
There, the interim phase appears at a higher temperature than the SM-like vacuum,
leading to a two-step transition (fig.~\ref{fig:2-step}).
For benchmarks of \ref{eq:case:2} at two loop level,
the interim phase and the SM-like vacuum are not distinct;
they are connected by a crossover.
Consequently, the system first transitions to the field-space
location of the interim phase, after which the minimum continuously evolves to the SM-like vacuum
(fig.~\ref{fig:1-step-2-loop}).
In the table~\ref{tab:2loop-benchmarks}, we consistently quote
$\bar{\alpha}$ and $\bar{\beta}/H_*$ for transition
$1$
(or $1'$ for benchmarks of \ref{eq:case:2}) at both orders 
{\tt[$V_3$@LO]} (fig.~\ref{fig:2-step}) and
{\tt[$V_3$@NLO]} (figs.~\ref{fig:2-step} and~\ref{fig:1-step-2-loop}).
This provides a
like-for-like comparison with the NLO result, isolating the
impact of higher-order corrections on the same transition path.
%
Below, we quantify relative uncertainties by 
\begin{align}
\label{eq:deltaX}
  \delta X &= \biggl|\frac{X_{1\ell} - X_{2\ell}}{X_{2\ell}}\biggr|
  \,,
\end{align}
for a quantity $X \in \{ \bar\alpha, \bar\beta/H_*, \Tp, \dots \}$ at
one-loop {\tt [3D@NLO $V_3$@LO]} ($1\ell$) and
two-loop {\tt [3D@NLO $V_3$@NLO]} ($2\ell$) order.

The percolation temperature shifts between the
one- and two-loop computations, which is
consistent with~\cite{Kainulainen:2019kyp}
\vtwo{or the singlet extension of the SM~\cite{Niemi:2021qvp}.}
Correspondingly,
the observe shifts in the transition strength parameter are 
$\delta \bar\alpha \simeq\mathcal{O}(0.1\text{--}1)$.
and
the inverse duration parameter are
$\delta \bar\beta \simeq \mathcal{O}(0.5\text{--}5)$
since
the uncertainty propagates exponentially via eq.~\eqref{eq:beta:def}.

Higher-order corrections can modify the character of the phase transition~\cite{Gould:2023jbz}.
While the full two-loop potential corrections are quantitatively non-negligible,
the qualitative features of the \vtwo{strongest} transitions remain unchanged,
and the predicted GW SNRs stay weak.
The benchmark points were chosen from transitions that are relatively strong.
By contrast, some weaker cases (e.g. {\tt BM4}--{\tt BM7}) switch from 2S to 1S,
with an additional crossover appearing at two loops,
which explains part of the larger differences between the one- and two-loop results
in the second half of table~\ref{tab:2loop-benchmarks}.

\subsection{Combining collider constraints with first-order phase transition scans}
\label{sec:collider:constraints}

\vtwo{To assess the viability of the regions identified above,
we perform an additional scan that explicitly incorporates collider constraints as implemented in
{\tt ScannerS}~\cite{Muhlleitner:2016mzt,Muhlleitner:2020wwk}.
{\tt ScannerS} automatically applies both theoretical and experimental constraints 
to each parameter point. Theoretical checks include perturbative unitarity, vacuum stability, and the boundedness of the scalar potential. 
Experimental limits combine collider exclusions from LEP, Tevatron, and the LHC (via interfaces to {\tt HiggsBounds} \cite{Bechtle:2008jh,Bechtle:2011sb,Bechtle:2013wla,Bechtle:2015pma,Bechtle:2020pkv} and {\tt HiggsSignals} \cite{Bechtle:2013xfa,Bechtle:2020uwn}), \vthree{where compatibility with Higgs signal-rate measurements is assessed using a
$\Delta\chi^2 = \chi^2_{\text{model}} - \chi^2_{\rmii{SM}} < 6.18$ criterion,
corresponding to a $2\sigma$ constraint for two degrees of freedom,} ensuring consistency  with measured Higgs signal rates and null searches for additional scalars. 
Electroweak precision observables are incorporated through the oblique parameters 
$S$, $T$, and $U$ evaluated at one loop,
following~\cite{Grimus:2007if,Muhlleitner:2016mzt,Muhlleitner:2020wwk}.
\vthree{In our numerical analysis,
we evaluate the oblique parameters for each parameter
point and construct a $\chi^2$ test using the experimentally determined central
values and covariance matrix provided by the Particle Data Group~\cite{ParticleDataGroup:2024cfk}.
Since $S$ and $T$ arise from dimension-six operators, while $U$ corresponds to a
dimension-eight operator and is typically much more weakly constrained, we adopt
a two-parameter $(S,T)$ fit with $U = 0$.
Parameter points are accepted if
$\Delta \chi^2_{ST} < 6.18$,
which (as above) corresponds to the $95\%$ confidence level for two degrees of freedom.}
Flavour constraints, as anticipated to be in moderate tension with this model \cite{Azevedo:2023zkg}, are not applied here.} 

\vtwo{
Guided by the diphoton excess reported near 95~GeV~\cite{Azevedo:2023zkg},
we scan the pseudoscalar mass within a $\pm 3\sigma$ window centered at $\mA=95$~GeV.
The remaining parameters are then varied within ranges indicated in table~\ref{tab:scan2}.
In comparison to the previous scan in sec.~\ref{sec:PT:scan},
we also varied $m_{12}^2$ slightly which, however,
has a negligible impact on the phase transition results.}
\begin{table}[t]
\centering
\small
\begin{tabular}{|l|ccccccc|} \hline
 & $\mh$~[GeV]
 & $\mH$~[GeV]
 & $\mA$~[GeV]
 & $\mHpm$~[GeV]
 & $c_{\beta-\alpha}$
 & $t_\beta$
 & $m_{12}^2$~[GeV] \\
\hline
\hline
Min & 125.09 & ~30 & 94 & 160 & $-0.3$ & 1.55 & $10^{-3}$ \\
Max & 125.09 & 300 & 97 & 320 & $+0.3$ & 1.60 & $1$ \\
\hline
\end{tabular}
\caption{%
  \vtwo{Ranges of different parameters
  for the collider scan in fig.~\ref{fig:collider:plot}.
  Individual points are generated using
  {\tt ScannerS}~\cite{Muhlleitner:2016mzt,Muhlleitner:2020wwk}.
  Here,
  $\mh$ and $\mH$ refer to the SM-like and
  non-SM-like CP-even scalars, respectively.
}
}
\label{tab:scan2}
\end{table}
\vtwo{This parameter choice closely follows our scans in the previous section,
but now allows a direct evaluation of which points remain phenomenologically viable.
In the unconstrained case, points with $|c_{\beta-\alpha}| \simeq 0$ tended
to produce crossovers rather than first-order transitions.
Conversely, points with more negative $c_{\beta-\alpha} < -0.2$ and
heavier charged Higgs masses,
which could yield stronger transitions,
are typically excluded once collider bounds are applied.
This is in alignment with the expectation,
that for a lighter scalar masses $\mH$,
parameter space points with a stronger gauge coupling to the SM-like Higgs
should be ruled out by collider constraints~\cite{Gunion:2002zf,Goncalves:2021egx,Azevedo:2023zkg}.
Viable parameter points tend to cluster near
$\mH \simeq \mHpm$,
which motivates scanning both masses over comparable ranges.
This ensures that the resulting parameter space respects
the precision electroweak bounds while still allowing for a meaningful exploration of the phase-transition dynamics.
}

\begin{figure}[t]
    \centering
    \includegraphics[width=\linewidth]{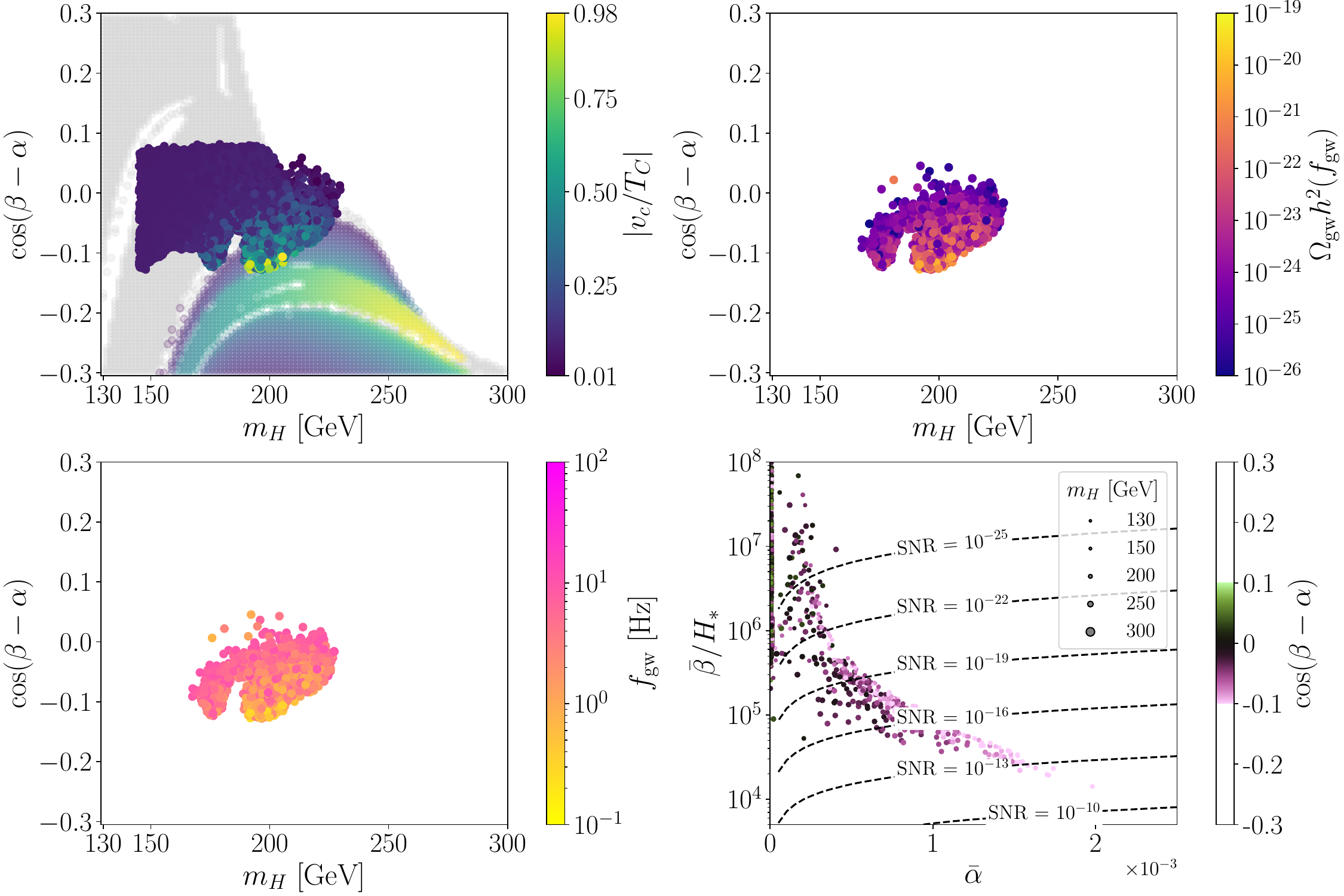}
    \caption{%
      \vtwo{%
      First-order phase transitions in the
      $\mH$--$\cos(\beta-\alpha)$ plane, showing
      transition strength $\vc/\Tc$ (top left),
      peak amplitude $\Omega_\text{gw}h^2(f_\text{gw})$ (top right),
      peak frequency $f_\text{gw}$ (bottom left), and
      phase transitions plotted against $\bar{\alpha}$ and $\bar{\beta}/H_*$ (bottom right).
      As in fig.~\ref{fig:multiplot}, but
      with the scan following table~\ref{tab:scan2}
      and applying constraints from
      {\tt ScannerS}~\cite{Muhlleitner:2016mzt,Muhlleitner:2020wwk}.
    }
    }
    \label{fig:collider:plot}
\end{figure}
\vtwo{The results of this collider-constrained scan are shown in
fig.~\ref{fig:collider:plot}.%
\footnote{
  \vthree{As a check, we also performed a similar scan
  where all three oblique parameters
  $S$, $T$, and $U$ were evaluated at the $2\sigma$ level,
  yielding no significant differences in the surviving parameter space.}
}
Comparing to fig.~\ref{fig:multiplot},
we see that the viable parameter space shrinks considerably.
The strongest transitions of
the surviving points cluster around the alignment limit
$|c_{\beta-\alpha}| \lesssim 0.1$ (consistent with \cite{ATLAS:2024lyh}),
with
an allowed variation of 
$|\mH - \mHpm|/\mH \lesssim \mathcal{O}(0.1)$ as expected.
The maximum transition strength is reduced to
$\bar{\alpha} \sim 0.0025$,
leading to even weaker GW signals.
}

%
\section{Conclusions}
\label{sec:conclusion}

In this paper, we identified the viable parameter space for implementing
the 95~GeV resonance in the real Type~I 2HDM, \vtwo{chosen} based on previous work
to ensure simultaneous best fits of the $\gamma\gamma$ and $\tau\tau$ resonances \vtwo{and to minimise the tension with the $b\to s\gamma$ branching ratio}.
The constraints leave $\mH$ and $c_{\beta-\alpha}$ as the most freely varying parameters.
After fixing the remaining model parameters, \vtwo{in a first scan we varied  just $\mH$ and $c_{\beta-\alpha}$ to obtain some primary insights in the structure of the vacuum transitions and the strength of the phase transition.} 
We employed the dimensionally reduced thermal effective potential
to compute the transition parameters
$\bar{\alpha}$ and
$\bar{\beta}/H_*$,
as well as the GW spectrum $\Omega_\text{gw}h^2(f)$, its
peak frequency $f_\text{gw}$, and
peak amplitude $\Omega_\text{gw}h^2(f_\text{gw})$.
\vtwo{Following this initial scan, we used {\tt ScannerS}~\cite{Muhlleitner:2016mzt,Muhlleitner:2020wwk} to explore how experimental tests further limit the parameter space. }

\vtwo{In our first scan in fig.~\ref{fig:multiplot},}
we find that the $\mH$--$c_{\beta-\alpha}$ parameter space
contains regions with crossovers, one-step, and
two-step first-order transitions.
\vtwo{In the experimentally constrained parameter space of fig.~\ref{fig:collider:plot}, 
the strongest transitions have a transition strength}
$\bar{\alpha}\sim \vtwo{0.0025}$,
corresponding to peak amplitudes of
$\Omega_\text{gw}h^2(f_\text{gw}) \sim \vtwo{10^{-20}}$ and frequencies
$f_\text{gw}\sim \vtwo{10^{-1}}$~Hz.
\vfour{The transition strength in terms of the conventional criterion for electroweak baryogenesis,
however, remains $\vc/\Tc \lesssim 1$.}
The resulting maximum LISA SNR is
around $\text{SNR} \sim \vtwo{10^{-10}}$, far below
the conventional detection threshold of $\text{SNR} \sim 10$.
We conclude that models simultaneously explaining the
$\gamma\gamma$ and $\tau\tau$ excesses within the Type~I 2HDM yield FOPTs
too weak for LISA detection,
requiring interferometers with significantly
greater sensitivity in the centihertz frequency band.
\vfour{Notably,
the final viable parameter space 
remains in tension with $\mathrm{BR}(b \to s\gamma)$
at the level of $\mathcal{O}(3\sigma)$.}

\vtwo{Additionally, we identified $\mathcal{O}(1)$ uncertainties
in thermodynamic transition parameters
arising from higher-order corrections in the effective potential
by comparing one-loop and two-loop potentials in the 3D EFT.
While the quantitative differences in transition parameters
can be significant,
the qualitative features of the phase transitions remain unchanged.
A further analysis of higher-order corrections,
including the full two-loop potential in the 3D EFT,
is left for future work.}

The modest EWPT strength observed in our scan can be attributed to
the radiative origin of the barrier separating the vacua.
In the parameter space compatible with identifying the 95~GeV excess
as a light pseudoscalar, the scalar mass spectrum is relatively compressed,
limiting the enhancement of thermal cubic terms that typically arise from large mass splittings.
Consequently, the barrier is primarily generated by gauge boson loops,
as in the Standard Model.
The light pseudoscalar itself contributes only weakly to
the thermal potential at high temperatures and
does not significantly enhance the barrier.
This scenario contrasts with regions of the \vtwo{Type~I} 2HDM where heavier scalars can radiatively strengthen
the transition or where tree-level terms provide a barrier already at zero temperature.
In particular, previous studies have shown that large mass splittings,
especially between the pseudoscalar and the heavy scalar or charged Higgs,
can amplify scalar contributions to the thermal potential and
lead to strong first-order transitions.
Consequently, while the transition is generically first-order in our setup,
it is not sufficiently strong to produce observable GW signals or
support electroweak baryogenesis
\vfour{since for the latter, we observe $\vc/\Tc \lesssim 1$.}
This conclusion is based on the standard criterion
$\vc/\Tc \gtrsim 1$ (cf.\ e.g.~\cite{Ahriche:2007jp,Basler:2016obg}) and
holds unless additional model ingredients are introduced.
\vtwo{See~\cite{Li:2025kyo} for a more refined and updated criterion
in the context of electroweak baryogenesis.}

Another possibility for generating a large barrier via significant mass splitting
is to treat the second, heavier scalar as a UV degree of freedom in the final EFT
and integrate it out during dimensional reduction.
However, implementing this approach would require dynamically switching between different
EFT hierarchies throughout the parameter scan~\cite{Gould:2023ovu}.
We have therefore chosen to focus on
the thermal mass hierarchy as outlined in our analysis
and defer the exploration of more intricate EFT hierarchies to future work.

Finally, we comment on how additional degrees of freedom could
alter the thermal history and potentially enhance the phase transition strength.
In the \vtwo{Type~I} 2HDM we have discussed,
the barrier between the high-temperature and low-temperature vacua is
generated radiatively.
Coupling the Higgs sector to a scalar gauge singlet can introduce
a tree-level barrier in the potential.
Alternatively, fermionic extensions or higher-dimensional operators
(e.g., $(H^\dagger H)^3/\Lambda^2$) in the UV of
the 4D theory~\cite{Croon:2020cgk} can enable viable electroweak baryogenesis.
Additionally, higher-dimensional operators in
the UV of the 3D EFT~\cite{Chala:2024xll,Chakrabortty:2024wto,Chala:2025aiz,Bernardo:2025vkz,Chala:2025oul}
can modify the shape of the effective potential at
finite temperature and thus the transition strength.
These extensions may also shift the transition dynamics into a more strongly supercooled regime,
lowering the percolation temperature and thereby
pushing the GW signal into
the most sensitive frequency band of space-based interferometers.
A systematic exploration of such scenarios within
the dimensional reduction framework,
including complementary constraints from colliders and
electroweak precision studies (e.g.,~\cite{Friedrich:2022cak,Ramsey-Musolf:2024ykk}),
is a promising direction for future work. 
%

%
\section*{Acknowledgements}

We generously thank
Lauri Niemi,
Francesco Rescigno,
Tuomas V.I.~Tenkanen,
and
Jorinde van de Vis
for illuminating discussions.
We also thank the authors of {\tt PhaseTracer2}~\cite{Athron:2020sbe,Athron:2024xrh} for their support
during the setup of the model parameters scans.
~AB and DC are supported by the STFC under Grant No.~ST/T001011/1.
PS was supported by
the Swiss National Science Foundation (SNSF) under grant
\href{https://data.snf.ch/grants/grant/215997}{\tt PZ00P2-215997}.

%
\appendix
\renewcommand{\thesection}{\Alph{section}}
\renewcommand{\thesubsection}{\Alph{section}.\arabic{subsection}}
\renewcommand{\theequation}{\Alph{section}.\arabic{equation}}

%
\section{Renormalisation of the \vtwo{Type~I} 2HDM}
\label{sec:renormalisaton:2hdm}

\subsection{Running and $\beta$-functions}
\label{app:betafunctions}

Electroweak resonances are typically measured at the $Z$-pole,
meaning that the physical mass inputs will exist at an energy scale $\mu = \mZ$.
Through the one-loop renormalisation relations,
we can relate this input to Lagrangian parameters that also exist at the same energy scale.
Next, we can renormalisation group evolve
the Lagrangian parameters via the beta functions to the 4D scale $\mu=\LamD$ of our theory,
where they can then act as input for our model.

The renormalisation group equations listed below are associated with
the parameters of the \vtwo{Type~I} 2HDM and encode
their running with respect to
the four-dimensional \MSbar{} renormalisation scale
$\LamD$ via the $\beta$-functions.
To this end, we use
\begin{equation}
\label{eq:rge:param}
t \equiv \ln\LamD
\;,
\end{equation}
where $\LamD^2 \equiv 4\pi e^{-\gammaE} \mu^2$,%
\footnote{This relates the \MSbar-scale with that of the MS scheme.}
and find at one-loop level:
\begin{align}
    \partial_t g_1^2 &= \frac{7}{8\pi^2}g_1^4
    \,,\\
    \partial_t g_2^2 &= -\frac{3}{8\pi^2}g_2^4 
    \,,\\
    \partial_t g_3^2 &= -\frac{7}{8\pi^2}g_3^4 
    \,,\\
    \partial_t \gY &= \frac{1}{192\pi^2}\gY\left(-17g_1^2 - 27g_2^2 -96g_3^2 + 54\gY^2)  \right)
    \,,\\
    \partial_t m_{11}^2 &= \frac{1}{32\pi^2}\left(-3m_{11}^2(g_1^2 +3 g_2^2 - 8\lambda_1) + 4m_{22}^2(2\lambda_3 +\lambda_4 )  \right)
    \,,\\
    \partial_t m_{22}^2 &= \frac{1}{32\pi^2}\left(-3m_{22}^2(g_1^2 +3 g_2^2 - 4\gY^2 -8\lambda_2) + 4m_{11}^2(2\lambda_3 +\lambda_4 )  \right)
    \,,\\
    \partial_t m_{12}^2 &= \frac{1}{32\pi^2}m_{12}^2\left(-3g_1^2 - 9 g_2^2 + 6\gY^2 + 4(\lambda_3 +2\lambda_4 + 3\lambda_5)  \right)
    \,,\\
    \partial_t \lambda_1 &= \frac{1}{128\pi^2}\Big(3g_1^4 + 9g_2^4 + 6g_1^2(g_2^2 - 4\lambda_1) - 72g_2^2\lambda_1
      \nn
    &\quad + 8(24\lambda_1^2 + 2\lambda_3^2 + 2\lambda_3\lambda_4 + \lambda_4^2 + \lambda_5^2) \Big) 
    \,,\\
    \partial_t \lambda_2 &= \frac{1}{128\pi^2}\Big(3g_1^4 + 9g_2^4 + 6g_1^2(g_2^2 - 4\lambda_2) - 72g_2^2\lambda_2
      \nn
    &\quad + 96\lambda_2(\gY^2+ 2\lambda_2) +8( -6\gY^4 + 2\lambda_3^2+ 2\lambda_3\lambda_4 + \lambda_4^2 + \lambda_5^2) \Big)
    \,,\\
    \partial_t \lambda_3 &= \frac{1}{64\pi^2}\Big(3g_1^4 + 9g_2^4 -36g_2^2\lambda_3 - 6g_1^2(g_2^2 + 2\lambda_3)
      \nn
    &\quad +8(\lambda_3(3\gY^2+ 6(\lambda_1+\lambda_2) + 2\lambda_3)+ 2(\lambda_1+\lambda_2)\lambda_4 + \lambda_4^2 + \lambda_5^2) \Big)
    \,,\\
    \partial_t \lambda_4 &= \frac{1}{16\pi^2}\Big(3g_1^2(g_2^2-\lambda_4) + 9g_2^2\lambda_4 +6\gY^2\lambda_4
      \nn
    &\quad +4\lambda_4(\lambda_1 + \lambda_2 + 2\lambda_3 + \lambda_4) + 8\lambda_5^2 \Big) 
    \,,\\
    \partial_t \lambda_5 &= \frac{1}{16\pi^2}\lambda_5\left(-3g_1^2 - 9 g_2^2 + 6\gY^2 + 4(\lambda_1 + \lambda_2 + 2\lambda_3 +3\lambda_4)  \right)
    \,,
\end{align}
The renormalisation scale of the thermal transition is chosen as
$\LamD = 4\pi e^{-\gammaE}T$,
which lies close to the thermal scale and suppresses its contribution to the thermal logarithms.
The scales of the soft and ultrasoft EFTs are set to
$\Lamd = \LamdUS = B T$,
and for simplicity we choose $B = 1$. We refer to this choice of the $\LamdUS$ scale as the `softer' scale.
In practice, the choice of $\Lamd$ can be made more rigorous by applying the
{\em principle of minimal sensitivity}~\cite{Stevenson:1981vj}.
This principle entails minimising the dependence of $\Lamd$ or $\LamdUS$,
for example,
in the effective potential,
to determine an optimal scale $\bar{\mu}_\rmi{opt}$,
as discussed in~\cite{Huang:1994cu,Laine:2005ai,Ghisoiu:2015uza}.

\subsection{Relations between \MSbar-parameters and physical observables}
\label{sec:vacuum:renormalisation}

The physical observables map to the \MSbar-parameters of the Lagrangian as,
\begin{equation}
\begin{gathered}
    (\mh, \mHpm, \mH, \mA,c_{\beta-\alpha},t_\beta,\mW, \mZ, \mt, \Gf, \alphas)
    \\
    \downmapsto\\
    (m_{11}^2,m_{22}^2,m_{12}^2,\lambda_1,\lambda_2,\lambda_3,\lambda_4,\lambda_5,g_1,g_2,g_3,\gY)
    \;.
\end{gathered}
\end{equation}
The physical observables, along with $m_{12}$, serve as input parameters measured at the $Z$-pole, $\mu = \mZ$.
We define the shorthand notation
$g_0^2 = 4\sqrt{2} \Gf \mW^2$
for the tree-level coupling.

At tree-level,
the vacuum relations for the gauge couplings are,
\begin{align}
\vtwo{g_2^2} &= g_0^2
\,,&
\vtwo{g_1^2} &= g_0^2 \Bigl(\Bigl(\frac{\mZ}{\mW}\Bigr)^2-1\Bigr)
\,,&
\vtwo{\gY^2} &= \frac{g_0^2}{2} \Bigl(\frac{\mt}{\mW}\Bigr)^2
\,.
\end{align}
For the other Lagrangian parameters, we list the tree-level relations given in Appendix B.2 of~\cite{Gorda:2018hvi}:
\begin{align}
m_{11}^2 &= m_{12}^2 t_\beta - \frac{1}{2} \left( \mH^2 + (\mH^2 - \mH^2) c_{\beta-\alpha} (c_{\beta-\alpha}+s_{\beta-\alpha} t_\beta) \right) \,,\\ 
m_{22}^2 &= m_{12}^2 t_\beta^{-1} - \frac{1}{2} \left(\mH^2 + (\mH^2 - \mH^2) c_{\beta-\alpha} (c_{\beta-\alpha}-s_{\beta-\alpha} t_\beta^{-1}) \right) \,,\\
v^2 \lambda_1 &= \frac{1}{2}\bigg( \mH^2 + \Omega^2 t_\beta^2 - (\mH^2 - \mH^2)\left(1-(s_{\beta-\alpha}+c_{\beta-\alpha} t_\beta^{-1})^2\right) t_\beta^2\bigg) \,,\\
v^2 \lambda_2 &= \frac{1}{2}\bigg( \mH^2 + \Omega^2 t_\beta^{-2} - (\mH^2 - \mH^2)\left(1-(s_{\beta-\alpha}-c_{\beta-\alpha} t_\beta)^2\right) t_\beta^{-2} \bigg) \,,\\
v^2 \lambda_3 &= 2 \mHpm + \Omega^2 -\mH^2 - (\mH^2 - \mH^2)\left(1+(s_{\beta-\alpha}+c_{\beta-\alpha} t_\beta^{-1})(s_{\beta-\alpha}-c_{\beta-\alpha} t_\beta)\right) \,,\\
v^2 \lambda_4 &= \mA^2 - 2 \mHpm + \mH^2 - \Omega^2 \,,\\
v^2 \lambda_5 &= \mH^2 - \mA^2 - \Omega^2\,,
\end{align}
where we have defined $\Omega^2 \equiv \mH^2-m_{12}^2(t_\beta + t_\beta^{-1})$ as in~\cite{Gorda:2018hvi},%
\footnote{Note that there is a difference in sign between our $m_{12}^2$ and the one defined in~\cite{Gorda:2018hvi}.}
and we use the notation $c_{\beta-\alpha} = \cos(\beta-\alpha)$ (and likewise for the other trigonometric functions and angles).
In appendix~A of~\cite{Kainulainen:2019kyp},
the one-loop $\overline{\text{MS}}$ renormalised expressions are provided with reference
to the self-energies for the gauge bosons, top quark, and the scalars $h$, $H$, $H^\pm$ and $A$.
We make use of the self energies calculated by the authors of~\cite{Kainulainen:2019kyp},
which have analytic expressions too unwieldy to quote here.
Explicit expressions for the self energies can be found in~\cite{Kanemura:2015mxa}. The one-loop renormalised Lagrangian parameters are also defined at $\mu = \mZ$, like the input physical observables. We then use the beta functions to run them to the thermal scale.

\section{Integration of DR EFT into {\tt PhaseTracer2}}
The output of
{\tt DRalgo}~\cite{Ekstedt:2022bff}
can be incredibly long and impractical for computation.
With increasing loop orders, the number of binary operations in
a single expression for the potential becomes too high for some compilers to parse and optimise.
Fortunately, techniques exist to optimise the expressions prior to inserting into code.
In Mathematica, after running the dimensional reduction in
{\tt DRalgo}, the {\tt Experimental\backtick OptimizeExpression} function
can simplify expressions by identifying common sub-expressions and creating
new `Compile' variables to save on evaluation time.
The subsequent expression that is composed of Compile variables is subsequently
far shorter and can be parsed by compilers.

Additionally, the {\tt CForm} function can allow for the conversion of this
optimised expression into {\tt C} code, with the {\tt StringReplace} function
allowing for the manipulation of the subsequent code strings so that they can
be adapted for use in {\tt Python}, {\tt C++}, or any other programming language.
Thus, a function can be created that compounds these operations together
such that {\tt DRalgo} output can readily be used in a programme.

When interfacing the dimensionally reduced output with {\tt PhaseTracer2}, particular care must be taken so that the package can handle divergences at low temperatures, as this is beyond the validity of the EFT (we remind the reader that the vacuum structure is encoded in the Lagrangian parameters through
the vacuum renormalisation of sec.~\ref{sec:vacuum:renormalisation}).
The 
\verb!set_t_low! and
\verb!set_t_high! functions in
the {\tt PhaseFinder} module allows for the temperature bounds to be set,
such that phases are only identified between those temperatures and numerical issues can be avoided.

Finally, regarding the parameter scans: {\tt PhaseTracer2} does not come with an interface to set them up. However, it is relatively straightforward to wrap the {\tt PhaseTracer2} objects in a class that can be instantiated with the a new set of parameters. Parallelisation of the scan across multiple threads allows for increased efficiency when computing results, particularly on machines with numerous cores.

{\small
%
\bibliographystyle{utphys}
\bibliography{ref}

@article{Misiak:2020vlo,
    author = "Misiak, M. and Rehman, Abdur and Steinhauser, Matthias",
    title = "{Towards $ \overline{B}\to {X}_s\gamma $ at the NNLO in QCD without interpolation in m$_{c}$}",
    eprint = "2002.01548",
    archivePrefix = "arXiv",
    primaryClass = "hep-ph",
    reportNumber = "TTP20-001, P3H-20-005, IFT-01/2020",
    doi = "10.1007/JHEP06(2020)175",
    journal = "JHEP",
    volume = "06",
    pages = "175",
    year = "2020"
}

@article{Misiak:2017bgg,
    author = "Misiak, Mikolaj and Steinhauser, Matthias",
    title = "{Weak radiative decays of the B meson and bounds on $M_{H^\pm }$ in the Two-Higgs-Doublet Model}",
    eprint = "1702.04571",
    archivePrefix = "arXiv",
    primaryClass = "hep-ph",
    reportNumber = "TTP17-004, IFT-1-2017",
    doi = "10.1140/epjc/s10052-017-4776-y",
    journal = "Eur. Phys. J. C",
    volume = "77",
    number = "3",
    pages = "201",
    year = "2017"
}

@article{ParticleDataGroup:2024cfk,
    author = "Navas, S. and others",
    collaboration = "Particle Data Group",
    title = "{Review of particle physics}",
    doi = "10.1103/PhysRevD.110.030001",
    journal = "Phys. Rev. D",
    volume = "110",
    number = "3",
    pages = "030001",
    year = "2024"
}

@article{Misiak:2015xwa,
    author = "Misiak, M. and others",
    title = "{Updated NNLO QCD predictions for the weak radiative B-meson decays}",
    eprint = "1503.01789",
    archivePrefix = "arXiv",
    primaryClass = "hep-ph",
    reportNumber = "TTP15-007, SFB-CPP-14-121, SI-HEP-2015-08, QFET-2015-09, TTK-15-09, IFT-2-2015",
    doi = "10.1103/PhysRevLett.114.221801",
    journal = "Phys. Rev. Lett.",
    volume = "114",
    number = "22",
    pages = "221801",
    year = "2015"
}

@article{Hermann:2012fc,
    author = "Hermann, Thomas and Misiak, Mikolaj and Steinhauser, Matthias",
    title = "{$\bar{B}\to X_s \gamma$ in the Two Higgs Doublet Model up to Next-to-Next-to-Leading Order in QCD}",
    eprint = "1208.2788",
    archivePrefix = "arXiv",
    primaryClass = "hep-ph",
    reportNumber = "SFB-CPP-12-60, TTP12-29, IFT-5-2012",
    doi = "10.1007/JHEP11(2012)036",
    journal = "JHEP",
    volume = "11",
    pages = "036",
    year = "2012"
}

@article{Mahmoudi:2009zx,
    author = "Mahmoudi, Farvah and Stal, Oscar",
    title = "{Flavor constraints on the two-Higgs-doublet model with general Yukawa couplings}",
    eprint = "0907.1791",
    archivePrefix = "arXiv",
    primaryClass = "hep-ph",
    doi = "10.1103/PhysRevD.81.035016",
    journal = "Phys. Rev. D",
    volume = "81",
    pages = "035016",
    year = "2010"
}

@article{Deschamps:2009rh,
    author = "Deschamps, O. and Descotes-Genon, S. and Monteil, S. and Niess, V. and T'Jampens, S. and Tisserand, V.",
    title = "{The Two Higgs Doublet of Type II facing flavour physics data}",
    eprint = "0907.5135",
    archivePrefix = "arXiv",
    primaryClass = "hep-ph",
    reportNumber = "LPC-CF-2009-05, LPT-ORSAY-2009-43, LAPP-EXP-2009-03",
    doi = "10.1103/PhysRevD.82.073012",
    journal = "Phys. Rev. D",
    volume = "82",
    pages = "073012",
    year = "2010"
}

@article{Bechtle:2011sb,
    author = "Bechtle, Philip and Brein, Oliver and Heinemeyer, Sven and Weiglein, Georg and Williams, Karina E.",
    title = "{HiggsBounds 2.0.0: Confronting Neutral and Charged Higgs Sector Predictions with Exclusion Bounds from LEP and the Tevatron}",
    eprint = "1102.1898",
    archivePrefix = "arXiv",
    primaryClass = "hep-ph",
    reportNumber = "FR-PHENO-2011-002, BONN-TH-2011-02, DESY-11-016",
    doi = "10.1016/j.cpc.2011.07.015",
    journal = "Comput. Phys. Commun.",
    volume = "182",
    pages = "2605--2631",
    year = "2011"
}

@article{Bechtle:2020pkv,
    author = "Bechtle, Philip and Dercks, Daniel and Heinemeyer, Sven and Klingl, Tobias and Stefaniak, Tim and Weiglein, Georg and Wittbrodt, Jonas",
    title = "{HiggsBounds-5: Testing Higgs Sectors in the LHC 13 TeV Era}",
    eprint = "2006.06007",
    archivePrefix = "arXiv",
    primaryClass = "hep-ph",
    reportNumber = "BONN-TH-2020-03, DESY 20-093, DESY-20-093, IFT-UAM/CSIC-20-072, LU 20-27",
    doi = "10.1140/epjc/s10052-020-08557-9",
    journal = "Eur. Phys. J. C",
    volume = "80",
    number = "12",
    pages = "1211",
    year = "2020"
}

@article{Bechtle:2013xfa,
    author = "Bechtle, Philip and Heinemeyer, Sven and St{\r{a}}l, Oscar and Stefaniak, Tim and Weiglein, Georg",
    title = "{$HiggsSignals$: Confronting arbitrary Higgs sectors with measurements at the Tevatron and the LHC}",
    eprint = "1305.1933",
    archivePrefix = "arXiv",
    primaryClass = "hep-ph",
    reportNumber = "BONN-TH-2013-07, DESY-13-078",
    doi = "10.1140/epjc/s10052-013-2711-4",
    journal = "Eur. Phys. J. C",
    volume = "74",
    number = "2",
    pages = "2711",
    year = "2014"
}

@article{Bechtle:2015pma,
    author = "Bechtle, Philip and Heinemeyer, Sven and Stal, Oscar and Stefaniak, Tim and Weiglein, Georg",
    title = "{Applying Exclusion Likelihoods from LHC Searches to Extended Higgs Sectors}",
    eprint = "1507.06706",
    archivePrefix = "arXiv",
    primaryClass = "hep-ph",
    reportNumber = "BONN-TH-2015-08, DESY-15-093, SCIPP-15-05",
    doi = "10.1140/epjc/s10052-015-3650-z",
    journal = "Eur. Phys. J. C",
    volume = "75",
    number = "9",
    pages = "421",
    year = "2015"
}

@article{Bechtle:2013wla,
    author = "Bechtle, Philip and Brein, Oliver and Heinemeyer, Sven and St{\r{a}}l, Oscar and Stefaniak, Tim and Weiglein, Georg and Williams, Karina E.",
    title = "{$\mathsf{HiggsBounds}-4$: Improved Tests of Extended Higgs Sectors against Exclusion Bounds from LEP, the Tevatron and the LHC}",
    eprint = "1311.0055",
    archivePrefix = "arXiv",
    primaryClass = "hep-ph",
    reportNumber = "BONN-TH-2013-21, DESY-13-110",
    doi = "10.1140/epjc/s10052-013-2693-2",
    journal = "Eur. Phys. J. C",
    volume = "74",
    number = "3",
    pages = "2693",
    year = "2014"
}

@article{Bechtle:2008jh,
    author = "Bechtle, Philip and Brein, Oliver and Heinemeyer, Sven and Weiglein, Georg and Williams, Karina E.",
    title = "{HiggsBounds: Confronting Arbitrary Higgs Sectors with Exclusion Bounds from LEP and the Tevatron}",
    eprint = "0811.4169",
    archivePrefix = "arXiv",
    primaryClass = "hep-ph",
    reportNumber = "DCPT-08-172, IPPP-08-86, BONN-TH-2008-17",
    doi = "10.1016/j.cpc.2009.09.003",
    journal = "Comput. Phys. Commun.",
    volume = "181",
    pages = "138--167",
    year = "2010"
}

@article{Bechtle:2020uwn,
    author = "Bechtle, Philip and Heinemeyer, Sven and Klingl, Tobias and Stefaniak, Tim and Weiglein, Georg and Wittbrodt, Jonas",
    title = "{HiggsSignals-2: Probing new physics with precision Higgs measurements in the LHC 13 TeV era}",
    eprint = "2012.09197",
    archivePrefix = "arXiv",
    primaryClass = "hep-ph",
    reportNumber = "BONN-TH-2020-09, DESY-20-228, DESY 20-228, IFT-UAM/CSIC-20-081, LU TP 20-53",
    doi = "10.1140/epjc/s10052-021-08942-y",
    journal = "Eur. Phys. J. C",
    volume = "81",
    number = "2",
    pages = "145",
    year = "2021"
}

@article{ATLAS:2024lyh,
    author = "Aad, Georges and others",
    collaboration = "ATLAS",
    title = "{Interpretations of the ATLAS measurements of Higgs boson production and decay rates and differential cross-sections in pp collisions at $ \sqrt{s} $ = 13 TeV}",
    eprint = "2402.05742",
    archivePrefix = "arXiv",
    primaryClass = "hep-ex",
    reportNumber = "CERN-EP-2024-017",
    doi = "10.1007/JHEP11(2024)097",
    journal = "JHEP",
    volume = "11",
    pages = "097",
    year = "2024"
}

@article{LEP_2006,
    author = "Schael, S. and others",
    collaboration = "ALEPH, DELPHI, L3, OPAL, LEP Working Group for Higgs Boson Searches",
    title = "{Search for neutral MSSM Higgs bosons at LEP}",
    eprint = "hep-ex/0602042",
    archivePrefix = "arXiv",
    reportNumber = "CERN-PH-EP-2006-001",
    doi = "10.1140/epjc/s2006-02569-7",
    journal = "Eur. Phys. J. C",
    volume = "47",
    pages = "547--587",
    year = "2006"
}

@article{CMS:2024yhz,
    author = "Hayrapetyan, Aram and others",
    collaboration = "CMS",
    title = "{Search for a standard model-like Higgs boson in the mass range between 70 and 110 GeV in the diphoton final state in proton-proton collisions at s=13 TeV}",
    eprint = "2405.18149",
    archivePrefix = "arXiv",
    primaryClass = "hep-ex",
    reportNumber = "CMS-HIG-20-002, CERN-EP-2024-088",
    doi = "10.1016/j.physletb.2024.139067",
    journal = "Phys. Lett. B",
    volume = "860",
    pages = "139067",
    year = "2025"
}

@article{Belyaev:2023xnv,
    author = "Belyaev, Alexander and Benbrik, Rachid and Boukidi, Mohammed and Chakraborti, Manimala and Moretti, Stefano and Semlali, Souad",
    title = "{Explanation of the hints for a 95 GeV Higgs boson within a 2-Higgs Doublet Model}",
    eprint = "2306.09029",
    archivePrefix = "arXiv",
    primaryClass = "hep-ph",
    doi = "10.1007/JHEP05(2024)209",
    journal = "JHEP",
    volume = "05",
    pages = "209",
    year = "2024"
}

@article{Witten:1984rs,
    author = "Witten, Edward",
    title = "{Cosmic Separation of Phases}",
    reportNumber = "PRINT-84-0400 (IAS,PRINCETON)",
    doi = "10.1103/PhysRevD.30.272",
    journal = "Phys. Rev. D",
    volume = "30",
    pages = "272--285",
    year = "1984"
}

@article{Hogan:1986qda,
    author = "Hogan, C. J.",
    title = "{Gravitational radiation from cosmological phase transitions}",
    journal = "Mon. Not. Roy. Astron. Soc.",
    volume = "218",
    pages = "629--636",
    year = "1986"
}

@article{Biekotter:2022kgf,
    author = {Biek\"otter, Thomas and Heinemeyer, Sven and No, Jos\'e Miguel and Olea-Romacho, Mar\'\i{}a Olalla and Weiglein, Georg},
    title = "{The trap in the early Universe: impact on the interplay between gravitational waves and LHC physics in the 2HDM}",
    eprint = "2208.14466",
    archivePrefix = "arXiv",
    primaryClass = "hep-ph",
    reportNumber = "DESY-22-127, IFT--UAM/CSIC--22-015",
    doi = "10.1088/1475-7516/2023/03/031",
    journal = "JCAP",
    volume = "03",
    pages = "031",
    year = "2023"
}

@article{Ramsey-Musolf:2024ykk,
    author = "Ramsey-Musolf, Michael J. and Tenkanen, Tuomas V. I. and Tran, Van Que",
    title = "{Refining Gravitational Wave and Collider Physics Dialogue via Singlet Scalar Extension}",
    eprint = "2409.17554",
    archivePrefix = "arXiv",
    primaryClass = "hep-ph",
    month = "9",
    year = "2024"
}

@article{Hindmarsh:2017gnf,
    author = "Hindmarsh, Mark and Huber, Stephan J. and Rummukainen, Kari and Weir, David J.",
    title = "{Shape of the acoustic gravitational wave power spectrum from a first order phase transition}",
    eprint = "1704.05871",
    archivePrefix = "arXiv",
    primaryClass = "astro-ph.CO",
    reportNumber = "HIP-2017-02-TH, HIP-2017-02/TH",
    doi = "10.1103/PhysRevD.96.103520",
    journal = "Phys. Rev. D",
    volume = "96",
    number = "10",
    pages = "103520",
    year = "2017",
    note = "[Erratum: Phys.Rev.D 101, 089902 (2020)]"
}

@article{Hindmarsh:2015qta,
    author = "Hindmarsh, Mark and Huber, Stephan J. and Rummukainen, Kari and Weir, David J.",
    title = "{Numerical simulations of acoustically generated gravitational waves at a first order phase transition}",
    eprint = "1504.03291",
    archivePrefix = "arXiv",
    primaryClass = "astro-ph.CO",
    reportNumber = "HIP-2015-13-TH",
    doi = "10.1103/PhysRevD.92.123009",
    journal = "Phys. Rev. D",
    volume = "92",
    number = "12",
    pages = "123009",
    year = "2015"
}

@article{Caprini:2009yp,
    author = "Caprini, Chiara and Durrer, Ruth and Servant, Geraldine",
    title = "{The stochastic gravitational wave background from turbulence and magnetic fields generated by a first-order phase transition}",
    eprint = "0909.0622",
    archivePrefix = "arXiv",
    primaryClass = "astro-ph.CO",
    doi = "10.1088/1475-7516/2009/12/024",
    journal = "JCAP",
    volume = "12",
    pages = "024",
    year = "2009"
}

@article{Carrington:1993ng,
    author = "Carrington, Margaret E. and Kapusta, Joseph I.",
    title = "{Dynamics of the electroweak phase transition}",
    reportNumber = "TPI-MINN-92-55-TA",
    doi = "10.1103/PhysRevD.47.5304",
    journal = "Phys. Rev. D",
    volume = "47",
    pages = "5304--5315",
    year = "1993"
}

@article{Aguilar-Saavedra:2023tql,
    author = "Aguilar-Saavedra, J. A. and C\^amara, H. B. and Joaquim, F. R. and Seabra, J. F.",
    title = "{Confronting the 95 GeV excesses
within the U(1)'-extended next-to-minimal 2HDM}",
    eprint = "2307.03768",
    archivePrefix = "arXiv",
    primaryClass = "hep-ph",
    reportNumber = "IFT-UAM-CSIC-23-86",
    doi = "10.1103/PhysRevD.108.075020",
    journal = "Phys. Rev. D",
    volume = "108",
    number = "7",
    pages = "075020",
    year = "2023"
}

@article{Arhrib:2025pxy,
    author = "Arhrib, Abdesslam and Phan, Khiem Hong and Tran, Van Que and Yuan, Tzu-Chiang",
    title = "{When the Standard Model Higgs meets its lighter 95 GeV twin}",
    doi = "10.1016/j.nuclphysb.2025.116909",
    journal = "Nucl. Phys. B",
    volume = "1015",
    pages = "116909",
    year = "2025"
}

@article{Benbrik:2024ptw,
    author = "Benbrik, Rachid and Boukidi, Mohammed and Moretti, Stefano",
    title = "{Superposition of CP-even and CP-odd Higgs resonances: Explaining the 95~GeV excesses within a two-Higgs-doublet model}",
    eprint = "2405.02899",
    archivePrefix = "arXiv",
    primaryClass = "hep-ph",
    doi = "10.1103/PhysRevD.110.115030",
    journal = "Phys. Rev. D",
    volume = "110",
    number = "11",
    pages = "115030",
    year = "2024"
}

@article{Biekotter:2023oen,
    author = {Biek\"otter, Thomas and Heinemeyer, Sven and Weiglein, Georg},
    title = "{95.4~GeV diphoton excess at ATLAS and CMS}",
    eprint = "2306.03889",
    archivePrefix = "arXiv",
    primaryClass = "hep-ph",
    reportNumber = "KA-TP-11-2023, DESY-23-071, IFT--UAM/CSIC-23-062",
    doi = "10.1103/PhysRevD.109.035005",
    journal = "Phys. Rev. D",
    volume = "109",
    number = "3",
    pages = "035005",
    year = "2024"
}

@article{Biekotter:2023jld,
    author = {Biek\"otter, Thomas and Heinemeyer, Sven and Weiglein, Georg},
    title = "{The CMS di-photon excess at 95 GeV in view of the LHC Run 2 results}",
    eprint = "2303.12018",
    archivePrefix = "arXiv",
    primaryClass = "hep-ph",
    reportNumber = "KA-TP-03-2023, DESY-23-033, IFT-UAM/CSIC-23-028",
    doi = "10.1016/j.physletb.2023.138217",
    journal = "Phys. Lett. B",
    volume = "846",
    pages = "138217",
    year = "2023"
}

@article{Biekotter:2022jyr,
    author = {Biek\"otter, Thomas and Heinemeyer, Sven and Weiglein, Georg},
    title = "{Mounting evidence for a 95 GeV Higgs boson}",
    eprint = "2203.13180",
    archivePrefix = "arXiv",
    primaryClass = "hep-ph",
    reportNumber = "DESY 22-057, IFT-UAM/CSIC-22-033, IFT-UAM/CSIC--22--033",
    doi = "10.1007/JHEP08(2022)201",
    journal = "JHEP",
    volume = "08",
    pages = "201",
    year = "2022"
}

@article{Abdelalim:2020xfk,
    author = "Abdelalim, Ahmed Ali and Das, Biswaranjan and Khalil, Shaaban and Moretti, Stefano",
    title = "{Di-photon decay of a light Higgs state in the BLSSM}",
    eprint = "2012.04952",
    archivePrefix = "arXiv",
    primaryClass = "hep-ph",
    doi = "10.1016/j.nuclphysb.2022.116013",
    journal = "Nucl. Phys. B",
    volume = "985",
    pages = "116013",
    year = "2022"
}

@article{Cacciapaglia:2016tlr,
    author = "Cacciapaglia, Giacomo and Deandrea, Aldo and Gascon-Shotkin, Suzanne and Le Corre, Sol\`ene and Lethuillier, Morgan and Tao, Junquan",
    title = "{Search for a lighter Higgs boson in Two Higgs Doublet Models}",
    eprint = "1607.08653",
    archivePrefix = "arXiv",
    primaryClass = "hep-ph",
    doi = "10.1007/JHEP12(2016)068",
    journal = "JHEP",
    volume = "12",
    pages = "068",
    year = "2016"
}

@article{Crivellin:2017upt,
    author = {Crivellin, Andreas and Heeck, Julian and M\"uller, Dario},
    title = "{Large $h\to b s$ in generic two-Higgs-doublet models}",
    eprint = "1710.04663",
    archivePrefix = "arXiv",
    primaryClass = "hep-ph",
    reportNumber = "ULB-TH-17-19, PSI-PR-17-15, ZH-TH-27-17",
    doi = "10.1103/PhysRevD.97.035008",
    journal = "Phys. Rev. D",
    volume = "97",
    number = "3",
    pages = "035008",
    year = "2018"
}

@article{Biekotter:2021ovi,
    author = {Biek\"otter, Thomas and Olea-Romacho, Mar\'\i{}a Olalla},
    title = "{Reconciling Higgs physics and pseudo-Nambu-Goldstone dark matter in the S2HDM using a genetic algorithm}",
    eprint = "2108.10864",
    archivePrefix = "arXiv",
    primaryClass = "hep-ph",
    reportNumber = "DESY-21-125",
    doi = "10.1007/JHEP10(2021)215",
    journal = "JHEP",
    volume = "10",
    pages = "215",
    year = "2021"
}

@article{Cline:2019okt,
    author = "Cline, James M. and Toma, Takashi",
    title = "{Pseudo-Goldstone dark matter confronts cosmic ray and collider anomalies}",
    eprint = "1906.02175",
    archivePrefix = "arXiv",
    primaryClass = "hep-ph",
    doi = "10.1103/PhysRevD.100.035023",
    journal = "Phys. Rev. D",
    volume = "100",
    number = "3",
    pages = "035023",
    year = "2019"
}

@article{Li:2022etb,
    author = "Li, Weichao and Qiao, Haoxue and Zhu, Jingya",
    title = "{Light Higgs boson in the NMSSM confronted with the CMS di-photon and di-tau excesses*}",
    eprint = "2212.11739",
    archivePrefix = "arXiv",
    primaryClass = "hep-ph",
    doi = "10.1088/1674-1137/acfaf1",
    journal = "Chin. Phys. C",
    volume = "47",
    number = "12",
    pages = "123102",
    year = "2023"
}

@article{Iguro:2022dok,
    author = "Iguro, Syuhei and Kitahara, Teppei and Omura, Yuji",
    title = "{Scrutinizing the 95\textendash{}100 GeV di-tau excess in the top associated process}",
    eprint = "2205.03187",
    archivePrefix = "arXiv",
    primaryClass = "hep-ph",
    reportNumber = "P3H-22-047, TTP22-027, KEK-TH-2424",
    doi = "10.1140/epjc/s10052-022-11028-y",
    journal = "Eur. Phys. J. C",
    volume = "82",
    number = "11",
    pages = "1053",
    year = "2022"
}

@article{Biekotter:2022abc,
    author = {Biek\"otter, Thomas and Heinemeyer, Sven and Weiglein, Georg},
    title = "{Excesses in the low-mass Higgs-boson search and the $W$-boson mass measurement}",
    eprint = "2204.05975",
    archivePrefix = "arXiv",
    primaryClass = "hep-ph",
    reportNumber = "DESY 22-067, IFT-UAM/CSIC--22--043",
    doi = "10.1140/epjc/s10052-023-11635-3",
    journal = "Eur. Phys. J. C",
    volume = "83",
    number = "5",
    pages = "450",
    year = "2023"
}

@article{Cao:2019ofo,
    author = "Cao, Junjie and Jia, Xinglong and Yue, Yuanfang and Zhou, Haijing and Zhu, Pengxuan",
    title = "{96 GeV diphoton excess in seesaw extensions of the natural NMSSM}",
    eprint = "1908.07206",
    archivePrefix = "arXiv",
    primaryClass = "hep-ph",
    doi = "10.1103/PhysRevD.101.055008",
    journal = "Phys. Rev. D",
    volume = "101",
    number = "5",
    pages = "055008",
    year = "2020"
}

@article{Biekotter:2019kde,
    author = {Biek\"otter, T. and Chakraborti, M. and Heinemeyer, S.},
    title = "{A 96 GeV Higgs boson in the N2HDM}",
    eprint = "1903.11661",
    archivePrefix = "arXiv",
    primaryClass = "hep-ph",
    reportNumber = "IFT-UAM/CSIC-19-034",
    doi = "10.1140/epjc/s10052-019-7561-2",
    journal = "Eur. Phys. J. C",
    volume = "80",
    number = "1",
    pages = "2",
    year = "2020"
}

@article{Biekotter:2021qbc,
    author = {Biek\"otter, Thomas and Grohsjean, Alexander and Heinemeyer, Sven and Schwanenberger, Christian and Weiglein, Georg},
    title = "{Possible indications for new Higgs bosons in the reach of the LHC: N2HDM and NMSSM interpretations}",
    eprint = "2109.01128",
    archivePrefix = "arXiv",
    primaryClass = "hep-ph",
    reportNumber = "IFT--UAM/CSIC--21-041, IFT-UAM/CSIC-21-041, DESY 21-132",
    doi = "10.1140/epjc/s10052-022-10099-1",
    journal = "Eur. Phys. J. C",
    volume = "82",
    number = "2",
    pages = "178",
    year = "2022"
}

@article{Cao:2016uwt,
    author = "Cao, Junjie and Guo, Xiaofei and He, Yangle and Wu, Peiwen and Zhang, Yang",
    title = "{Diphoton signal of the light Higgs boson in natural NMSSM}",
    eprint = "1612.08522",
    archivePrefix = "arXiv",
    primaryClass = "hep-ph",
    doi = "10.1103/PhysRevD.95.116001",
    journal = "Phys. Rev. D",
    volume = "95",
    number = "11",
    pages = "116001",
    year = "2017"
}

@article{Heinemeyer:2021msz,
    author = "Heinemeyer, S. and Li, C. and Lika, F. and Moortgat-Pick, G. and Paasch, S.",
    title = "{Phenomenology of a 96~GeV Higgs boson in the 2HDM with an additional singlet}",
    eprint = "2112.11958",
    archivePrefix = "arXiv",
    primaryClass = "hep-ph",
    reportNumber = "DESY 21-230, IFT-UAM/CSIC-21-158",
    doi = "10.1103/PhysRevD.106.075003",
    journal = "Phys. Rev. D",
    volume = "106",
    number = "7",
    pages = "075003",
    year = "2022"
}

@article{Sumner:2019qmx,
    author = "Sumner, Timothy J. and Mueller, Guido and Conklin, John W. and Wass, Peter J. and Hollington, Daniel",
    title = "{Charge Induced Acceleration Noise in the LISA Gravitational Reference Sensor}",
    eprint = "1909.12608",
    archivePrefix = "arXiv",
    primaryClass = "physics.ins-det",
    doi = "10.1088/1361-6382/ab5f6e",
    journal = "Class. Quant. Grav.",
    volume = "37",
    number = "4",
    pages = "045010",
    year = "2020"
}

@article{Smith:2019wny,
    author = "Smith, Tristan L. and Smith, Tristan L. and Caldwell, Robert R. and Caldwell, Robert",
    title = "{LISA for Cosmologists: Calculating the Signal-to-Noise Ratio for Stochastic and Deterministic Sources}",
    eprint = "1908.00546",
    archivePrefix = "arXiv",
    primaryClass = "astro-ph.CO",
    doi = "10.1103/PhysRevD.100.104055",
    journal = "Phys. Rev. D",
    volume = "100",
    number = "10",
    pages = "104055",
    year = "2019",
    note = "[Erratum: Phys.Rev.D 105, 029902 (2022)]"
}

@article{Krajewski:2024gma,
    author = "Krajewski, Tomasz and Lewicki, Marek and Zych, Mateusz",
    title = "{Bubble-wall velocity in local thermal equilibrium: hydrodynamical simulations vs analytical treatment}",
    eprint = "2402.15408",
    archivePrefix = "arXiv",
    primaryClass = "astro-ph.CO",
    doi = "10.1007/JHEP05(2024)011",
    journal = "JHEP",
    volume = "05",
    pages = "011",
    year = "2024"
}

@article{Hindmarsh:2013xza,
    author = "Hindmarsh, Mark and Huber, Stephan J. and Rummukainen, Kari and Weir, David J.",
    title = "{Gravitational waves from the sound of a first order phase transition}",
    eprint = "1304.2433",
    archivePrefix = "arXiv",
    primaryClass = "hep-ph",
    reportNumber = "HIP-2013-07-TH",
    doi = "10.1103/PhysRevLett.112.041301",
    journal = "Phys. Rev. Lett.",
    volume = "112",
    pages = "041301",
    year = "2014"
}

@article{Ai:2021kak,
    author = "Ai, Wen-Yuan and Garbrecht, Bjorn and Tamarit, Carlos",
    title = "{Bubble wall velocities in local equilibrium}",
    eprint = "2109.13710",
    archivePrefix = "arXiv",
    primaryClass = "hep-ph",
    reportNumber = "CP3-21-53, TUM-HEP-1365-21",
    doi = "10.1088/1475-7516/2022/03/015",
    journal = "JCAP",
    volume = "03",
    number = "03",
    pages = "015",
    year = "2022"
}

@article{Ai:2023see,
    author = "Ai, Wen-Yuan and Laurent, Benoit and van de Vis, Jorinde",
    title = "{Model-independent bubble wall velocities in local thermal equilibrium}",
    eprint = "2303.10171",
    archivePrefix = "arXiv",
    primaryClass = "astro-ph.CO",
    reportNumber = "KCL-PH-TH/2023-19",
    doi = "10.1088/1475-7516/2023/07/002",
    journal = "JCAP",
    volume = "07",
    pages = "002",
    year = "2023"
}

@article{Laine:1993ey,
    author = "Laine, M.",
    title = "{Bubble growth as a detonation}",
    eprint = "hep-ph/9309242",
    archivePrefix = "arXiv",
    reportNumber = "HU-TFT-93-44",
    doi = "10.1103/PhysRevD.49.3847",
    journal = "Phys. Rev. D",
    volume = "49",
    pages = "3847--3853",
    year = "1994"
}

@article{Steinhardt:1981ct,
    author = "Steinhardt, Paul Joseph",
    title = "{Relativistic Detonation Waves and Bubble Growth in False Vacuum Decay}",
    reportNumber = "UPR-0181T",
    doi = "10.1103/PhysRevD.25.2074",
    journal = "Phys. Rev. D",
    volume = "25",
    pages = "2074",
    year = "1982"
}

@article{Wang:2024slx,
    author = "Wang, Xiao and Tian, Chi and Bal\'azs, Csaba",
    title = "{Self-consistent prediction of gravitational waves from cosmological phase transitions}",
    eprint = "2409.06599",
    archivePrefix = "arXiv",
    primaryClass = "hep-ph",
    month = "9",
    year = "2024"
}

@article{Tian:2024ysd,
    author = "Tian, Chi and Wang, Xiao and Bal{\'a}zs, Csaba",
    title = "{Gravitational waves from cosmological first-order phase transitions with precise hydrodynamics}",
    eprint = "2409.14505",
    archivePrefix = "arXiv",
    primaryClass = "hep-ph",
    doi = "10.1140/epjc/s10052-025-14826-2",
    journal = "Eur. Phys. J. C",
    volume = "85",
    number = "10",
    pages = "1091",
    year = "2025"
}

@article{LISACosmologyWorkingGroup:2022jok,
    author = "Auclair, Pierre and others",
    collaboration = "LISA Cosmology Working Group",
    title = "{Cosmology with the Laser Interferometer Space Antenna}",
    eprint = "2204.05434",
    archivePrefix = "arXiv",
    primaryClass = "astro-ph.CO",
    reportNumber = "LISA CosWG-22-03, FERMILAB-PUB-22-349-SCD",
    doi = "10.1007/s41114-023-00045-2",
    journal = "Living Rev. Rel.",
    volume = "26",
    number = "1",
    pages = "5",
    year = "2023"
}

@article{Coleman:1977py,
    author = "Coleman, Sidney R.",
    title = "{The Fate of the False Vacuum. 1. Semiclassical Theory}",
    reportNumber = "HUTP-77-A004",
    doi = "10.1103/PhysRevD.16.1248",
    journal = "Phys. Rev. D",
    volume = "15",
    pages = "2929--2936",
    year = "1977",
    note = "[Erratum: Phys.Rev.D 16, 1248 (1977)]"
}

@article{Branco:2011iw,
    author = "Branco, G. C. and Ferreira, P. M. and Lavoura, L. and Rebelo, M. N. and Sher, Marc and Silva, Joao P.",
    title = "{Theory and phenomenology of two-Higgs-doublet models}",
    eprint = "1106.0034",
    archivePrefix = "arXiv",
    primaryClass = "hep-ph",
    doi = "10.1016/j.physrep.2012.02.002",
    journal = "Phys. Rept.",
    volume = "516",
    pages = "1--102",
    year = "2012"
}

@article{Athron:2024xrh,
    author = "Athron, Peter and Balazs, Csaba and Fowlie, Andrew and Morris, Lachlan and Searle, William and Xiao, Yang and Zhang, Yang",
    title = "{PhaseTracer2: from the effective potential to gravitational waves}",
    eprint = "2412.04881",
    archivePrefix = "arXiv",
    primaryClass = "astro-ph.CO",
    doi = "10.1140/epjc/s10052-025-14258-y",
    journal = "Eur. Phys. J. C",
    volume = "85",
    number = "5",
    pages = "559",
    year = "2025"
}

@article{Athron:2020sbe,
    author = "Athron, Peter and Bal\'azs, Csaba and Fowlie, Andrew and Zhang, Yang",
    title = "{PhaseTracer: tracing cosmological phases and calculating transition properties}",
    eprint = "2003.02859",
    archivePrefix = "arXiv",
    primaryClass = "hep-ph",
    reportNumber = "CoEPP-MN-20-3",
    doi = "10.1140/epjc/s10052-020-8035-2",
    journal = "Eur. Phys. J. C",
    volume = "80",
    number = "6",
    pages = "567",
    year = "2020"
}

@article{Ekstedt:2022bff,
    author = "Ekstedt, Andreas and Schicho, Philipp and Tenkanen, Tuomas V. I.",
    title = "{DRalgo: A package for effective field theory approach for thermal phase transitions}",
    eprint = "2205.08815",
    archivePrefix = "arXiv",
    primaryClass = "hep-ph",
    reportNumber = "HIP-2022-11/TH, NORDITA 2022-030",
    doi = "10.1016/j.cpc.2023.108725",
    journal = "Comput. Phys. Commun.",
    volume = "288",
    pages = "108725",
    year = "2023"
}

@article{Croon:2020cgk,
    author = "Croon, Djuna and Gould, Oliver and Schicho, Philipp and Tenkanen, Tuomas V. I. and White, Graham",
    title = "{Theoretical uncertainties for cosmological first-order phase transitions}",
    eprint = "2009.10080",
    archivePrefix = "arXiv",
    primaryClass = "hep-ph",
    reportNumber = "HIP-2020-26/TH",
    doi = "10.1007/JHEP04(2021)055",
    journal = "JHEP",
    volume = "04",
    pages = "055",
    year = "2021"
}

@article{Belyaev:2024lah,
    author = "Belyaev, A. and Benbrik, R. and Boukidi, M. and Chakraborti, M. and Moretti, S. and Semlali, S.",
    title = "{Probing 95 GeV Higgs in the 2HDM Type-III}",
    eprint = "2402.03998",
    archivePrefix = "arXiv",
    primaryClass = "hep-ph",
    doi = "10.22323/1.450.0299",
    journal = "PoS",
    volume = "LHCP2023",
    pages = "299",
    year = "2024"
}

@article{Kanemura:2015mxa,
    author = "Kanemura, Shinya and Kikuchi, Mariko and Yagyu, Kei",
    title = "{Fingerprinting the extended Higgs sector using one-loop corrected Higgs boson couplings and future precision measurements}",
    eprint = "1502.07716",
    archivePrefix = "arXiv",
    primaryClass = "hep-ph",
    reportNumber = "UT-HET-099",
    doi = "10.1016/j.nuclphysb.2015.04.015",
    journal = "Nucl. Phys. B",
    volume = "896",
    pages = "80--137",
    year = "2015"
}

@article{Laine:2017hdk,
    author = "Laine, M. and Meyer, M. and Nardini, G.",
    title = "{Thermal phase transition with full 2-loop effective potential}",
    eprint = "1702.07479",
    archivePrefix = "arXiv",
    primaryClass = "hep-ph",
    doi = "10.1016/j.nuclphysb.2017.04.023",
    journal = "Nucl. Phys. B",
    volume = "920",
    pages = "565--600",
    year = "2017"
}

@article{Gorda:2018hvi,
    author = "Gorda, Tyler and Helset, Andreas and Niemi, Lauri and Tenkanen, Tuomas V. I. and Weir, David J.",
    title = "{Three-dimensional effective theories for the two Higgs doublet model at high temperature}",
    eprint = "1802.05056",
    archivePrefix = "arXiv",
    primaryClass = "hep-ph",
    reportNumber = "HIP-2018-6/TH, HIP-2018-6-TH",
    doi = "10.1007/JHEP02(2019)081",
    journal = "JHEP",
    volume = "02",
    pages = "081",
    year = "2019"
}

@article{CMS:2018cyk,
    author = "Sirunyan, Albert M and others",
    collaboration = "CMS",
    title = "{Search for a standard model-like Higgs boson in the mass range between 70 and 110 GeV in the diphoton final state in proton-proton collisions at $\sqrt{s}=$ 8 and 13 TeV}",
    eprint = "1811.08459",
    archivePrefix = "arXiv",
    primaryClass = "hep-ex",
    reportNumber = "CMS-HIG-17-013, CERN-EP-2018-207",
    doi = "10.1016/j.physletb.2019.03.064",
    journal = "Phys. Lett. B",
    volume = "793",
    pages = "320--347",
    year = "2019"
}

@article{CMS:2022goy,
    author = "Tumasyan, Armen and others",
    collaboration = "CMS",
    title = "{Searches for additional Higgs bosons and for vector leptoquarks in $\tau\tau$ final states in proton-proton collisions at $\sqrt{s}$ = 13 TeV}",
    eprint = "2208.02717",
    archivePrefix = "arXiv",
    primaryClass = "hep-ex",
    reportNumber = "CMS-HIG-21-001, CERN-EP-2022-137",
    doi = "10.1007/JHEP07(2023)073",
    journal = "JHEP",
    volume = "07",
    pages = "073",
    year = "2023"
}

@techreport{ATLAS-CONF-2023-035,
      collaboration = "ATLAS",
      title         = "{Search for diphoton resonances in the 66 to 110 GeV mass
                       range using 140 fb$^{-1}$ of 13 TeV $pp$ collisions
                       collected with the ATLAS detector}",
      institution   = "CERN",
      reportNumber  = "ATLAS-CONF-2023-035",
      address       = "Geneva",
      year          = "2023",
      url           = "https://cds.cern.ch/record/2862024",
      note          = "All figures including auxiliary figures are available at
                       https://atlas.web.cern.ch/Atlas/GROUPS/PHYSICS/CONFNOTES/ATLAS-CONF-2023-035",
}

@article{Fox:2017uwr,
    author = "Fox, Patrick J. and Weiner, Neal",
    title = "{Light Signals from a Lighter Higgs}",
    eprint = "1710.07649",
    archivePrefix = "arXiv",
    primaryClass = "hep-ph",
    reportNumber = "FERMILAB-PUB-17-468-T",
    doi = "10.1007/JHEP08(2018)025",
    journal = "JHEP",
    volume = "08",
    pages = "025",
    year = "2018"
}

@article{Azevedo:2023zkg,
    author = {Azevedo, Duarte and Biek\"otter, Thomas and Ferreira, P. M.},
    title = "{2HDM interpretations of the CMS diphoton excess at 95 GeV}",
    eprint = "2305.19716",
    archivePrefix = "arXiv",
    primaryClass = "hep-ph",
    reportNumber = "KA-TP-10-2023",
    doi = "10.1007/JHEP11(2023)017",
    journal = "JHEP",
    volume = "11",
    pages = "017",
    year = "2023"
}

@mastersthesis{Helset:2017esj,
    author = "Helset, Andreas",
    title = "{Dimensional reduction of the Two-Higgs Doublet Model with a softly broken Z2 symmetry at one-loop}",
    school = "Norwegian U. Sci. Tech.",
    month = "6",
    year = "2017"
}

@article{Kainulainen:2019kyp,
    author = "Kainulainen, Kimmo and Keus, Venus and Niemi, Lauri and Rummukainen, Kari and Tenkanen, Tuomas V. I. and Vaskonen, Ville",
    title = "{On the validity of perturbative studies of the electroweak phase transition in the Two Higgs Doublet model}",
    eprint = "1904.01329",
    archivePrefix = "arXiv",
    primaryClass = "hep-ph",
    doi = "10.1007/JHEP06(2019)075",
    journal = "JHEP",
    volume = "06",
    pages = "075",
    year = "2019"
}

@article{Stevenson:1981vj,
    author = "Stevenson, Paul M.",
    title = "{Optimized Perturbation Theory}",
    reportNumber = "DOE-ER-00881-185",
    doi = "10.1103/PhysRevD.23.2916",
    journal = "Phys. Rev. D",
    volume = "23",
    pages = "2916",
    year = "1981"
}

@article{Laine:2005ai,
    author = "Laine, M. and Schroder, Y.",
    title = "{Two-loop QCD gauge coupling at high temperatures}",
    eprint = "hep-ph/0503061",
    archivePrefix = "arXiv",
    reportNumber = "BI-TP-2005-07",
    doi = "10.1088/1126-6708/2005/03/067",
    journal = "JHEP",
    volume = "03",
    pages = "067",
    year = "2005"
}

@article{Ghisoiu:2015uza,
    author = "Ghisoiu, Ioan and Moller, Jan and Schroder, York",
    title = "{Debye screening mass of hot Yang-Mills theory to three-loop order}",
    eprint = "1509.08727",
    archivePrefix = "arXiv",
    primaryClass = "hep-ph",
    doi = "10.1007/JHEP11(2015)121",
    journal = "JHEP",
    volume = "11",
    pages = "121",
    year = "2015"
}

@article{Huang:1994cu,
    author = "Huang, Su-zhou and Lissia, Marcello",
    title = "{The Relevant scale parameter in the high temperature phase of QCD}",
    eprint = "hep-ph/9411293",
    archivePrefix = "arXiv",
    reportNumber = "MIT-CTP-2360, INFN-CA-TH-94-24",
    doi = "10.1016/0550-3213(95)00007-F",
    journal = "Nucl. Phys. B",
    volume = "438",
    pages = "54--66",
    year = "1995"
}

@article{Croon:2023zay,
    author = "Croon, Djuna",
    title = "{TASI lectures on Phase Transitions, Baryogenesis, and Gravitational Waves}",
    eprint = "2307.00068",
    archivePrefix = "arXiv",
    primaryClass = "hep-ph",
    doi = "10.22323/1.439.0003",
    journal = "PoS",
    volume = "TASI2022",
    pages = "003",
    year = "2024"
}

@article{Andreassen:2016cvx,
    author = "Andreassen, Anders and Farhi, David and Frost, William and Schwartz, Matthew D.",
    title = "{Precision decay rate calculations in quantum field theory}",
    eprint = "1604.06090",
    archivePrefix = "arXiv",
    primaryClass = "hep-th",
    doi = "10.1103/PhysRevD.95.085011",
    journal = "Phys. Rev. D",
    volume = "95",
    number = "8",
    pages = "085011",
    year = "2017"
}

@article{Guth:1979bh,
    author = "Guth, Alan H. and Tye, S. H. H.",
    title = "{Phase Transitions and Magnetic Monopole Production in the Very Early Universe}",
    reportNumber = "SLAC-PUB-2448, CLNS-79-441",
    doi = "10.1103/PhysRevLett.44.631",
    journal = "Phys. Rev. Lett.",
    volume = "44",
    pages = "631",
    year = "1980",
    note = "[Erratum: Phys.Rev.Lett. 44, 963 (1980)]"
}

@article{Guth:1981uk,
    author = "Guth, Alan H. and Weinberg, Erick J.",
    title = "{Cosmological Consequences of a First Order Phase Transition in the SU(5) Grand Unified Model}",
    reportNumber = "CU-TP-183",
    doi = "10.1103/PhysRevD.23.876",
    journal = "Phys. Rev. D",
    volume = "23",
    pages = "876",
    year = "1981"
}

@article{Enqvist:1991xw,
    author = "Enqvist, K. and Ignatius, J. and Kajantie, K. and Rummukainen, K.",
    title = "{Nucleation and bubble growth in a first order cosmological electroweak phase transition}",
    reportNumber = "HU-TFT-91-35",
    doi = "10.1103/PhysRevD.45.3415",
    journal = "Phys. Rev. D",
    volume = "45",
    pages = "3415--3428",
    year = "1992"
}

@article{Ekstedt:2024fyq,
    author = "Ekstedt, Andreas and Gould, Oliver and Hirvonen, Joonas and Laurent, Benoit and Niemi, Lauri and Schicho, Philipp and van de Vis, Jorinde",
    title = "{How fast does the WallGo? A package for computing wall velocities in first-order phase transitions}",
    eprint = "2411.04970",
    archivePrefix = "arXiv",
    primaryClass = "hep-ph",
    reportNumber = "CERN-TH-2024-174, DESY-24-162, HIP-2024-21/TH",
    doi = "10.1007/JHEP04(2025)101",
    journal = "JHEP",
    volume = "04",
    pages = "101",
    year = "2025"
}

@article{Laurent:2022jrs,
    author = "Laurent, Benoit and Cline, James M.",
    title = "{First principles determination of bubble wall velocity}",
    eprint = "2204.13120",
    archivePrefix = "arXiv",
    primaryClass = "hep-ph",
    doi = "10.1103/PhysRevD.106.023501",
    journal = "Phys. Rev. D",
    volume = "106",
    number = "2",
    pages = "023501",
    year = "2022"
}

@article{Kajantie:1995dw,
    author = "Kajantie, K. and Laine, M. and Rummukainen, K. and Shaposhnikov, Mikhail E.",
    title = "{Generic rules for high temperature dimensional reduction and their application to the standard model}",
    eprint = "hep-ph/9508379",
    archivePrefix = "arXiv",
    reportNumber = "CERN-TH-95-226, HU-TFT-95-50, IUHET-312",
    doi = "10.1016/0550-3213(95)00549-8",
    journal = "Nucl. Phys. B",
    volume = "458",
    pages = "90--136",
    year = "1996"
}

@article{Braaten:1995cm,
    author = "Braaten, Eric and Nieto, Agustin",
    title = "{Effective field theory approach to high temperature thermodynamics}",
    eprint = "hep-ph/9501375",
    archivePrefix = "arXiv",
    reportNumber = "NUHEP-TH-95-2",
    doi = "10.1103/PhysRevD.51.6990",
    journal = "Phys. Rev. D",
    volume = "51",
    pages = "6990--7006",
    year = "1995"
}

@article{Gould:2023ovu,
    author = "Gould, Oliver and Tenkanen, Tuomas V. I.",
    title = "{Perturbative effective field theory expansions for cosmological phase transitions}",
    eprint = "2309.01672",
    archivePrefix = "arXiv",
    primaryClass = "hep-ph",
    reportNumber = "NORDITA 2023-037",
    doi = "10.1007/JHEP01(2024)048",
    journal = "JHEP",
    volume = "01",
    pages = "048",
    year = "2024"
}

@article{Schicho:2022wty,
    author = "Schicho, Philipp and Tenkanen, Tuomas V. I. and White, Graham",
    title = "{Combining thermal resummation and gauge invariance for electroweak phase transition}",
    eprint = "2203.04284",
    archivePrefix = "arXiv",
    primaryClass = "hep-ph",
    reportNumber = "HIP-2022-2/TH, NORDITA 2022-009",
    doi = "10.1007/JHEP11(2022)047",
    journal = "JHEP",
    volume = "11",
    pages = "047",
    year = "2022"
}

@article{Lewicki:2024xan,
    author = "Lewicki, Marek and Merchand, Marco and Sagunski, Laura and Schicho, Philipp and Schmitt, Daniel",
    title = "{Impact of theoretical uncertainties on model parameter reconstruction from GW signals sourced by cosmological phase transitions}",
    eprint = "2403.03769",
    archivePrefix = "arXiv",
    primaryClass = "hep-ph",
    doi = "10.1103/PhysRevD.110.023538",
    journal = "Phys. Rev. D",
    volume = "110",
    number = "2",
    pages = "023538",
    year = "2024"
}

@article{Danzmann:2000yvf,
    author = "Danzmann, K.",
    collaboration = "LISA  Study  Team",
    title = "{LISA mission overview}",
    doi = "10.1016/S0273-1177(99)00973-4",
    journal = "Adv. Space Res.",
    volume = "25",
    pages = "1129--1136",
    year = "2000"
}

@article{Gould:2021ccf,
    author = "Gould, Oliver and Hirvonen, Joonas",
    title = "{Effective field theory approach to thermal bubble nucleation}",
    eprint = "2108.04377",
    archivePrefix = "arXiv",
    primaryClass = "hep-ph",
    reportNumber = "HIP-2020-19/TH",
    doi = "10.1103/PhysRevD.104.096015",
    journal = "Phys. Rev. D",
    volume = "104",
    number = "9",
    pages = "096015",
    year = "2021"
}

@article{Cline:1995dg,
    author = "Cline, James M. and Kainulainen, Kimmo and Vischer, Axel P.",
    title = "{Dynamics of two Higgs doublet CP violation and baryogenesis at the electroweak phase transition}",
    eprint = "hep-ph/9506284",
    archivePrefix = "arXiv",
    reportNumber = "MCGILL-95-16, CERN-TH-95-136, UMN-TH-1343-94",
    doi = "10.1103/PhysRevD.54.2451",
    journal = "Phys. Rev. D",
    volume = "54",
    pages = "2451--2472",
    year = "1996"
}

@article{Dorsch:2013wja,
    author = "Dorsch, G. C. and Huber, S. J. and No, J. M.",
    title = "{A strong electroweak phase transition in the 2HDM after LHC8}",
    eprint = "1305.6610",
    archivePrefix = "arXiv",
    primaryClass = "hep-ph",
    doi = "10.1007/JHEP10(2013)029",
    journal = "JHEP",
    volume = "10",
    pages = "029",
    year = "2013"
}

@article{Andersen:2017ika,
    author = "Andersen, Jens O. and Gorda, Tyler and Helset, Andreas and Niemi, Lauri and Tenkanen, Tuomas V. I. and Tranberg, Anders and Vuorinen, Aleksi and Weir, David J.",
    title = "{Nonperturbative Analysis of the Electroweak Phase Transition in the Two Higgs Doublet Model}",
    eprint = "1711.09849",
    archivePrefix = "arXiv",
    primaryClass = "hep-ph",
    reportNumber = "HIP-2017-26/TH, HIP-2017-26-TH",
    doi = "10.1103/PhysRevLett.121.191802",
    journal = "Phys. Rev. Lett.",
    volume = "121",
    number = "19",
    pages = "191802",
    year = "2018"
}

@article{Andersen:1998br,
    author = "Andersen, Jens O.",
    title = "{Dimensional reduction of the two Higgs doublet model at high temperature}",
    eprint = "hep-ph/9804280",
    archivePrefix = "arXiv",
    reportNumber = "OHSTPY-HEP-T-98-007",
    doi = "10.1007/s100520050655",
    journal = "Eur. Phys. J. C",
    volume = "11",
    pages = "563--570",
    year = "1999"
}

@article{Losada:1996ju,
    author = "Losada, Marta",
    title = "{High temperature dimensional reduction of the MSSM and other multiscalar models}",
    eprint = "hep-ph/9605266",
    archivePrefix = "arXiv",
    reportNumber = "RU-96-25",
    doi = "10.1103/PhysRevD.56.2893",
    journal = "Phys. Rev. D",
    volume = "56",
    pages = "2893--2913",
    year = "1997"
}

@article{Kajantie:1996mn,
    author = "Kajantie, K. and Laine, M. and Rummukainen, K. and Shaposhnikov, Mikhail E.",
    title = "{Is there a~ hot electroweak phase transition at $m_H \gtrsim m_W$?}",
    eprint = "hep-ph/9605288",
    archivePrefix = "arXiv",
    reportNumber = "CERN-TH-96-126, HD-THEP-96-15, IUHET-333",
    doi = "10.1103/PhysRevLett.77.2887",
    journal = "Phys. Rev. Lett.",
    volume = "77",
    pages = "2887--2890",
    year = "1996"
}

@article{Kajantie:1996qd,
    author = "Kajantie, K. and Laine, M. and Rummukainen, K. and Shaposhnikov, Mikhail E.",
    title = "{A Nonperturbative analysis of the finite T phase transition in SU(2) x U(1) electroweak theory}",
    eprint = "hep-lat/9612006",
    archivePrefix = "arXiv",
    reportNumber = "BI-TP-96-54, CERN-TH-96-334A, HD-THEP-96-48",
    doi = "10.1016/S0550-3213(97)00164-8",
    journal = "Nucl. Phys. B",
    volume = "493",
    pages = "413--438",
    year = "1997"
}

@article{Gurtler:1997hr,
    author = "Gurtler, M. and Ilgenfritz, Ernst-Michael and Schiller, A.",
    title = "{Where the electroweak phase transition ends}",
    eprint = "hep-lat/9704013",
    archivePrefix = "arXiv",
    reportNumber = "UL-NTZ-10-97, HUB-EP-97-24, DESY-97-086",
    doi = "10.1103/PhysRevD.56.3888",
    journal = "Phys. Rev. D",
    volume = "56",
    pages = "3888--3895",
    year = "1997"
}

@article{Csikor:1998eu,
    author = "Csikor, F. and Fodor, Z. and Heitger, J.",
    title = "{Endpoint of the hot electroweak phase transition}",
    eprint = "hep-ph/9809291",
    archivePrefix = "arXiv",
    reportNumber = "ITP-BUDAPEST-541, KEK-TH-580, KEK-PREPRINT-98-160, MS-TPI-98-16",
    doi = "10.1103/PhysRevLett.82.21",
    journal = "Phys. Rev. Lett.",
    volume = "82",
    pages = "21--24",
    year = "1999"
}

@article{DOnofrio:2015gop,
    author = "D'Onofrio, Michela and Rummukainen, Kari",
    title = "{Standard model cross-over on the lattice}",
    eprint = "1508.07161",
    archivePrefix = "arXiv",
    primaryClass = "hep-ph",
    reportNumber = "HIP-2015-30-TH",
    doi = "10.1103/PhysRevD.93.025003",
    journal = "Phys. Rev. D",
    volume = "93",
    number = "2",
    pages = "025003",
    year = "2016"
}

@article{Niemi:2020hto,
    author = "Niemi, Lauri and Ramsey-Musolf, Michael J. and Tenkanen, Tuomas V. I. and Weir, David J.",
    title = "{Thermodynamics of a Two-Step Electroweak Phase Transition}",
    eprint = "2005.11332",
    archivePrefix = "arXiv",
    primaryClass = "hep-ph",
    reportNumber = "HIP-2020-11/TH, ACFI-T20-05",
    doi = "10.1103/PhysRevLett.126.171802",
    journal = "Phys. Rev. Lett.",
    volume = "126",
    number = "17",
    pages = "171802",
    year = "2021"
}

@article{Ellis:2019oqb,
    author = "Ellis, John and Lewicki, Marek and No, Jos\'e Miguel and Vaskonen, Ville",
    title = "{Gravitational wave energy budget in strongly supercooled phase transitions}",
    eprint = "1903.09642",
    archivePrefix = "arXiv",
    primaryClass = "hep-ph",
    reportNumber = "KCL-PH-TH/2019-32, CERN-TH-2019-032, IFT-UAM/CSIC-19-32",
    doi = "10.1088/1475-7516/2019/06/024",
    journal = "JCAP",
    volume = "06",
    pages = "024",
    year = "2019"
}

@article{Kierkla:2023von,
    author = "Kierkla, Maciej and Swiezewska, Bogumila and Tenkanen, Tuomas V. I. and van de Vis, Jorinde",
    title = "{Gravitational waves from supercooled phase transitions: dimensional transmutation meets dimensional reduction}",
    eprint = "2312.12413",
    archivePrefix = "arXiv",
    primaryClass = "hep-ph",
    doi = "10.1007/JHEP02(2024)234",
    journal = "JHEP",
    volume = "02",
    pages = "234",
    year = "2024"
}

@article{Kierkla:2025qyz,
    author = "Kierkla, Maciej and Schicho, Philipp and Swiezewska, Bogumila and Tenkanen, Tuomas V. I. and van de Vis, Jorinde",
    title = "{Finite-temperature bubble nucleation with shifting scale hierarchies}",
    eprint = "2503.13597",
    archivePrefix = "arXiv",
    primaryClass = "hep-ph",
    reportNumber = "CERN-TH-2025-046, HIP-2024-27/TH",
    doi = "10.1007/JHEP07(2025)153",
    journal = "JHEP",
    volume = "07",
    pages = "153",
    year = "2025"
}

@article{Kierkla:2022odc,
    author = "Kierkla, Maciej and Karam, Alexandros and Swiezewska, Bogumila",
    title = "{Conformal model for gravitational waves and dark matter: a status update}",
    eprint = "2210.07075",
    archivePrefix = "arXiv",
    primaryClass = "astro-ph.CO",
    doi = "10.1007/JHEP03(2023)007",
    journal = "JHEP",
    volume = "03",
    pages = "007",
    year = "2023"
}

@article{Lyth:1998xn,
    author = "Lyth, David H. and Riotto, Antonio",
    title = "{Particle physics models of inflation and the cosmological density perturbation}",
    eprint = "hep-ph/9807278",
    archivePrefix = "arXiv",
    reportNumber = "LANCS-TH-9720, FERMILAB-PUB-97-292-A, CERN-TH-97-383, OUTP-98-39-P",
    doi = "10.1016/S0370-1573(98)00128-8",
    journal = "Phys. Rept.",
    volume = "314",
    pages = "1--146",
    year = "1999"
}

@article{Sagunski:2023ynd,
    author = "Sagunski, Laura and Schicho, Philipp and Schmitt, Daniel",
    title = "{Supercool exit: Gravitational waves from QCD-triggered conformal symmetry breaking}",
    eprint = "2303.02450",
    archivePrefix = "arXiv",
    primaryClass = "hep-ph",
    doi = "10.1103/PhysRevD.107.123512",
    journal = "Phys. Rev. D",
    volume = "107",
    number = "12",
    pages = "123512",
    year = "2023"
}

@article{Caprini:2019egz,
    author = "Caprini, Chiara and others",
    title = "{Detecting gravitational waves from cosmological phase transitions with LISA: an update}",
    eprint = "1910.13125",
    archivePrefix = "arXiv",
    primaryClass = "astro-ph.CO",
    reportNumber = "DESY-19-159, IPPP/19/27, HIP-2019-14/TH, MITP/19-066, IFT-UAM/CSIC-19-139",
    doi = "10.1088/1475-7516/2020/03/024",
    journal = "JCAP",
    volume = "03",
    pages = "024",
    year = "2020"
}

@article{Niemi:2021qvp,
    author = "Niemi, Lauri and Schicho, Philipp and Tenkanen, Tuomas V. I.",
    title = "{Singlet-assisted electroweak phase transition at two loops}",
    eprint = "2103.07467",
    archivePrefix = "arXiv",
    primaryClass = "hep-ph",
    reportNumber = "HIP-2021-8/TH, NORDITA 2021-011",
    doi = "10.1103/PhysRevD.103.115035",
    journal = "Phys. Rev. D",
    volume = "103",
    number = "11",
    pages = "115035",
    year = "2021",
    note = "[Erratum: Phys.Rev.D 109, 039902 (2024)]"
}

@article{ATLAS:2012yve,
    author = "Aad, Georges and others",
    collaboration = "ATLAS",
    title = "{Observation of a new particle in the search for the Standard Model Higgs boson with the ATLAS detector at the LHC}",
    eprint = "1207.7214",
    archivePrefix = "arXiv",
    primaryClass = "hep-ex",
    reportNumber = "CERN-PH-EP-2012-218",
    doi = "10.1016/j.physletb.2012.08.020",
    journal = "Phys. Lett. B",
    volume = "716",
    pages = "1--29",
    year = "2012"
}

@article{CMS:2012qbp,
    author = "Chatrchyan, Serguei and others",
    collaboration = "CMS",
    title = "{Observation of a New Boson at a Mass of 125 GeV with the CMS Experiment at the LHC}",
    eprint = "1207.7235",
    archivePrefix = "arXiv",
    primaryClass = "hep-ex",
    reportNumber = "CMS-HIG-12-028, CERN-PH-EP-2012-220",
    doi = "10.1016/j.physletb.2012.08.021",
    journal = "Phys. Lett. B",
    volume = "716",
    pages = "30--61",
    year = "2012"
}

@article{Ivanov:2008er,
    author = "Ivanov, I. P.",
    title = "{Thermal evolution of the ground state of the most general 2HDM}",
    eprint = "0812.4984",
    archivePrefix = "arXiv",
    primaryClass = "hep-ph",
    journal = "Acta Phys. Polon. B",
    volume = "40",
    pages = "2789--2807",
    year = "2009"
}

@article{Ginzburg:2009dp,
    author = "Ginzburg, I. F. and Ivanov, I. P. and Kanishev, K. A.",
    title = "{The Evolution of vacuum states and phase transitions in 2HDM during cooling of Universe}",
    eprint = "0911.2383",
    archivePrefix = "arXiv",
    primaryClass = "hep-ph",
    doi = "10.1103/PhysRevD.81.085031",
    journal = "Phys. Rev. D",
    volume = "81",
    pages = "085031",
    year = "2010"
}

@article{Zhou:2020irf,
    author = "Zhou, Ruiyu and Bian, Ligong",
    title = "{Gravitational wave and electroweak baryogenesis with two Higgs doublet models}",
    eprint = "2001.01237",
    archivePrefix = "arXiv",
    primaryClass = "hep-ph",
    doi = "10.1016/j.physletb.2022.137105",
    journal = "Phys. Lett. B",
    volume = "829",
    pages = "137105",
    year = "2022"
}

@article{Goncalves:2023svb,
    author = "Gon\c{c}alves, Dorival and Kaladharan, Ajay and Wu, Yongcheng",
    title = "{Gravitational waves, bubble profile, and baryon asymmetry in the complex 2HDM}",
    eprint = "2307.03224",
    archivePrefix = "arXiv",
    primaryClass = "hep-ph",
    doi = "10.1103/PhysRevD.108.075010",
    journal = "Phys. Rev. D",
    volume = "108",
    number = "7",
    pages = "075010",
    year = "2023"
}

@article{Lee:2025hgb,
    author = "Lee, Soojin and Kim, Dongjoo and Cho, Jin-Hwan and Kim, Jinheung and Song, Jeonghyeon",
    title = "{Multistep strong first-order electroweak phase transitions in the inverted type-I 2HDM: Parameter space, gravitational waves, and collider phenomenology}",
    eprint = "2506.03260",
    archivePrefix = "arXiv",
    primaryClass = "hep-ph",
    reportNumber = "KIAS-P25030",
    doi = "10.1103/cbgr-w9cb",
    journal = "Phys. Rev. D",
    volume = "112",
    number = "5",
    pages = "055035",
    year = "2025"
}

@article{Gould:2023jbz,
    author = "Gould, Oliver and Xie, Cheng",
    title = "{Higher orders for cosmological phase transitions: a global study in a Yukawa model}",
    eprint = "2310.02308",
    archivePrefix = "arXiv",
    primaryClass = "hep-ph",
    doi = "10.1007/JHEP12(2023)049",
    journal = "JHEP",
    volume = "12",
    pages = "049",
    year = "2023"
}

@article{Ekstedt:2021kyx,
    author = "Ekstedt, Andreas",
    title = "{Higher-order corrections to the bubble-nucleation rate at finite temperature}",
    eprint = "2104.11804",
    archivePrefix = "arXiv",
    primaryClass = "hep-ph",
    doi = "10.1140/epjc/s10052-022-10130-5",
    journal = "Eur. Phys. J. C",
    volume = "82",
    number = "2",
    pages = "173",
    year = "2022"
}

@article{Bernon:2015qea,
    author = "Bernon, J\'er\'emy and Gunion, John F. and Haber, Howard E. and Jiang, Yun and Kraml, Sabine",
    title = "{Scrutinizing the alignment limit in two-Higgs-doublet models: m$_h$=125  GeV}",
    eprint = "1507.00933",
    archivePrefix = "arXiv",
    primaryClass = "hep-ph",
    doi = "10.1103/PhysRevD.92.075004",
    journal = "Phys. Rev. D",
    volume = "92",
    number = "7",
    pages = "075004",
    year = "2015"
}

@article{Paschos:1976ay,
    author = "Paschos, E. A.",
    title = "{Diagonal Neutral Currents}",
    reportNumber = "BNL-21870",
    doi = "10.1103/PhysRevD.15.1966",
    journal = "Phys. Rev. D",
    volume = "15",
    pages = "1966",
    year = "1977"
}

@article{Glashow:1976nt,
    author = "Glashow, Sheldon L. and Weinberg, Steven",
    title = "{Natural Conservation Laws for Neutral Currents}",
    reportNumber = "HUTP-76-A158",
    doi = "10.1103/PhysRevD.15.1958",
    journal = "Phys. Rev. D",
    volume = "15",
    pages = "1958",
    year = "1977"
}

@article{Bernal:2022wct,
    author = "Bernal, A. and Casas, J. A. and Moreno, J. M.",
    title = "{Fine-tuning in the 2HDM}",
    eprint = "2202.09103",
    archivePrefix = "arXiv",
    primaryClass = "hep-ph",
    doi = "10.1140/epjc/s10052-022-10886-w",
    journal = "Eur. Phys. J. C",
    volume = "82",
    number = "10",
    pages = "950",
    year = "2022"
}

@article{RoperPol:2019wvy,
    author = "Roper Pol, Alberto and Mandal, Sayan and Brandenburg, Axel and Kahniashvili, Tina and Kosowsky, Arthur",
    title = "{Numerical simulations of gravitational waves from early-universe turbulence}",
    eprint = "1903.08585",
    archivePrefix = "arXiv",
    primaryClass = "astro-ph.CO",
    reportNumber = "NORDITA-2019-024",
    doi = "10.1103/PhysRevD.102.083512",
    journal = "Phys. Rev. D",
    volume = "102",
    number = "8",
    pages = "083512",
    year = "2020"
}

@article{Jiang:2022btc,
    author = "Jiang, Siyu and Huang, Fa Peng and Wang, Xiao",
    title = "{Bubble wall velocity during electroweak phase transition in the inert doublet model}",
    eprint = "2211.13142",
    archivePrefix = "arXiv",
    primaryClass = "hep-ph",
    doi = "10.1103/PhysRevD.107.095005",
    journal = "Phys. Rev. D",
    volume = "107",
    number = "9",
    pages = "095005",
    year = "2023"
}

@article{Banerjee:2024fam,
    author = "Banerjee, Indra Kumar and Dey, Ujjal Kumar and Khalil, Shaaban",
    title = "{Primordial Black Holes and Gravitational Waves in the U(1)$_{B-L}$ extended inert doublet model: a first-order phase transition perspective}",
    eprint = "2406.12518",
    archivePrefix = "arXiv",
    primaryClass = "hep-ph",
    doi = "10.1007/JHEP12(2024)009",
    journal = "JHEP",
    volume = "12",
    pages = "009",
    year = "2024"
}

@article{Jung:2010ik,
    author = "Jung, Martin and Pich, Antonio and Tuzon, Paula",
    title = "{Charged-Higgs phenomenology in the Aligned two-Higgs-doublet model}",
    eprint = "1006.0470",
    archivePrefix = "arXiv",
    primaryClass = "hep-ph",
    doi = "10.1007/JHEP11(2010)003",
    journal = "JHEP",
    volume = "11",
    pages = "003",
    year = "2010"
}

@article{Haber:1999zh,
    author = "Haber, Howard E. and Logan, Heather E.",
    title = "{Radiative corrections to the Z b anti-b vertex and constraints on extended Higgs sectors}",
    eprint = "hep-ph/9909335",
    archivePrefix = "arXiv",
    reportNumber = "SCIPP-98-47",
    doi = "10.1103/PhysRevD.62.015011",
    journal = "Phys. Rev. D",
    volume = "62",
    pages = "015011",
    year = "2000"
}

@article{Funk:2011ad,
    author = "Funk, Gerhardt and O'Neil, Deva and Winters, R. Michael",
    title = "{What the Oblique Parameters S, T, and U and Their Extensions Reveal About the 2HDM: A Numerical Analysis}",
    eprint = "1110.3812",
    archivePrefix = "arXiv",
    primaryClass = "hep-ph",
    doi = "10.1142/S0217751X12500212",
    journal = "Int. J. Mod. Phys. A",
    volume = "27",
    pages = "1250021",
    year = "2012"
}

@article{Haber:2010bw,
    author = "Haber, Howard E. and O'Neil, Deva",
    title = "{Basis-independent methods for the two-Higgs-doublet model III: The CP-conserving limit, custodial symmetry, and the oblique parameters S, T, U}",
    eprint = "1011.6188",
    archivePrefix = "arXiv",
    primaryClass = "hep-ph",
    reportNumber = "SCIPP-10-18",
    doi = "10.1103/PhysRevD.83.055017",
    journal = "Phys. Rev. D",
    volume = "83",
    pages = "055017",
    year = "2011"
}

@article{Grimus:2007if,
    author = "Grimus, W. and Lavoura, L. and Ogreid, O. M. and Osland, P.",
    title = "{A Precision constraint on multi-Higgs-doublet models}",
    eprint = "0711.4022",
    archivePrefix = "arXiv",
    primaryClass = "hep-ph",
    reportNumber = "UWTHPH-2007-28",
    doi = "10.1088/0954-3899/35/7/075001",
    journal = "J. Phys. G",
    volume = "35",
    pages = "075001",
    year = "2008"
}

@article{Basler:2016obg,
    author = "Basler, P. and Krause, M. and Muhlleitner, M. and Wittbrodt, J. and Wlotzka, A.",
    title = "{Strong First Order Electroweak Phase Transition in the CP-Conserving 2HDM Revisited}",
    eprint = "1612.04086",
    archivePrefix = "arXiv",
    primaryClass = "hep-ph",
    doi = "10.1007/JHEP02(2017)121",
    journal = "JHEP",
    volume = "02",
    pages = "121",
    year = "2017"
}

@article{Ramsey-Musolf:2024zex,
    author = "Ramsey-Musolf, Michael J. and Tran, Van Que and Yuan, Tzu-Chiang",
    title = "{Gravitational waves and dark matter in the gauged two-Higgs doublet model}",
    eprint = "2408.05167",
    archivePrefix = "arXiv",
    primaryClass = "hep-ph",
    doi = "10.1007/JHEP01(2025)129",
    journal = "JHEP",
    volume = "01",
    pages = "129",
    year = "2025"
}

@article{Caprini:2015zlo,
    author = "Caprini, Chiara and others",
    title = "{Science with the space-based interferometer eLISA. II: Gravitational waves from cosmological phase transitions}",
    eprint = "1512.06239",
    archivePrefix = "arXiv",
    primaryClass = "astro-ph.CO",
    reportNumber = "DESY-15-246",
    doi = "10.1088/1475-7516/2016/04/001",
    journal = "JCAP",
    volume = "04",
    pages = "001",
    year = "2016"
}

@article{Harry:2006fi,
    author = "Harry, G. M. and Fritschel, P. and Shaddock, D. A. and Folkner, W. and Phinney, E. S.",
    title = "{Laser interferometry for the big bang observer}",
    doi = "10.1088/0264-9381/23/15/008",
    journal = "Class. Quant. Grav.",
    volume = "23",
    pages = "4887--4894",
    year = "2006",
    note = "[Erratum: Class.Quant.Grav. 23, 7361 (2006)]"
}

@article{Kawamura:2011zz,
    author = "Kawamura, Seiji and others",
    editor = "Buchman, Sasha and Sun, Ke-Xun",
    title = "{The Japanese space gravitational wave antenna: DECIGO}",
    doi = "10.1088/0264-9381/28/9/094011",
    journal = "Class. Quant. Grav.",
    volume = "28",
    pages = "094011",
    year = "2011"
}

@article{Ruan:2018tsw,
    author = "Ruan, Wen-Hong and Guo, Zong-Kuan and Cai, Rong-Gen and Zhang, Yuan-Zhong",
    title = "{Taiji program: Gravitational-wave sources}",
    eprint = "1807.09495",
    archivePrefix = "arXiv",
    primaryClass = "gr-qc",
    doi = "10.1142/S0217751X2050075X",
    journal = "Int. J. Mod. Phys. A",
    volume = "35",
    number = "17",
    pages = "2050075",
    year = "2020"
}

@article{Friedrich:2022cak,
    author = "Friedrich, Leon S. and Ramsey-Musolf, Michael J. and Tenkanen, Tuomas V. I. and Tran, Van Que",
    title = "{Addressing the Gravitational Wave - Collider Inverse Problem}",
    eprint = "2203.05889",
    archivePrefix = "arXiv",
    primaryClass = "hep-ph",
    reportNumber = "NORDITA 2022-010",
    month = "3",
    year = "2022"
}

@article{Bernardo:2025vkz,
    author = "Bernardo, Fabio and Klose, Philipp and Schicho, Philipp and Tenkanen, Tuomas V. I.",
    title = "{Higher-dimensional operators at finite temperature affect gravitational-wave predictions}",
    eprint = "2503.18904",
    archivePrefix = "arXiv",
    primaryClass = "hep-ph",
    reportNumber = "HIP-2025-6/TH",
    doi = "10.1007/JHEP08(2025)109",
    journal = "JHEP",
    volume = "08",
    pages = "109",
    year = "2025"
}

@article{Chala:2024xll,
    author = "Chala, Mikael and Criado, Juan Carlos and Gil, Luis and Miras, Javier L\'opez",
    title = "{Higher-order-operator corrections to phase-transition parameters in dimensional reduction}",
    eprint = "2406.02667",
    archivePrefix = "arXiv",
    primaryClass = "hep-ph",
    doi = "10.1007/JHEP10(2024)025",
    journal = "JHEP",
    volume = "10",
    pages = "025",
    year = "2024"
}

@article{Chala:2025aiz,
    author = "Chala, Mikael and Guedes, Guilherme",
    title = "{The high-temperature limit of the SM(EFT)}",
    eprint = "2503.20016",
    archivePrefix = "arXiv",
    primaryClass = "hep-ph",
    doi = "10.1007/JHEP07(2025)085",
    journal = "JHEP",
    volume = "07",
    pages = "085",
    year = "2025"
}

@article{Gould:2021oba,
    author = "Gould, Oliver and Tenkanen, Tuomas V. I.",
    title = "{On the perturbative expansion at high temperature and implications for cosmological phase transitions}",
    eprint = "2104.04399",
    archivePrefix = "arXiv",
    primaryClass = "hep-ph",
    reportNumber = "NORDITA 2021-010",
    doi = "10.1007/JHEP06(2021)069",
    journal = "JHEP",
    volume = "06",
    pages = "069",
    year = "2021"
}

@article{Grojean:2006bp,
    author = "Grojean, Christophe and Servant, Geraldine",
    title = "{Gravitational Waves from Phase Transitions at the Electroweak Scale and Beyond}",
    eprint = "hep-ph/0607107",
    archivePrefix = "arXiv",
    reportNumber = "CERN-PH-TH-2006-125",
    doi = "10.1103/PhysRevD.75.043507",
    journal = "Phys. Rev. D",
    volume = "75",
    pages = "043507",
    year = "2007"
}

@article{Delaunay:2007wb,
    author = "Delaunay, Cedric and Grojean, Christophe and Wells, James D.",
    title = "{Dynamics of Non-renormalizable Electroweak Symmetry Breaking}",
    eprint = "0711.2511",
    archivePrefix = "arXiv",
    primaryClass = "hep-ph",
    reportNumber = "CERN-PH-TH-2007-219, MCTP-07-31, SACLAY-T07-141",
    doi = "10.1088/1126-6708/2008/04/029",
    journal = "JHEP",
    volume = "04",
    pages = "029",
    year = "2008"
}

@article{Dorsch:2016nrg,
    author = "Dorsch, G. C. and Huber, S. J. and Konstandin, T. and No, J. M.",
    title = "{A Second Higgs Doublet in the Early Universe: Baryogenesis and Gravitational Waves}",
    eprint = "1611.05874",
    archivePrefix = "arXiv",
    primaryClass = "hep-ph",
    reportNumber = "DESY-16-213",
    doi = "10.1088/1475-7516/2017/05/052",
    journal = "JCAP",
    volume = "05",
    pages = "052",
    year = "2017"
}

@article{Ellis:2018mja,
    author = "Ellis, John and Lewicki, Marek and No, Jos\'e Miguel",
    title = "{On the Maximal Strength of a First-Order Electroweak Phase Transition and its Gravitational Wave Signal}",
    eprint = "1809.08242",
    archivePrefix = "arXiv",
    primaryClass = "hep-ph",
    reportNumber = "KCL-PH-TH/2018-46, CERN-TH/2018-197, IFT-UAM/CSIC-18-94, CERN-TH-2018-197",
    doi = "10.1088/1475-7516/2019/04/003",
    journal = "JCAP",
    volume = "04",
    pages = "003",
    year = "2019"
}

@article{Ginsparg:1980ef,
    author = "Ginsparg, Paul H.",
    title = "{First Order and Second Order Phase Transitions in Gauge Theories at Finite Temperature}",
    reportNumber = "SACLAY-DPh-T 80/27",
    doi = "10.1016/0550-3213(80)90418-6",
    journal = "Nucl. Phys. B",
    volume = "170",
    pages = "388--408",
    year = "1980"
}

@article{Appelquist:1981vg,
    author = "Appelquist, Thomas and Pisarski, Robert D.",
    title = "{High-Temperature Yang-Mills Theories and Three-Dimensional Quantum Chromodynamics}",
    reportNumber = "Print-81-0020 (YALE), YTP-81-01, COO-3075-203",
    doi = "10.1103/PhysRevD.23.2305",
    journal = "Phys. Rev. D",
    volume = "23",
    pages = "2305",
    year = "1981"
}

@article{Nadkarni:1982kb,
    author = "Nadkarni, Sudhir",
    title = "{Dimensional Reduction in Hot QCD}",
    reportNumber = "YTP-82-21-REV, YTP-82-21",
    doi = "10.1103/PhysRevD.27.917",
    journal = "Phys. Rev. D",
    volume = "27",
    pages = "917",
    year = "1983"
}

@article{Landsman:1989be,
    author = "Landsman, N. P.",
    title = "{Limitations to Dimensional Reduction at High Temperature}",
    reportNumber = "ITFA-89-101",
    doi = "10.1016/0550-3213(89)90424-0",
    journal = "Nucl. Phys. B",
    volume = "322",
    pages = "498--530",
    year = "1989"
}

@article{Braaten:1995jr,
    author = "Braaten, Eric and Nieto, Agustin",
    title = "{Free energy of QCD at high temperature}",
    eprint = "hep-ph/9510408",
    archivePrefix = "arXiv",
    reportNumber = "OHSTPY-HEP-T-95-020",
    doi = "10.1103/PhysRevD.53.3421",
    journal = "Phys. Rev. D",
    volume = "53",
    pages = "3421--3437",
    year = "1996"
}

@article{Laine:2000rm,
    author = "Laine, M. and Rummukainen, K.",
    title = "{Two Higgs doublet dynamics at the electroweak phase transition: A Nonperturbative study}",
    eprint = "hep-lat/0009025",
    archivePrefix = "arXiv",
    reportNumber = "CERN-TH-2000-226, NORDITA-2000-80-HE",
    doi = "10.1016/S0550-3213(00)00736-7",
    journal = "Nucl. Phys. B",
    volume = "597",
    pages = "23--69",
    year = "2001"
}

@article{Matsubara:1955ws,
    author = "Matsubara, Takeo",
    title = "{A New approach to quantum statistical mechanics}",
    doi = "10.1143/PTP.14.351",
    journal = "Prog. Theor. Phys.",
    volume = "14",
    pages = "351--378",
    year = "1955"
}

@article{Hirvonen:2022jba,
    author = "Hirvonen, J.",
    title = "{Intuitive method for constructing effective field theories}",
    eprint = "2205.02687",
    archivePrefix = "arXiv",
    primaryClass = "hep-ph",
    reportNumber = "HIP-2022-6/TH",
    month = "5",
    year = "2022"
}

@article{Linde:1980ts,
    author = "Linde, Andrei D.",
    title = "{Infrared Problem in Thermodynamics of the Yang-Mills Gas}",
    reportNumber = "LEBEDEV-80-106",
    doi = "10.1016/0370-2693(80)90769-8",
    journal = "Phys. Lett. B",
    volume = "96",
    pages = "289--292",
    year = "1980"
}

@article{Ruijl:2017dtg,
    author = "Ruijl, Ben and Ueda, Takahiro and Vermaseren, Jos",
    title = "{FORM version 4.2}",
    eprint = "1707.06453",
    archivePrefix = "arXiv",
    primaryClass = "hep-ph",
    month = "7",
    year = "2017"
}

@article{Nogueira:1991ex,
    author = "Nogueira, Paulo",
    title = "{Automatic Feynman Graph Generation}",
    reportNumber = "IFM-7-91",
    doi = "10.1006/jcph.1993.1074",
    journal = "J. Comput. Phys.",
    volume = "105",
    pages = "279--289",
    year = "1993"
}

@article{Ekstedt:2022zro,
    author = {Ekstedt, Andreas and Gould, Oliver and L\"ofgren, Johan},
    title = "{Radiative first-order phase transitions to next-to-next-to-leading order}",
    eprint = "2205.07241",
    archivePrefix = "arXiv",
    primaryClass = "hep-ph",
    doi = "10.1103/PhysRevD.106.036012",
    journal = "Phys. Rev. D",
    volume = "106",
    number = "3",
    pages = "036012",
    year = "2022",
    note = "[Erratum: Phys.Rev.D 110, 019901 (2024)]"
}

@article{Ekstedt:2024etx,
    author = "Ekstedt, Andreas and Schicho, Philipp and Tenkanen, Tuomas V. I.",
    title = "{Cosmological phase transitions at three loops: The final verdict on perturbation theory}",
    eprint = "2405.18349",
    archivePrefix = "arXiv",
    primaryClass = "hep-ph",
    reportNumber = "HIP-2024-15/TH",
    doi = "10.1103/PhysRevD.110.096006",
    journal = "Phys. Rev. D",
    volume = "110",
    number = "9",
    pages = "096006",
    year = "2024"
}

@article{Giese:2020rtr,
    author = "Giese, Felix and Konstandin, Thomas and van de Vis, Jorinde",
    title = "{Model-independent energy budget of cosmological first-order phase transitions\textemdash{}A sound argument to go beyond the bag model}",
    eprint = "2004.06995",
    archivePrefix = "arXiv",
    primaryClass = "astro-ph.CO",
    reportNumber = "DESY-20-064",
    doi = "10.1088/1475-7516/2020/07/057",
    journal = "JCAP",
    volume = "07",
    number = "07",
    pages = "057",
    year = "2020"
}

@article{Tenkanen:2022tly,
    author = "Tenkanen, Tuomas V. I. and van de Vis, Jorinde",
    title = "{Speed of sound in cosmological phase transitions and effect on gravitational waves}",
    eprint = "2206.01130",
    archivePrefix = "arXiv",
    primaryClass = "hep-ph",
    reportNumber = "NORDITA 2022-031, DESY-22-091",
    doi = "10.1007/JHEP08(2022)302",
    journal = "JHEP",
    volume = "08",
    pages = "302",
    year = "2022"
}

@article{Kosowsky:1992vn,
    author = "Kosowsky, Arthur and Turner, Michael S.",
    title = "{Gravitational radiation from colliding vacuum bubbles: envelope approximation to many bubble collisions}",
    eprint = "astro-ph/9211004",
    archivePrefix = "arXiv",
    reportNumber = "FERMILAB-PUB-92-295-A",
    doi = "10.1103/PhysRevD.47.4372",
    journal = "Phys. Rev. D",
    volume = "47",
    pages = "4372--4391",
    year = "1993"
}

@article{Coleman:1973jx,
    author = "Coleman, Sidney R. and Weinberg, Erick J.",
    title = "{Radiative Corrections as the Origin of Spontaneous Symmetry Breaking}",
    doi = "10.1103/PhysRevD.7.1888",
    journal = "Phys. Rev. D",
    volume = "7",
    pages = "1888--1910",
    year = "1973"
}

@article{Fromme:2006cm,
    author = "Fromme, Lars and Huber, Stephan J. and Seniuch, Michael",
    title = "{Baryogenesis in the two-Higgs doublet model}",
    eprint = "hep-ph/0605242",
    archivePrefix = "arXiv",
    reportNumber = "CERN-PH-TH-2006-094, BI-TP-2006-18",
    doi = "10.1088/1126-6708/2006/11/038",
    journal = "JHEP",
    volume = "11",
    pages = "038",
    year = "2006"
}

@article{Giese:2020znk,
    author = "Giese, Felix and Konstandin, Thomas and Schmitz, Kai and van de Vis, Jorinde",
    title = "{Model-independent energy budget for LISA}",
    eprint = "2010.09744",
    archivePrefix = "arXiv",
    primaryClass = "astro-ph.CO",
    reportNumber = "DESY-20-173, DESY 20-173, CERN-TH-2020-170",
    doi = "10.1088/1475-7516/2021/01/072",
    journal = "JCAP",
    volume = "01",
    pages = "072",
    year = "2021"
}

@article{Espinosa:2010hh,
    author = "Espinosa, Jose R. and Konstandin, Thomas and No, Jose M. and Servant, Geraldine",
    title = "{Energy Budget of Cosmological First-order Phase Transitions}",
    eprint = "1004.4187",
    archivePrefix = "arXiv",
    primaryClass = "hep-ph",
    reportNumber = "CERN-PH-TH-2010-027",
    doi = "10.1088/1475-7516/2010/06/028",
    journal = "JCAP",
    volume = "06",
    pages = "028",
    year = "2010"
}

@article{Ai:2024btx,
    author = "Ai, Wen-Yuan and Laurent, Benoit and van de Vis, Jorinde",
    title = "{Bounds on the bubble wall velocity}",
    eprint = "2411.13641",
    archivePrefix = "arXiv",
    primaryClass = "hep-ph",
    reportNumber = "CERN-TH-2024-198, KCL-PH-TH/2024-57",
    doi = "10.1007/JHEP02(2025)119",
    journal = "JHEP",
    volume = "02",
    pages = "119",
    year = "2025"
}

@article{Hindmarsh:2020hop,
    author = {Hindmarsh, Mark B. and L\"uben, Marvin and Lumma, Johannes and Pauly, Martin},
    title = "{Phase transitions in the early universe}",
    eprint = "2008.09136",
    archivePrefix = "arXiv",
    primaryClass = "astro-ph.CO",
    reportNumber = "MPP-2020-163, HIP-2020-27/TH",
    doi = "10.21468/SciPostPhysLectNotes.24",
    journal = "SciPost Phys. Lect. Notes",
    volume = "24",
    pages = "1",
    year = "2021"
}

@article{Ekstedt:2023sqc,
    author = "Ekstedt, Andreas and Gould, Oliver and Hirvonen, Joonas",
    title = "{BubbleDet: a Python package to compute functional determinants for bubble nucleation}",
    eprint = "2308.15652",
    archivePrefix = "arXiv",
    primaryClass = "hep-ph",
    doi = "10.1007/JHEP12(2023)056",
    journal = "JHEP",
    volume = "12",
    pages = "056",
    year = "2023"
}

@article{Branchina:2025jou,
    author = "Branchina, Carlo and Conaci, Angela and Delle Rose, Luigi and De Curtis, Stefania",
    title = "{Electroweak phase transition and bubble wall velocity in local thermal equilibrium}",
    eprint = "2504.21213",
    archivePrefix = "arXiv",
    primaryClass = "hep-ph",
    doi = "10.1103/lkht-1nv1",
    journal = "Phys. Rev. D",
    volume = "112",
    number = "9",
    pages = "095008",
    year = "2025"
}

@article{DeCurtis:2022hlx,
    author = "De Curtis, Stefania and Rose, Luigi Delle and Guiggiani, Andrea and Muyor, {\'A}ngel Gil and Panico, Giuliano",
    title = "{Bubble wall dynamics at the electroweak phase transition}",
    eprint = "2201.08220",
    archivePrefix = "arXiv",
    primaryClass = "hep-ph",
    doi = "10.1007/JHEP03(2022)163",
    journal = "JHEP",
    volume = "03",
    pages = "163",
    year = "2022"
}

@article{DeCurtis:2023hil,
    author = "De Curtis, Stefania and Delle Rose, Luigi and Guiggiani, Andrea and Gil Muyor, {\'A}ngel and Panico, Giuliano",
    title = "{Collision integrals for cosmological phase transitions}",
    eprint = "2303.05846",
    archivePrefix = "arXiv",
    primaryClass = "hep-ph",
    doi = "10.1007/JHEP05(2023)194",
    journal = "JHEP",
    volume = "05",
    pages = "194",
    year = "2023"
}

@article{DeCurtis:2024hvh,
    author = "De Curtis, Stefania and Delle Rose, Luigi and Guiggiani, Andrea and Gil Muyor, {\'A}ngel and Panico, Giuliano",
    title = "{Non-linearities in cosmological bubble wall dynamics}",
    eprint = "2401.13522",
    archivePrefix = "arXiv",
    primaryClass = "hep-ph",
    doi = "10.1007/JHEP05(2024)009",
    journal = "JHEP",
    volume = "05",
    pages = "009",
    year = "2024"
}

@article{Banik:2024ugs,
    author = "Banik, Sumit and Coloretti, Guglielmo and Crivellin, Andreas and Haber, Howard E.",
    title = "{Correlating A{\textrightarrow}{\ensuremath{\gamma}}{\ensuremath{\gamma}} with electric dipole moments in the two Higgs doublet model in light of the diphoton excesses at 95~GeV and 152~GeV}",
    eprint = "2412.00523",
    archivePrefix = "arXiv",
    primaryClass = "hep-ph",
    reportNumber = "PSI-PR-24-24, ZU-TH 58/24",
    doi = "10.1103/PhysRevD.111.075021",
    journal = "Phys. Rev. D",
    volume = "111",
    number = "7",
    pages = "075021",
    year = "2025"
}

@article{Cao:2023gkc,
    author = "Cao, Junjie and Jia, Xinglong and Lian, Jingwei and Meng, Lei",
    title = "{95~GeV diphoton and bb{\textasciimacron} excesses in the general next-to-minimal supersymmetric standard model}",
    eprint = "2310.08436",
    archivePrefix = "arXiv",
    primaryClass = "hep-ph",
    doi = "10.1103/PhysRevD.109.075001",
    journal = "Phys. Rev. D",
    volume = "109",
    number = "7",
    pages = "075001",
    year = "2024"
}

@article{Lian:2024smg,
    author = "Lian, Jingwei",
    title = "{95~GeV excesses in the Z3-symmetric next-to-minimal supersymmetric standard model}",
    eprint = "2406.10969",
    archivePrefix = "arXiv",
    primaryClass = "hep-ph",
    doi = "10.1103/PhysRevD.110.115018",
    journal = "Phys. Rev. D",
    volume = "110",
    number = "11",
    pages = "115018",
    year = "2024"
}

@article{Cao:2024axg,
    author = "Cao, Junjie and Jia, Xinglong and Lian, Jingwei",
    title = "{Unified interpretation of the muon g-2 anomaly, the 95~GeV diphoton, and bb\textasciimacron{} excesses in the general next-to-minimal supersymmetric standard model}",
    eprint = "2402.15847",
    archivePrefix = "arXiv",
    primaryClass = "hep-ph",
    doi = "10.1103/PhysRevD.110.115039",
    journal = "Phys. Rev. D",
    volume = "110",
    number = "11",
    pages = "115039",
    year = "2024"
}

@article{Coutinho:2024zyp,
    author = "Coutinho, Antonio M. and Karan, Anirban and Miralles, V{\'\i}ctor and Pich, Antonio",
    title = "{Light scalars within the $ \mathcal{CP} $-conserving Aligned-two-Higgs-doublet model}",
    eprint = "2412.14906",
    archivePrefix = "arXiv",
    primaryClass = "hep-ph",
    doi = "10.1007/JHEP02(2025)057",
    journal = "JHEP",
    volume = "02",
    pages = "057",
    year = "2025"
}

@article{Ahriche:2007jp,
    author = "Ahriche, Amine",
    title = "{What is the criterion for a strong first order electroweak phase transition in singlet models?}",
    eprint = "hep-ph/0701192",
    archivePrefix = "arXiv",
    doi = "10.1103/PhysRevD.75.083522",
    journal = "Phys. Rev. D",
    volume = "75",
    pages = "083522",
    year = "2007"
}

@article{Chala:2025oul,
    author = "Chala, Mikael and Gil, Luis and Ren, Zhe",
    title = "{Phase transitions in dimensional reduction up to three loops*}",
    eprint = "2505.14335",
    archivePrefix = "arXiv",
    primaryClass = "hep-ph",
    doi = "10.1088/1674-1137/adf322",
    journal = "Chin. Phys.",
    volume = "49",
    number = "12",
    pages = "123105",
    year = "2025"
}

@article{Chakrabortty:2024wto,
    author = "Chakrabortty, Joydeep and Mohanty, Subhendra",
    title = "{One Loop Thermal Effective Action}",
    eprint = "2411.14146",
    archivePrefix = "arXiv",
    primaryClass = "hep-th",
    doi = "10.1016/j.nuclphysb.2025.117165",
    journal = "Nucl. Phys. B",
    volume = "1020",
    pages = "117165",
    year = "2025"
}

@article{Kierkla:2025vwp,
    author = "Kierkla, Maciej and Ramberg, Nicklas and Schicho, Philipp and Schmitt, Daniel",
    title = "{Theoretical uncertainties for primordial black holes from cosmological phase transitions}",
    eprint = "2506.15496",
    archivePrefix = "arXiv",
    primaryClass = "hep-ph",
    reportNumber = "SISSA 07/2025/FISI",
    month = "6",
    year = "2025"
}

@article{langer1973hydrodynamic,
  title         = {{Hydrodynamic model of the condensation of a vapor near its
                  critical point}},
  author        = {Langer, JS and Turski, LA},
  journal       = {Physical Review A},
  volume        = {8},
  number        = {6},
  pages         = {3230},
  year          = {1973},
  publisher     = {APS},
}

@article{Li:2025kyo,
    author = "Li, Xu-Xiang and Ramsey-Musolf, Michael J. and Tenkanen, Tuomas V. I. and Wu, Yanda",
    title = "{An Effective Sphaleron Awakens}",
    eprint = "2506.01585",
    archivePrefix = "arXiv",
    primaryClass = "hep-ph",
    reportNumber = "HIP-2025-8/TH",
    month = "6",
    year = "2025"
}

@article{Muhlleitner:2020wwk,
    author = {M{\"u}hlleitner, Margarete and Sampaio, Marco O. P. and Santos, Rui and Wittbrodt, Jonas},
    title = "{ScannerS: parameter scans in extended scalar sectors}",
    eprint = "2007.02985",
    archivePrefix = "arXiv",
    primaryClass = "hep-ph",
    reportNumber = "KA-TP-05-2020, LU TP 20-38",
    doi = "10.1140/epjc/s10052-022-10139-w",
    journal = "Eur. Phys. J. C",
    volume = "82",
    number = "3",
    pages = "198",
    year = "2022"
}

@article{Muhlleitner:2016mzt,
    author = "Muhlleitner, Margarete and Sampaio, Marco O. P. and Santos, Rui and Wittbrodt, Jonas",
    title = "{The N2HDM under Theoretical and Experimental Scrutiny}",
    eprint = "1612.01309",
    archivePrefix = "arXiv",
    primaryClass = "hep-ph",
    doi = "10.1007/JHEP03(2017)094",
    journal = "JHEP",
    volume = "03",
    pages = "094",
    year = "2017"
}

@article{Sirlin:1980nh,
    author = "Sirlin, A.",
    title = "{Radiative Corrections in the SU(2)-L x U(1) Theory: A Simple Renormalization Framework}",
    reportNumber = "PRINT-80-0267 (IAS,PRINCETON)",
    doi = "10.1103/PhysRevD.22.971",
    journal = "Phys. Rev. D",
    volume = "22",
    pages = "971--981",
    year = "1980"
}

@article{Sirlin:1983ys,
    author = "Sirlin, A.",
    title = "{On the O(alpha**2) Corrections to tau (mu), m (W), m (Z) in the SU(2)-L x U(1) Theory}",
    reportNumber = "RU 82/B/60",
    doi = "10.1103/PhysRevD.29.89",
    journal = "Phys. Rev. D",
    volume = "29",
    pages = "89",
    year = "1984"
}

@article{Bohm:1986rj,
    author = "Bohm, M. and Spiesberger, H. and Hollik, W.",
    title = "{On the One Loop Renormalization of the Electroweak Standard Model and Its Application to Leptonic Processes}",
    doi = "10.1002/prop.19860341102",
    journal = "Fortsch. Phys.",
    volume = "34",
    pages = "687--751",
    year = "1986"
}

@article{Hollik:1988ii,
    author = "Hollik, W. F. L.",
    title = "{Radiative Corrections in the Standard Model and their Role for Precision Tests of the Electroweak Theory}",
    reportNumber = "DESY-88-188",
    doi = "10.1002/prop.2190380302",
    journal = "Fortsch. Phys.",
    volume = "38",
    pages = "165--260",
    year = "1990"
}

@article{Baak:2014ora,
    author = {Baak, M. and C{\'u}th, J. and Haller, J. and Hoecker, A. and Kogler, R. and M{\"o}nig, K. and Schott, M. and Stelzer, J.},
    collaboration = "Gfitter Group",
    title = "{The global electroweak fit at NNLO and prospects for the LHC and ILC}",
    eprint = "1407.3792",
    archivePrefix = "arXiv",
    primaryClass = "hep-ph",
    reportNumber = "DESY-14-124",
    doi = "10.1140/epjc/s10052-014-3046-5",
    journal = "Eur. Phys. J. C",
    volume = "74",
    pages = "3046",
    year = "2014"
}

@article{vandeVis:2025plm,
    author = "van de Vis, Jorinde and Schicho, Philipp and Niemi, Lauri and Laurent, Benoit and Hirvonen, Joonas and Gould, Oliver",
    title = "{WallGo investigates: Theoretical uncertainties in the bubble wall velocity}",
    eprint = "2510.27691",
    archivePrefix = "arXiv",
    primaryClass = "hep-ph",
    reportNumber = "CERN-TH-2025-221",
    month = "10",
    year = "2025"
}

@article{Altarelli:1990zd,
    author = "Altarelli, Guido and Barbieri, Riccardo",
    title = "{Vacuum polarization effects of new physics on electroweak processes}",
    reportNumber = "CERN-TH-5863-90",
    doi = "10.1016/0370-2693(91)91378-9",
    journal = "Phys. Lett. B",
    volume = "253",
    pages = "161--167",
    year = "1991"
}

@article{Peskin:1990zt,
    author = "Peskin, Michael E. and Takeuchi, Tatsu",
    title = "{A New constraint on a strongly interacting Higgs sector}",
    reportNumber = "SLAC-PUB-5272",
    doi = "10.1103/PhysRevLett.65.964",
    journal = "Phys. Rev. Lett.",
    volume = "65",
    pages = "964--967",
    year = "1990"
}

@article{Peskin:1991sw,
    author = "Peskin, Michael E. and Takeuchi, Tatsu",
    title = "{Estimation of oblique electroweak corrections}",
    reportNumber = "SLAC-PUB-5618",
    doi = "10.1103/PhysRevD.46.381",
    journal = "Phys. Rev. D",
    volume = "46",
    pages = "381--409",
    year = "1992"
}

@article{Maksymyk:1993zm,
    author = "Maksymyk, I. and Burgess, C. P. and London, David",
    title = "{Beyond S, T and U}",
    eprint = "hep-ph/9306267",
    archivePrefix = "arXiv",
    reportNumber = "MCGILL-93-13, UDEM-LPN-TH-93-151",
    doi = "10.1103/PhysRevD.50.529",
    journal = "Phys. Rev. D",
    volume = "50",
    pages = "529--535",
    year = "1994"
}

@article{Dolle:2009fn,
    author = "Dolle, Ethan M. and Su, Shufang",
    title = "{The Inert Dark Matter}",
    eprint = "0906.1609",
    archivePrefix = "arXiv",
    primaryClass = "hep-ph",
    doi = "10.1103/PhysRevD.80.055012",
    journal = "Phys. Rev. D",
    volume = "80",
    pages = "055012",
    year = "2009"
}

@article{Goncalves:2021egx,
    author = "Gon{\c{c}}alves, Dorival and Kaladharan, Ajay and Wu, Yongcheng",
    title = "{Electroweak phase transition in the 2HDM: Collider and gravitational wave complementarity}",
    eprint = "2108.05356",
    archivePrefix = "arXiv",
    primaryClass = "hep-ph",
    doi = "10.1103/PhysRevD.105.095041",
    journal = "Phys. Rev. D",
    volume = "105",
    number = "9",
    pages = "095041",
    year = "2022"
}

@article{Su:2020pjw,
    author = "Su, Wei and Williams, Anthony G. and Zhang, Mengchao",
    title = "{Strong first order electroweak phase transition in 2HDM confronting future Z {\&} Higgs factories}",
    eprint = "2011.04540",
    archivePrefix = "arXiv",
    primaryClass = "hep-ph",
    reportNumber = "ADP-20-31/T1141",
    doi = "10.1007/JHEP04(2021)219",
    journal = "JHEP",
    volume = "04",
    pages = "219",
    year = "2021"
}

@article{Gunion:2002zf,
    author = "Gunion, John F. and Haber, Howard E.",
    title = "{The CP conserving two Higgs doublet model: The Approach to the decoupling limit}",
    eprint = "hep-ph/0207010",
    archivePrefix = "arXiv",
    reportNumber = "SCIPP-02-10",
    doi = "10.1103/PhysRevD.67.075019",
    journal = "Phys. Rev. D",
    volume = "67",
    pages = "075019",
    year = "2003"
}

@article{Linde:1981zj,
    author = "Linde, Andrei D.",
    title = "{Decay of the False Vacuum at Finite Temperature}",
    reportNumber = "LEBEDEV-81-265",
    doi = "10.1016/0550-3213(83)90072-X",
    journal = "Nucl. Phys. B",
    volume = "216",
    pages = "421",
    year = "1983",
    note = "[Erratum: Nucl.Phys.B 223, 544 (1983)]"
}

@article{Linde:1980tt,
    author = "Linde, Andrei D.",
    title = "{Fate of the False Vacuum at Finite Temperature: Theory and Applications}",
    reportNumber = "LEBEDEV-80-92",
    doi = "10.1016/0370-2693(81)90281-1",
    journal = "Phys. Lett. B",
    volume = "100",
    pages = "37--40",
    year = "1981"
}

}

\end{document}